\documentclass[sn-mathphys]{sn-jnl}

\jyear{2021}%
\usepackage{graphicx,lineno}
\usepackage{dcolumn}
\usepackage{bm}
\usepackage{hyperref}
\usepackage{xcolor}
\usepackage{nicefrac}
\newcommand{\bseq}{\begin{subequations}}
\newcommand{\eseq}{\end{subequations}}
\newcommand{\beq}{\begin{equation}}
\newcommand{\eeq}{\end{equation}}
\newcommand{\Lim}[1]{\raisebox{0.5ex}{\scalebox{0.8}{$\displaystyle \lim_{#1}\;$}}}
\theoremstyle{thmstyleone}%
\theoremstyle{thmstyletwo}%

\theoremstyle{thmstylethree}%

\begin{document}
\title[Article Title]{Spatiotemporal linear stability of viscoelastic subdiffusive channel flows: a fractional calculus
framework}


\author[1]{\fnm{Tanisha} \sur{Chauhan}}\email{tanishai@iiitd.ac.in}
\equalcont{These authors contributed equally to this work.}

\author[1]{\fnm{Diksha} \sur{Bansal}}\email{dikshab@iiitd.ac.in}
\equalcont{These authors contributed equally to this work.}

\author*[1]{\fnm{Sarthok} \sur{Sircar}}\email{sarthok@iiitd.ac.in}

\affil*[1]{\orgdiv{Dept. Mathematics}, \orgname{IIIT Delhi}, \orgaddress{\street{Okhla Phase III}, \city{New Delhi}, \postcode{110020}, \state{Delhi}, \country{India}}}




\abstract{The temporal and spatiotemporal linear stability analyses of viscoelastic, subdiffusive, plane Poiseuille and Couette flows obeying the Fractional Upper Convected Maxwell (FUCM) equation in the limit of low to moderate Reynolds number ($Re$) and Weissenberg number ($We$), is reported to identify the regions of topological transition of the advancing flow interface. In particular, we demonstrate how the exponent in the subdiffusive power{\color{black}-}law scaling ($t^\alpha$, with $0 {\color{black}   < } \alpha \le 1$) of the mean square displacement of the tracer particle, in the microscale [Mason and Weitz, Phys. Rev. Lett. {\bf 74}, 1250-1253 (1995)] is related to the fractional order of the derivative, $\alpha$, of the corresponding non-linear stress constitutive equation in the continuum. The stability studies are limited to two exponents: monomer diffusion in Rouse chain melts, $\alpha=\nicefrac{1}{2}$, and in Zimm chain solutions, $\alpha=\nicefrac{2}{3}$. The temporal stability analysis indicates that with decreasing order of the fractional derivative: (a) the most unstable mode decreases, (b) the peak of the most unstable mode shifts to lower values of $Re$, and (c) the peak of the most unstable mode, for the Rouse model precipitates towards the limit $Re \rightarrow 0$. The Briggs idea of analytic continuation is deployed to classify regions of temporal stability, absolute and convective instabilities and evanescent modes. The spatiotemporal phase diagram indicates an abnormal region of temporal stability at high fluid inertia, revealing the presence of a non-homogeneous environment with hindered flow, thus highlighting the potential of the model to effectively capture certain experimentally observed, flow-instability transition in subdiffusive flows.}

\keywords{Spatiotemporal stability, anomalous diffusion, non-Markovian processes, Caputo integral, Upper Convected Maxwell model}



\maketitle

\section{Introduction}\label{sec:intro}
The subject of anomalous diffusion has received tremendous attention over the last half{\color{black}-}century{\color{black},} ranging from physics~\cite{Goychuk2017,Goychuk2020,Goychuk2021}, biology~\cite{Lai2009} to quantitative finance~\cite{Coffey2004}. Some of the most significant and profoundly published experimental results are better rationalized within the viscoelastic subdiffusive approach in random environments such as the cytosol and the plasma membrane of biological cells~\cite{Rubenstein2003}, crowded complex fluids and polymer solutions~\cite{Levine2001}, dense colloidal suspensions~\cite{Kremer1990} and single-file diffusion in colloidal systems~\cite{Kou2004}. The observed (anomalous) subdiffusion often combines features of ergodic fractional Brownian motion (reflecting viscoelasticity) and the nonergodic jumplike non-Markovian diffusional processes (reflecting disorder)~\cite{Fricks2009, Morgado2002}. The subdiffusive object is considered primarily as being elastic and structurally robust, although it requires `fluidity' and flexibility besides its elasticity for a proper functioning, e.~g., consider a viscoelastic nanoscaled polymer drop armed with a rigid backbone that can take on different macroscopic conformations~\cite{Vainstein2008}. In this article, we demonstrate how the dynamics of the subdiffusive fluids at microscale (often represented via a generalized Langevin equation (GLE) at the molecular level, using a dissipative memory kernel) is `upscaled' to a fractional viscoelastic stress constitutive equation at the continuum level~\cite{Adelman1976}. 

Fractional calculus serves as a powerful tool for modeling the constitutive relations in the linear~\cite{gemant1936} as well as nonlinear viscoelasticity theory~\cite{Kremer1990} and to explain certain paradoxical experimental findings~\cite{Goychuk2020,Goychuk2021} (one such experimentally abnormal feature captured by our model, namely the occurrence of temporal stability at high fluid inertia, is mentioned in Section~\ref{subsec:STSA}). Gemant highlighted the relaxation curves for some viscoelastic fluids by employing a fractional viscoelastic model for the first time~\cite{gemant1938}. Scott-Blair developed a new constitutive law (known as the `fractional Newton model') to describe the experimental outcome of Gemant on stress relaxation~\cite{Blair1944,Blair1947}. Caputo introduced the fractional Voigt model to simulate the dissipation in seismology~\cite{caputo1967}. Bagley and Torvik~\cite{bagley1983} showed that there exists a quantitative connection between the fractional viscoelastic model (at the macroscopic level) and the molecular theory of Rouse's polymer chain melts~\cite{Rouse1953}. With the development of the fractional viscoelastic model, the flow of the fractional viscoelastic fluid has been extensively investigated~\cite{Tan2002,Qi2009,Fetecau2009,Zheng2012,Zhao2016}. Tan and Xu~\cite{Tan2002} used Laplace transforms to obtain the analytical solution for velocity and stress of the plane surface flow of a fractional Maxwell fluid. Qi and Xu~\cite{Qi2009} studied the plane Poiseuille flow and plane Couette flow of a generalized Oldroyd-B fluid with fractional derivative. Zheng et al.~\cite{Zheng2012} found the analytical solution for velocity and stress of magnetohydrodynamic flow of a generalized Oldroyd-B fluid generated with an accelerating plate, with fractional derivatives. {\color{black}R}ecently, Zhao et al.~\cite{Zhao2016} considered the natural convection heat transfer of viscoelastic fluid with fractional {\color{black}the} Maxwell model over a vertical plate. More recent applications of the fractional viscoelastic flows include the study of the stability of coastal morphodynamics and seafloor topology~\cite{Ancey2019}, regulation of the tissue morphodynamics~\cite{Siedlik2015} and capturing spatiotemporal disorder in anomalous transport of viscous flows~\cite{Zaks2018}.

The detailed exploration of the existing literature serves as a clear motivation for the work reported here, which is to provide a comprehensive picture of the stability of the two-dimensional, viscoelastic, subdiffusive, fully developed, Poiseuille and Couette flows. The present work significantly differs from the existing studies in the sense that we analyse the linear stability of viscoelastic, subdiffusive, channel flows through a combined temporal and spatiotemporal stability analysis (rather than only a temporal stability analysis of the classical (or integer order) viscoelastic channel flows~\cite{Khalid2021}) and the aim is to address the following intriguing questions: What is the critical flow/polymer relaxation condition for the onset of instability? {\color{black}A}nd more crucially, what is the linear spatiotemporal, time asymptotic response of the flow at the critical value of the material parameters, leading to the topological transition of the advancing flow interface of the subdiffusive channel flows?

While the molecular theory of polymer dynamics has already established the correspondence between subdiffusive dynamics and linear viscoelastic relaxation of polymer melts and solutions.~\cite{zwanzig1970,Mason1995}, we `upscale' these ideas at the continuum mechanical scale. In particular, we highlight how the exponent in the subdiffusive power{\color{black}-}law timescale, $t^\alpha$~\cite{Mason1996}, is related to the fractional order, $\alpha$, of the corresponding non-linear stress constitutive equations in the continuum (refer Section~\ref{subsec:MM}). 
The temporal and spatiotemporal stability of two specific cases of monomer diffusion in Rouse chain melts ($\alpha = \nicefrac{1}{2}$)~\cite{Rouse1953}, and in Zimm chain solution ($\alpha = \nicefrac{2}{3}$)~\cite{zimm1956} are reported in detail (Section~\ref{subsec:TSA}, Section~\ref{subsec:STSA}). 
The Rouse model predicts that the viscoelastic properties of the polymer chain can be described by a generalized Maxwell model, where the elasticity is governed by a single relaxation time, which is independent of the number of Maxwell elements (or the so{\color{black}-}called `submolecules'). In contrast, the Zimm's model predicts the (`shear rate and polymer concentration independent') viscosity of the polymer solution by calculating the hydrodynamic interaction of flexible polymers (an idea which was originally proposed by Kirkwood~\cite{Kirkwood1954}) by approximating the chains using a bead-spring setup.

\section{Problem formulation: Mathematical model, linear stability analysis and numerical method}\label{sec:math}
In this study, the linearized stability analyses of the fully developed, planar Poiseuille and Couette flows inside an infinitely long channel of width $H$ (i.~e., $0 \le y \le H$ such that $\nicefrac{x}{y} \gg 1$, where $x$ and $y$ are the flow and the shear gradient direction, respectively) is reported.  

\subsection{Mathematical model}\label{subsec:MM}
We consider a viscoelastic fluid subject to a shear deformation. Then, an infinitesimal elastic stress, {\color{black}$\tau_{xy}$} at time $t$ arising from a small strain increment $d \gamma$ at an earlier time $t'$ is given by,
\beq
{\color{black} d \tau_{xy}} = G(t - t') \dot{\gamma}(t') dt',
\label{eqn:Step1} 
\eeq
where the relaxation modulus, $G(t)$, represents the influence of the dissipative processes of the surrounding concentrated fluid medium~\cite{Makris2021}. Assuming linearity, the Boltzmann superposition principle may be utilized to construct the {\color{black} elastic} stress at time $t$ by summing up all of the infinitesimal contributions over the entire flow history, {\color{black} which is extended into the infinite past}~\cite{Brader2010},
\beq
{\color{black} \tau_{xy}} = \int^t_{{\color{black} -\infty}} G(t - t') \dot{\gamma}(t') dt'.
\label{eqn:Step2} 
\eeq

In their seminal work on passive micro-rheology, Mason and co-workers~\cite{Mason1995} have identified an approximate relation between the time-dependent memory kernel describing the viscous damping of the tracer particle at micro-scale (and which obeys the GLE, e.~g., see equation~(15) in~\cite{Mason1996}), $\zeta(t)$, and the stress relaxation modulus, $G(t)$, i.~e.,
\beq
\zeta(t) = 6 \pi a G(t),
\label{eqn:Step3} 
\eeq
where $a$ is the radius of the tracer particle (assumed spherical). In the regime of linear viscoelasticity, one of the most commonly used three-parameter family of memory kernel is the generalized Rouse kernel for an equally weighted sum of negatively decaying exponential functions~\cite{Mckinley2009},
\beq
\zeta(t) = \frac{1}{N} \sum^{N-1}_{k=0} e^{-(\frac{k}{N})^{\color{black}\frac{1}{\alpha}} (\frac{t}{\lambda_0})},
\label{eqn:Step4} 
\eeq
for a number of kernels determining the length of the subdiffusive phase, $N$, relaxation time, $\lambda_0$, and a subdiffusive exponent, $\alpha \in {\color{black}(}0 \,\, 1]$. Since the polymeric liquids of our interest~\cite{Sircar2010,Sircar2010eLC,Sircar2010IJEFMS,Sircar2015,Sircar2015JTB,Sircar2016JMB,Sircar2016EPJE,Sircar2019,Sircar2020} show subdiffusive behavior on all length scales, we consider the case when $N \rightarrow \infty$ in the prony series~\eqref{eqn:Step4}. For $t > \lambda_0$, this limiting behavior leads to the relation
\beq
\zeta(t) = \frac{\tilde{G}}{\Gamma(1-\alpha)} \left( \frac{t}{\lambda_0} \right)^{-\alpha},
\label{eqn:Step5} 
\eeq
where $\Gamma(x)$ is the complete gamma function and $\tilde{G} = \Gamma(1+\alpha) \Gamma(1-\alpha)$, is a constant. In equation~\eqref{eqn:Step5}, we have used the fact that the Riemann sum on an infinite interval,
\beq
\Lim{N \rightarrow \infty} \frac{1}{N} \sum^{N-1}_{k = 0} f{\color{black}\left({\color{black}\frac{k}{N}}\right)} = \int_0^\infty f(x) d x.
\label{eqn:Step5a}
\eeq
Using equations~(\ref{eqn:Step2}, \ref{eqn:Step3}, \ref{eqn:Step5}), one arrives at,
\beq
{\color{black} \tau_{xy}} = \frac{\mathcal{G}\lambda^\alpha_0}{\Gamma(1-\alpha)} \int^t_{{\color{black} -\infty}} dt' (t - t')^{-\alpha} \frac{d \gamma(t')}{dt'},
\label{eqn:Step6}
\eeq
where the constant, $\mathcal{G} = \frac{\tilde{G}}{6\pi a}$. We remark that $\mathcal{G}$ is no longer a constant (typically $\mathcal{G} = \mathcal{G}(t)$) when the concentration effects, such as the bond and entanglement effects, are considered~\cite{Kremer1990}. The right-hand side of equation~\eqref{eqn:Step6} represents a fractional integral corresponding to the {\color{black}Caputo} formalism~\cite{Glockle1991,Glockle1994},
\beq
_{-\infty}{D^{-\beta}_t} f(t) = \frac{1}{\Gamma(\beta)} \int^t_{{\color{black} -\infty}} \frac{dt'}{(t - t')^{1-\beta}} {\color{black}\frac{d f(t')}{d t'}}.
\label{eqn:Step7}
\eeq
Utilizing equations~(\ref{eqn:Step6}, \ref{eqn:Step7}), we arrive at the basic equation governing stress-strain relation in linear viscoelastic subdiffusive media,
\beq
{\color{black} \tau_{xy}} = \mathcal{G} \lambda^\alpha_0 \frac{d^{\alpha-1}}{d t^{\alpha-1}} \frac{d \gamma(t)}{d t} = \mathcal{G} \lambda^\alpha_0 \frac{d^\alpha \gamma}{d t^\alpha},
\label{eqn:Step8}
\eeq
including the limiting cases of a purely elastic solid ($\alpha {\color{black} \rightarrow} 0$ or a Hookean spring) and a purely viscous fluid ($\alpha = 1$ or a dashpot)~\cite{Blair1944}. Through combinations of springs and dashpots, one arrives at standard linear viscoelastic models, including the Maxwell, Kelvin-Voigt, Zener, Poynting-Thomson and Burgers' model and others~\cite{Blair1947}. The problem is that the corresponding differential equations have a relatively restricted class of solutions, which are too limited to provide an adequate description for the class of complex fluids discussed in Section~\ref{sec:intro}. To overcome this shortcoming, one can relate the stress and strain through the fractional equation~\eqref{eqn:Step8}, which allows a smooth interpolation between a purely elastic behavior and a purely viscous pattern. In the present analysis, we have selected the Fractional Upper Convected Maxwell equation (FUCM) to describe the nonlinear viscoelastic response of the subdiffusive media, derived next.

\begin{figure}
\centering
\includegraphics[width=0.75\linewidth, height=0.55\linewidth]{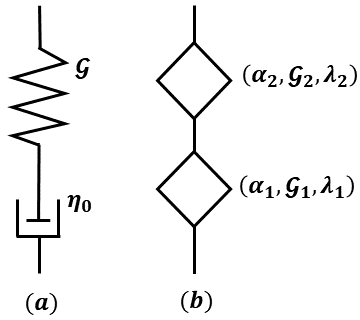}
\caption{(a) The Maxwell element and (b) its fractional generalization.}
\label{fig1}
\end{figure}
Figure~\ref{fig1}a depicts the standard Maxwell model in which a spring and a dashpot are connected in series~\cite{Makris2021}. We generalize this model by replacing these elements with their corresponding fractional elements: $(\alpha_i, \mathcal{G}_i, \lambda_i), \,\, i = 1, 2$ (figure~\ref{fig1}b). Because of the sequential construction, {\color{black}the stress, $\tau$,} is the same for both elements and their respective stress-strain relations are
\beq
\gamma_i = \mathcal{G}^{-1}_i \lambda_i^{-\alpha_i} \frac{d^{-\alpha_i} {\color{black} \tau_{xy}}}{d t^{-\alpha_i}}, \quad \text{i = 1, 2},
\label{eqn:Step9}
\eeq
where both expressions follow from equation~\eqref{eqn:Step8}. Due to the construction of the generalized Maxwell model, we have $\gamma(t) = \gamma_1(t) + \gamma_2(t)$, from which it follows,
\beq
{\color{black} \tau_{xy}} + \frac{\mathcal{G}_1 \lambda^{\alpha_1}_1}{\mathcal{G}_2 \lambda^{\alpha_2}_2} \frac{d^{\alpha_1-\alpha_2} {\color{black} \tau_{xy}}}{d t^{\alpha_1-\alpha_2}} = \mathcal{G}_1 \lambda^{\alpha_1}_1 \frac{d^{\alpha_1} \gamma}{d t^{\alpha_1}}.
\label{eqn:Step10}
\eeq
Equation~\eqref{eqn:Step10} can be simplified by setting $\lambda = (\mathcal{G}_1 \lambda^{\alpha_1}_1 / \mathcal{G}_2 \lambda^{\alpha_2}_2)^{1/(\alpha_1-\alpha_2)}$ and $E = \mathcal{G}_1 (\lambda_1/\lambda)^{\alpha_1}$. Without loss of generality, we assume $\alpha_2 = 0$ and $\alpha = \alpha_1 (> 0)$, and arrive at
\beq
{\color{black} \tau_{xy}} + \lambda^\alpha \frac{d^\alpha {\color{black} \tau_{xy}}}{d t^\alpha} = {\color{black} \eta_p} \frac{d^\alpha \gamma}{d t^\alpha},
\label{eqn:Step11}
\eeq
where the constant, ${\color{black} \eta_p} = E \lambda^\alpha$. We can extend equation~\eqref{eqn:Step11} to three dimensions by replacing the {\color{black}elastic} stress, ${\color{black} \tau_{xy}}$, with the stress tensor, ${\color{black} {\bf \tau}}$, and the derivative, $\frac{d^\alpha \gamma}{d t^\alpha}$, with the rate of strain tensor, ${\color{black} {\bf D} = (\nabla {\bf v} + (\nabla {\bf v})^T)}$ (where the operator $\nabla (\cdot) = \frac{\partial}{\partial {\bf x}} (\cdot)$), to arrive at
\beq
{\color{black} {\bf \tau}} + \lambda^\alpha \frac{d^\alpha {\color{black} {\bf \tau}}}{d t^\alpha} = {\color{black} \eta_p {\bf D}},
\label{eqn:Step11a}
\eeq
using the definition of {\color{black} fractional} velocity, ${\bf v} = \frac{d^\alpha {\bf x}}{d t^\alpha}${\color{black}~\cite{Prodanov2016}, which has a dimension of $\frac{H}{T^\alpha}$ (refer Section~\ref{subsec:LSA} for the discussion on non-dimensionalization). Fractional
velocities are defined as limits of the difference quotients of a fractional power and they generalize the notion of a local derivative~\cite{Prodanov2017}. These derivatives are frequently used, for example, to model instantaneous interactions in Langevin dynamics~\cite{Prodanov2018}.} 

Equation~\eqref{eqn:Step11a} is the rheological constitutive equation of the fractional Maxwell model describing the linear viscoelastic media. The simplest way to combine rheological nonlinearity is to replace the (fractional) material time derivative in equation~\eqref{eqn:Step11a} with the (fractional) frame invariant, upper-convected time derivative~\cite{Macosko1994,Spagnolie2015}, which leads us to FUCM, as follows,
\beq
{\color{black}{\bf \tau} + \lambda^\alpha \overset{\triangledown}{{\bf \tau}} =\eta_p {\bf D}},
\label{eqn:FUCM}
\eeq
where the fractional upper-convected time derivative of the tensor ${\color{black} {\bf \tau}}$ is defined as,
\beq
{\color{black} \overset{\triangledown}{{\bf \tau}}} = \frac{\partial^\alpha {\color{black} {\bf \tau}}}{\partial t^\alpha} + {\bf v}\cdot \nabla {\color{black} {\bf \tau}} - (\nabla {\bf v})^T {\color{black} {\bf \tau}} -{\color{black} {\bf \tau}}\nabla {\bf v}.
\label{eqn:FUCTD}
\eeq
The fractional time derivative, $\frac{\partial^\alpha }{\partial t^\alpha}$, in equation~(\ref{eqn:FUCM}, \ref{eqn:FUCTD}) is based on the {\color{black}Caputo} definition~\eqref{eqn:Step7}.

The continuity and the momentum equations for an incompressible, subdiffusive flow (consistent with the stress constitutive relation~\eqref{eqn:FUCM}) are,
\beq
\nabla\cdot {\bf v} = 0, \qquad \rho \left[ \frac{\partial^\alpha {\bf v}}{\partial t^\alpha} + {\bf v} \cdot \nabla {\bf v} \right] = -\nabla p  {\color{black} + \eta_s \nabla \cdot {\bf D}} + \nabla \cdot \tau,  
\label{eqn:MainModel} 
\eeq
where $\rho$ is the density {\color{black}and} $p$ is the isotropic pressure. Equations~(\ref{eqn:FUCM},\ref{eqn:MainModel}) represent the equations of motion describing the flow-instability of the subdiffusive viscoelastic fluids.

As a result of the dissipative processes, viscoelastic materials have memory, that is, their actual mechanical response is modulated by the past~\cite{Jimenez2002}. The fractional derivative operators account for the complete history to obtain the derivative at an instant. Unlike the classical Maxwell model~\cite{Sircar2019,Bansal2021} which account{\color{black}s} for only the elastic (or stored) part of the deformation work, the fractional Maxwell model accounts for both forms (stored and dissipated) of energy at any time point. Although Mckinley pointed out that the fractional Maxwell model generally cannot capture polymer shear-thinning~\cite{Jaishankar2014}, the fractional version provides a better fit of the relaxation and creep behavior for a significantly large class of viscoelastic materials using fewer parameters than the classical version~\cite{Jimenez2002}.

\subsection{Linear stability analysis}\label{subsec:LSA}
Using the following scales for non-dimensionalizing the governing equations: the height of the channel $H$ for {\color{black}length}, the timescale $T$ corresponding to maximum base flow velocity (i.~e., $T = (H / \mathcal{U}_0)^{1/\alpha}$) for time and $\rho \mathcal{U}^2_0$ for {\color{black}pressure} and stresses, we characterize equations~(\ref{eqn:FUCM}, \ref{eqn:MainModel}), rephrased as follows,
\bseq \label{eqn:FullSystem}
\begin{align}
&\nabla \cdot {\bf v} = 0, \label{eqn:Continuity} \\
&Re \left[ \frac{\partial^\alpha {\bf v}}{\partial t^\alpha} + {\bf v} \cdot \nabla {\bf v} \right] = -\nabla p + \nu \nabla \cdot {\bf D} + (1-\nu) \nabla \cdot {\bf A}, \label{eqn:Momentum} \\
&\frac{\partial^\alpha {\bf A}}{\partial t^\alpha} + {\bf v}\cdot \nabla {\bf A} - (\nabla {\bf v})^T {\bf A} -{\bf A}\nabla {\bf v} = \frac{{\bf D} - {\bf A}}{We}, \label{eqn:ExtraStress}
\end{align}
\eseq
using the dimensionless groups, $Re = \frac{\rho \mathcal{U}_0 H}{\eta_0}$ (Reynolds number), $We = \frac{\lambda^\alpha \mathcal{U}_0}{H}$ (Weissenberg number) {\color{black}and where $\eta_s, \eta_p$, $\eta_0 (= \eta_s + \eta_p)$ and $\nu (= \eta_s/\eta_0)$ are the solvent viscosity, the polymeric contribution to the shear viscosity, the total viscosity and the viscous contribution to the total viscosity of the fluid, respectively. In equation~\eqref{eqn:FullSystem}, the elastic stress is represented as $\tau = (1-\nu){\bf A}$.} The current analysis deploys fractional derivative of exponentials~\cite{Glockle1991,Glockle1994} given as, 
\beq 
\frac{d^\alpha (e^{i a t})}{d t^\alpha}  = (i a)^\alpha e^{i a t},
\label{eqn:LSD}
\eeq
Let us denote the mean flow variables with capital letters and with a subscript `0'. We assume that the mean flow is two-dimensional, quasiparallel with its variation entirely in the shear gradient direction. Then, the (non-dimensional) velocity can be written as follows,
\beq
{\bf U}_0 = \left( (y - y^2) + \delta y \right) {\bf e_x},
\label{eqn:MV}
\eeq
where ${\bf e_x}$ is the unit vector along the x-direction. Flow-instability studies of two specific forms of channel flows are considered in this article: plane Poiseuille flow ($\delta=0$) and the plane Couette flow ($\delta=1.0$). The other mean flow variables satisfying equation~\eqref{eqn:FullSystem}, including the mean pressure, $P_0$, and the base state elastic stress tensor, ${\bf A}_0 = [A_{0_{i j}}]$, is given by,
\bseq \label{eqn:NormalMode}
\begin{align}
&P_0  = - {\color{black}2} x  - {\color{black} 8} We (1-\nu)\left(y-y^2+\delta y \right), \label{eqn:P} \\
&A_{0_{11}} = 0, \label{eqn:A11} \\
&A_{0_{12}} = A_{0_{21}} = {\color{black}\left(1+ \delta - 2 y\right)}, \label{eqn:A12} \\
&A_{0_{22}} ={\color{black} 2} We \left( 1-2y + \delta \right)^2, \label{eqn:A22}
\end{align}
\eseq
and whose linearized stability analysis is presented next.

The viscoelastic version of the Squire's theorem for plane parallel, classical Oldroyd-B fluids~\cite{Bistagnino2007} indicate{\color{black}s} that it is possible to restrict our stability analysis to the case when the disturbances are two-dimensional. Assuming an independent fate of each wavenumber, $k$ (whose real part is chosen to be positive) and frequency, $\omega$, it is natural to consider disturbances in the form of a normal mode expansion, such that the total velocity, pressure and stress are expressed in terms of their mean values and perturbations amplitudes (denoted by $\mathring{(\cdot)}$), as follows,
\beq
{\bf r} = {\bf R}_0 + \epsilon \mathring{r} e^{i (k x - \omega t)}
\label{eqn:NormalMode}
\eeq
where $\epsilon \ll 1$ and ${\bf r} = [{\bf v} \, p \, A_{11} \, A_{12} \, A_{22}]^T$, ${\bf R}_0 = [{\bf U}_0 \, P_0 \, A_{0_{11}} \, A_{0_{12}} \, A_{0_{22}}]^T$ and $\mathring{r} =[X_0y(1-y) \, X_1y(1-y) \, X_2 \, X_3 \, X_4 \, X_5]^T$ represent the total, the mean flow variables and the disturbance amplitudes, respectively. The disturbance amplitude, $\mathring{r}$, is chosen such that it satisfies the no-slip condition on the channel walls. Substituting the solution form~\eqref{eqn:NormalMode} in equations~(\ref{eqn:Continuity}-\ref{eqn:ExtraStress}) and retaining the $\mathcal{O}(\epsilon)$ terms to arrive at the linearized equation governing conservation of mass,
\beq
X_0 \left[i k (y-y^2)\right] + X_1 \left[ 1 - 2y\right] = 0,
\label{eqn:L1}
\eeq
the linearized equation describing the conservation of momentum in the $x-$direction,
\begin{align}
& X_0 \left[ (y-y^2) \left(Re  (-i \omega)^\alpha  + i k Re (y-y^2 + \delta y)  - {\color{black} 2} \nu(i k)^{2}\right) +
\right. \nonumber \\
& {\color{black}2 }\left. \nu  \right]   + X_1 \left[ Re ( 1-2y+\delta) (y-y^2) - {\color{black} \nu} (1-2y) (i k) \right] +  i k 
\nonumber \\
& X_2 - i k (1-\nu) X_3 = 0,
\label{eqn:L2}
\end{align}
and the one governing the conservation of momentum in the $y-$direction,
\begin{align}
&X_0 \left[ - {\color{black} i k \nu}  \left(1-2y\right) \right] + X_1 \left[(y-y^2)\left( Re  (-i \omega)^\alpha + i k Re 
\right. \right. \nonumber \\
& \left. \left. (y-y^2 + \delta y) - {\color{black} \nu}  (ik)^{2}\right) + {\color{black} 4} \nu \right] - (1-\nu)i k  X_4  =0.
\label{eqn:L3}
\end{align}
The linearized equation for the elastic stress component $A_{11}$,
\begin{align}
&X_0 \left[ \frac{-{\color{black} 2}}{We} (y-y^2) (ik)\right] + X_1 \left[- {\color{black} 2} i k \left(1-2y+\delta \right) (y-y^2) \right] +
\nonumber \\
& X_3 \left[(-i \omega)^\alpha  + i k (y-y^2+\delta y) + \frac{1}{We}\right] = 0,
\label{eqn:L4}
\end{align}
for the component $A_{12}$ (or $A_{21}$),
\begin{align}
&X_0 \left[ {\color{black} -i k} \left(1-2y+\delta \right)(y-y^2) -{\color{black} \frac{1}{ We}} \left( 1-2y \right)\right] + X_1
\nonumber \\
& \left[- {\color{black} 2} (y-y^2) - {\color{black} 2} i k We (1-2y+\delta)^2 (y-y^2) - {\color{black} \left(1-2y+\delta \right)}
\right. \nonumber \\
& \left. (1-2y) - {\color{black}\frac{1}{ We}} (y-y^2)(i k ) \right] + X_3 \left[ -(1-2y+\delta)\right] + X_4 
\nonumber \\
& \left[(-i \omega)^\alpha  + i k (y-y^2+\delta y) + \frac{1}{We}\right] = 0,
\label{eqn:L5}
\end{align}
and for the component $A^{22}$,
\begin{align}
&X_0 \left[{\color{black} 2}\left(1 - 2y + \delta \right) (2y-1) \right] \!+\! X_1 \left[ - {\color{black} 8} We \left(1-2y+\delta \right) (y\!-\!y^2) 
\right. \nonumber \\
& \left. - {\color{black} 4} We (1-2y + \delta)^2 (1-2y) -\frac{{\color{black} 2} (1-2y)}{We} \right] + X_4 \left[-2(1-2y
\right. \nonumber \\
& \left.
+\delta)\right] + X_5 \left[(-i \omega)^\alpha  + i k (y-y^2+\delta y) + \frac{1}{We}\right] = 0.
\label{eqn:L6}
\end{align}
Equations~(\ref{eqn:L1}-\ref{eqn:L6}) may be written in a matrix-vector format as follows,
\beq
\begin{bmatrix}
i k (y\!\!-\!\!y^2) & 1\!\!-\!\!2y &0 &0 &0 &0\\[1.1ex]
M_1 & M_2 & i k & -i k(1\!\!-\!\!\nu)& 0 &0\\[1.1ex]
{\color{black}\nu i k} (2y\!\!-\!\!1) & M_3 & 0 & 0 & -i k(1\!\!-\!\!\nu) &0  \\[1.1ex]
\frac{{\color{black} 2} i k}{We}(y^2\!\!-\!\!y) & {\color{black} 2}i k (1\!\!-\!\!2y\!\!+\!\!\delta)(y^2\!\!-\!\!y) & 0 & M_4 & 0& 0\\[1.1ex]
M_5 & M_6 & 0 & -(1\!\!-\!\!2y\!\!+\!\!\delta) & M_4 & 0\\[1.1ex]
{\color{black} 2} (1\!\!-\!\!2y\!\!+\!\!\delta)(2y\!\!-\!\!1) & M_7  & 0 & 0& -2(1\!\!-\!\!2y\!\!+\!\!\delta) & M_4
\end{bmatrix}
\begin{bmatrix}
X_0\\[1.1ex]
X_1\\[1.1ex]
X_2\\[1.1ex]
X_3\\[1.1ex]
X_4\\[1.1ex]
X_5
\end{bmatrix} = \begin{bmatrix}
0\\[1.1ex]
0\\[1.1ex]
0\\[1.1ex]
0\\[1.1ex]
0\\[1.1ex]
0
\end{bmatrix},
\label{eqn:DRPMatrix}
\eeq
where the expressions $M_i\,\, (i = 1, \ldots 7)$ are listed in Section~\ref{sec:appA}. A nontrivial solution for the system~\eqref{eqn:DRPMatrix}, imposes a zero determinant condition on the coefficient matrix which leads to the dispersion relation, $D(k, \omega) = 0$, given by,
\begin{align}
&{\color{black} \dfrac{1}{We} k^2 M_4 \left(2 i k \left(-1 + \nu\right) \left(1 + \delta - 2 y\right) \left(-1 + y\right) y \left(1 + \left(-2
\right. \right.\right. }\nonumber \\
& {\color{black} \left. \left. \left.
+ i k We \left(1 + \delta - 2 y\right) \left(-1 + y\right)\right) y\right) + M_4^2 We \left(\nu \left(1 - 2 y\right)^2
\right. \right. } \nonumber \\
& {\color{black} \left. \left.
- M_3 \left(-1 + y\right) y\right) + M_4 \left(-1 + \nu\right) We \left(M_5 - 2 M_5 y + i k
\right. \right.} \nonumber \\
& {\color{black} \left. \left.
M_6 \left(-1 + y\right) y\right)\right) = 0.}
\label{eqn:DRP}
\end{align}

\subsection{Numerical method}\label{subsec:NM}
In the ensuing description, we denote real/imaginary components with subscript {\it r}/{\it i}, respectively. The zeros of the dispersion relation (equation~\eqref{eqn:DRP}) were explored within the complex $k - \omega$ plane inside the region $\omega_r \in [-1700, \, 0.2]$, $\omega_i \in [-5000, \, 100]$, $k_r \in [0, \, 5]$ and $k_i \in [-0.1, \, 0.2]$. For a real wavenumber $k$, the procedure for finding the most unstable mode (which is the largest positive imaginary component of any root of the dispersion relation or the temporal growth rate, $\omega_{\text{Temp}}$, refer Section~\ref{subsec:TSA}), consists of detecting the admissible saddle points ($\omega \in \mathbb{C}, k \in \mathbb{R}$) satisfying the equations~\cite{Huerre1990},
\bseq \label{eqn:wTemp}
\begin{align}
&D(k, \omega) = 0, \label{eqn:wTemp1} \\
&\frac{\partial \omega_i}{\partial k} = \frac{\nicefrac{\partial D}{\partial k}}{\nicefrac{\partial D}{\partial \omega_i}} = 0, \label{eqn:wTemp2}
\end{align}
\eseq
and then (among all the possible roots of equation~\eqref{eqn:wTemp}) identifying those roots with the largest positive imaginary component of the frequency. Equation~\eqref{eqn:wTemp} is solved using a multivariate Newton-Raphson algorithm (refer Author's previously published results~\cite{Sircar2019,Bansal2021} for a detailed outline of this method).

Next, in the spatiotemporal analysis, eigenpairs with complex wavenumbers and frequencies are permitted in the solution of equation~\eqref{eqn:wTemp}. The necessary (but not sufficient) condition for the presence of absolute instability is the vanishing characteristic of the group velocity of the flow, ${\it v}_g$, at the saddle point in the $k$-plane or the branch point in the $\omega$-plane (${\it v}_g = \frac{\partial \omega}{\partial k} = \nicefrac{(\frac{\partial D}{\partial k})}{(\frac{\partial D}{\partial \omega})} = 0$, such that $\omega=D(k)$ satisfies the dispersion relation). But the group velocity is zero at every saddle point, in particular where the two $k$-branches meet, independent of whether the branches originate from the same half of the $k$-plane (i.~e., when evanescent modes are detected) or not. To overcome this inadequacy, Briggs~\cite{Briggs1964} devised the idea of analytic continuation in which the Laplace contour is deformed towards the $\omega_r$ axis of the complex $\omega$-plane, with the simultaneous adjustment of the Fourier contour in the $k$-plane to maintain the separation of the $k$-branches; those which originate from the top half (the upstream modes with $k_i > 0$) from those which originate from the bottom half of the $k$-plane (or the downstream modes). The deformation of the Fourier contour (while preserving causality) is inhibited, however, when the paths of the two $k$-branches originating from the opposite halves of the $k$-plane intersect each other, leading to the appearance of saddle points which are the {\it pinch point}, $k^{\text pinch}$. The concurrent branch point appearance in the $\omega$-plane is the {\it cusp point}, $\omega^{\text cusp}$ (i. e., $D(k^{\text pinch}, \omega^{\text cusp})\!=\!\frac{\partial D(k^{\text pinch}, \omega^{\text cusp})}{\partial k}\!=\!0$ but $\frac{\partial^2 D(k^{\text pinch}, \omega^{\text cusp})}{\partial k^2}\!\ne \!0$). Kupfer~\cite{Kupfer1987} employed a local mapping procedure to conceptualize the stability characteristics of this branch point. Near a `reasonably close' neighborhood of the pinch point, a local Taylor series expansion yields a dispersion relation {\color{black}that} has a second-order algebraic form in the $\omega$-plane (and which is a first-order saddle point in the $k$-plane), i.~e., $(\omega - \omega^{\text cusp}) \sim (k - k^{\text pinch})^2$. This period-doubling characteristic of the map causes the $k_i$-contours to `rotate' around $\omega^{\text cusp}$, forming a cusp. In the $\omega$-plane, we draw a ray parallel to the $\omega_i$-axis from the cusp point such that it intersects the image of the Fourier contour (or $k_i = 0$ curve) and count the number of intersections (consequently, count the number of times both $k$-branches cross the $k_r$-axis before forming a pinch point in the $k$-plane. If the ray drawn from the cusp point intersects the image of the Fourier contour in the $\omega$-plane (or if either one or both the $k$-branches cross the $k_r$-axis) even number of times, then the flow dynamics correspond to an evanescent mode. Otherwise, in the case of odd intersections{\color{black},} the observed cusp point is genuine, leading to either {\color{black}an} absolutely unstable system (in the upper half of the $\omega$-plane) or {\color{black}a} convectively unstable system (in the lower half of the $\omega$-plane); provided the system is temporally unstable.

Under the assumption that dispersion relation is a complex analytic function satisfying Cauchy-Riemann relations, the expressions are chosen preferentially to numerically evaluate the derivatives in equation~\eqref{eqn:wTemp2}. The numerical continuation of the temporal growth rate (Section~\ref{subsec:TSA}) and the absolute growth rate curves (Section~\ref{subsec:STSA}) were realized within the range $Re \in [10^{-6}, \, 10^2]$, $We \in [0, \, 10^3]$ and at two specific values of $\nu=0.05$ (the elastic stress{\color{black}-}dominated case) and $\nu=0.3$ (the viscous stress{\color{black}-}dominated case), using a discrete step-size of $\triangle Re = 10^{-6}$ and $\triangle We = 10^{-4}$, respectively. While the (non-dimensional) physical domain spans within the range, $x \in (-\infty, \infty); y \in [0, 1]$, the temporal growth rate ($\omega^{\text{Temp}}_i$) and the absolute growth rate ($\omega^{\text{cusp}}_i$) of the perturbations are probed at four discrete, transverse spatial locations of the advancing interface: $y = 0.2, 0.9, 0.7, 0.5$. While the former two values of $y$ are chosen qualitatively to probe the near-wall effects, the last value is selected to understand the development of the centerline instability.

\section{Model validation}\label{sec:MV}
\begin{figure}[htbp]
\centering
\includegraphics[width=0.9\linewidth, height=0.6\linewidth]{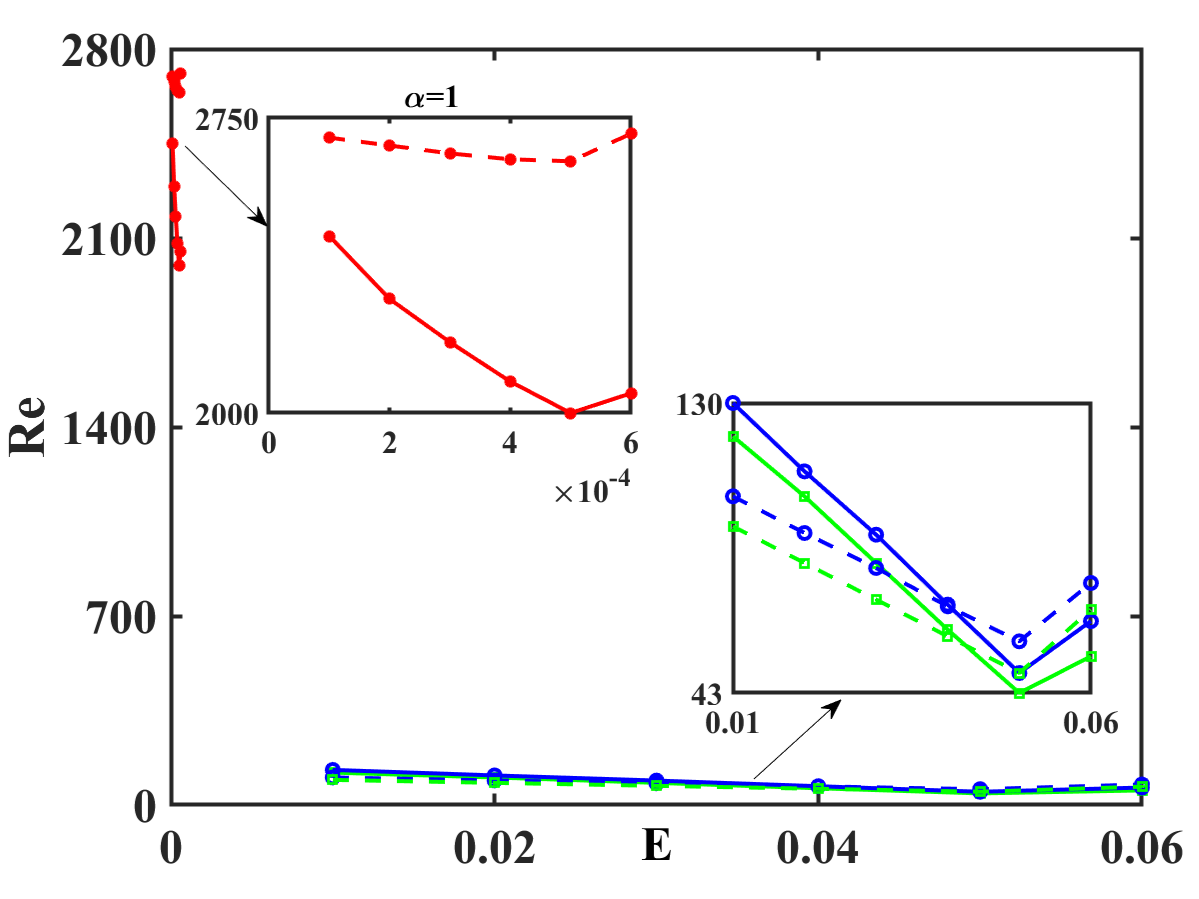}
\caption{Neutral stability curves at the centerline ($y = 0$) for plane Poiseulle viscoelastic flow of a classical fluid (red curves, source: figure 6 in \cite{Atalik2002}) versus subdiffusive fluid at $\alpha = 0.99$ (blue curves), $\alpha = 0.95$ (green curves) and at viscosity ratio, $\nu = 0.01$ (solid curves), $\nu = 0.9$ (dashed curves), projected onto the $Re-E$ plane.}
\label{fig1a}
\end{figure}

The model and the numerical method outlined in Section~\ref{sec:math} is validated by reproducing the neutral stability curves for a plane Poiseuille of a classical Oldroyd-B fluid, as investigated by Atalik~\cite{Atalik2002} ($\alpha = 1.0$ or the red curves in figure~\ref{fig1a}, see figure 6 in~\cite{Atalik2002}). The neutral stability curves for two fractional orders ($\alpha = 0.99$ (blue curves) and $\alpha = 0.95$ (green curves)) are also shown for comparison. The locus of neutrally stable points are found after selecting $\omega_i = 0$ in the dispersion relation~\eqref{eqn:DRP} and solving for the unknowns $(\omega_r, k)$, at fixed values of Reynolds and elasticity number ($E = \frac{We}{Re}$). 

Two conclusions can be deduced from figure~\ref{fig1a}. First, notice that the minimum value of the critical Reynolds number predicting a temporal instability {\color{black} increases, with increasing viscosity ratio}, both for the classical case (a result identical to the one predicted by Atalik~\cite{Atalik2002}) as well as the subdiffusive case. Second, observe that this minimum value of $Re$ is significantly lower and appears at significantly larger values of $E${\color{black},} for the subdiffusive fluid. These two observations indicate that the transition to instability are primarily driven by elasticity (rather than fluid inertia) for subdiffusive fluids. A more detailed outlook of the influence of elasticity is acquired by examining the temporal growth rates, described next.

\section{Temporal stability analysis}\label{subsec:TSA}
\begin{figure*}[htbp]
\centering
\includegraphics[width=0.45\linewidth, height=0.305\linewidth]{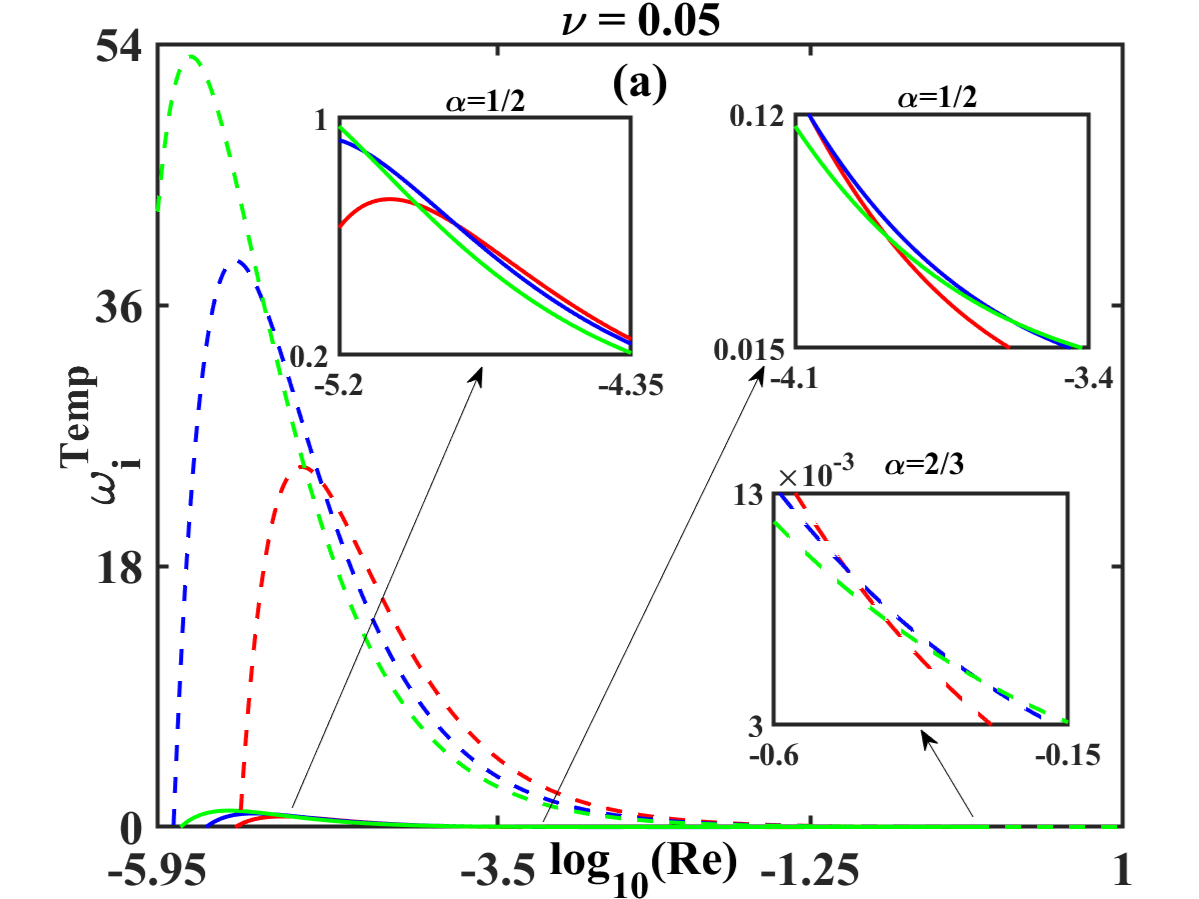}
\includegraphics[width=0.45\linewidth, height=0.305\linewidth]{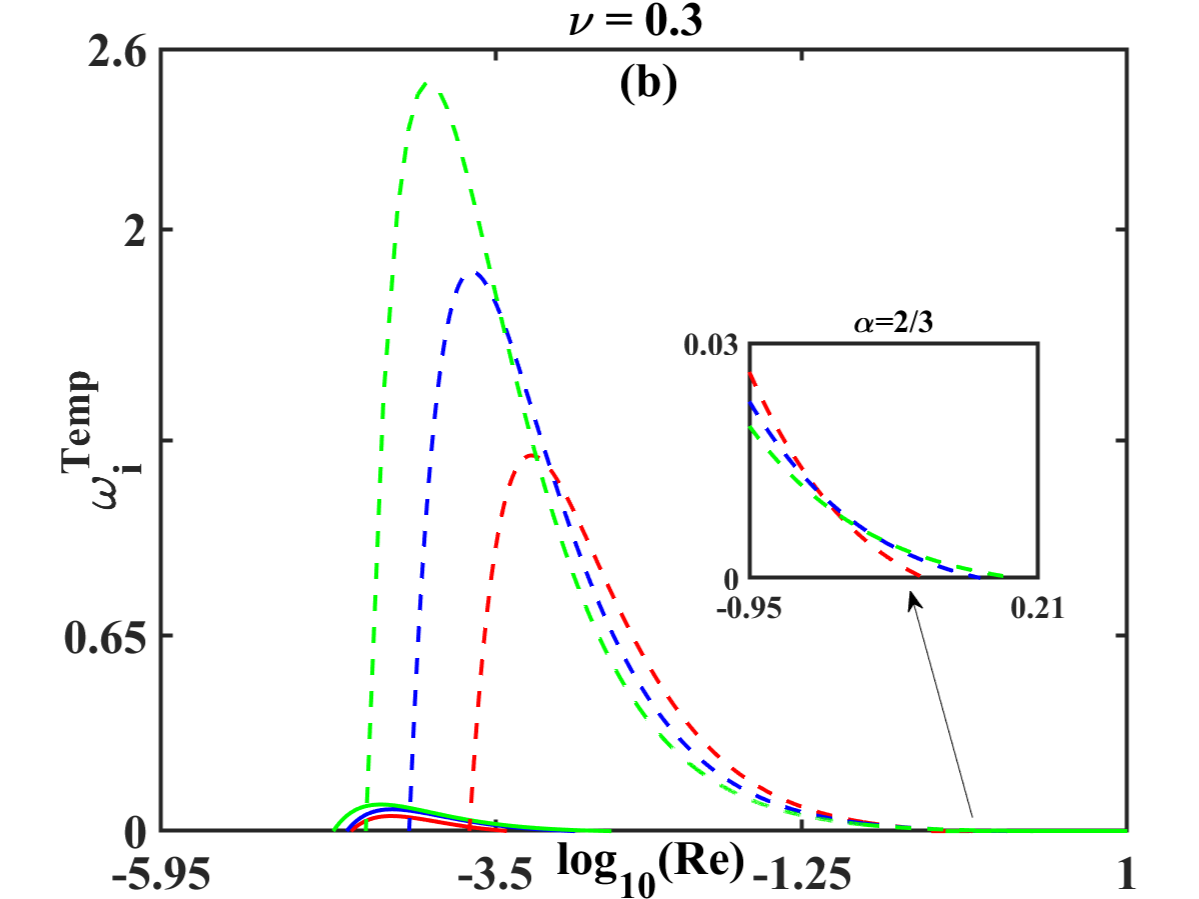}
\vskip -1pt
\includegraphics[width=0.45\linewidth, height=0.305\linewidth]{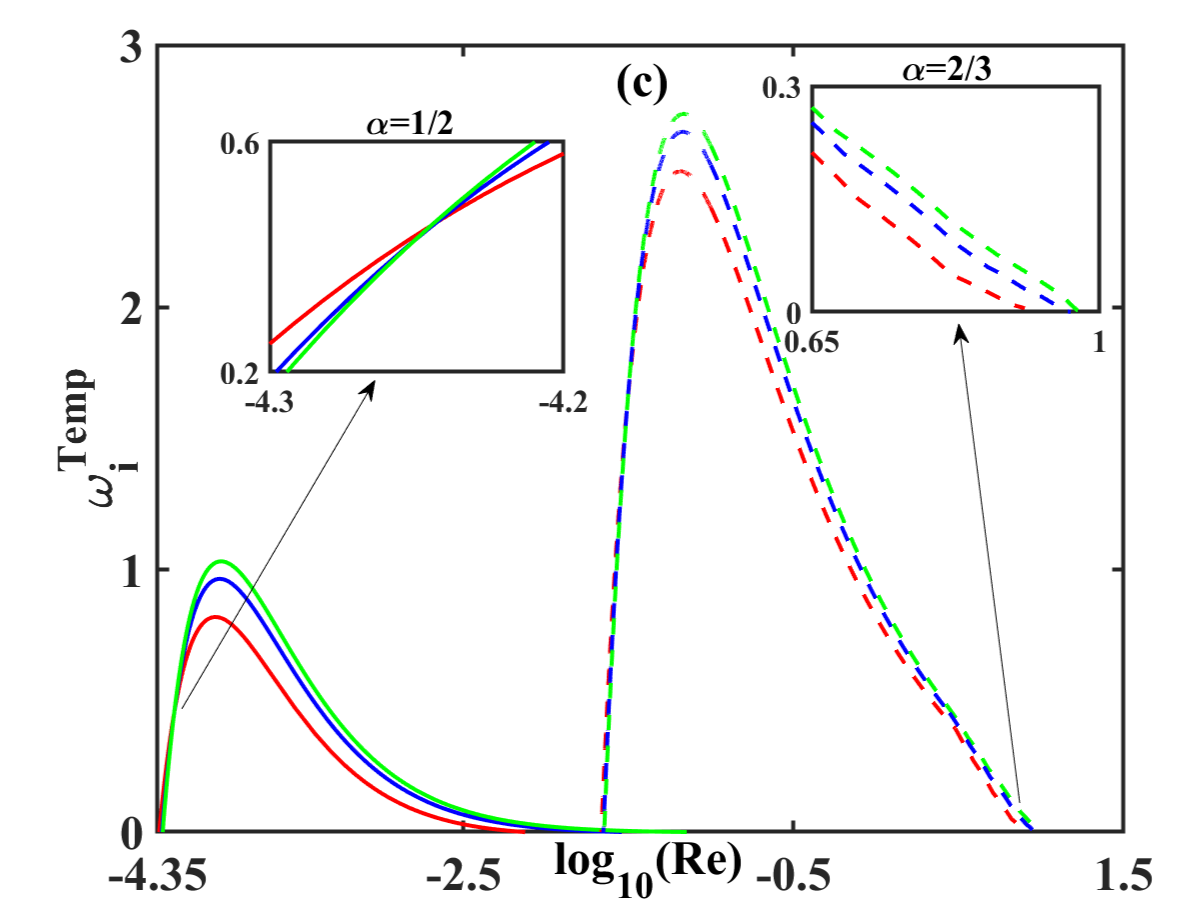}
\includegraphics[width=0.45\linewidth, height=0.305\linewidth]{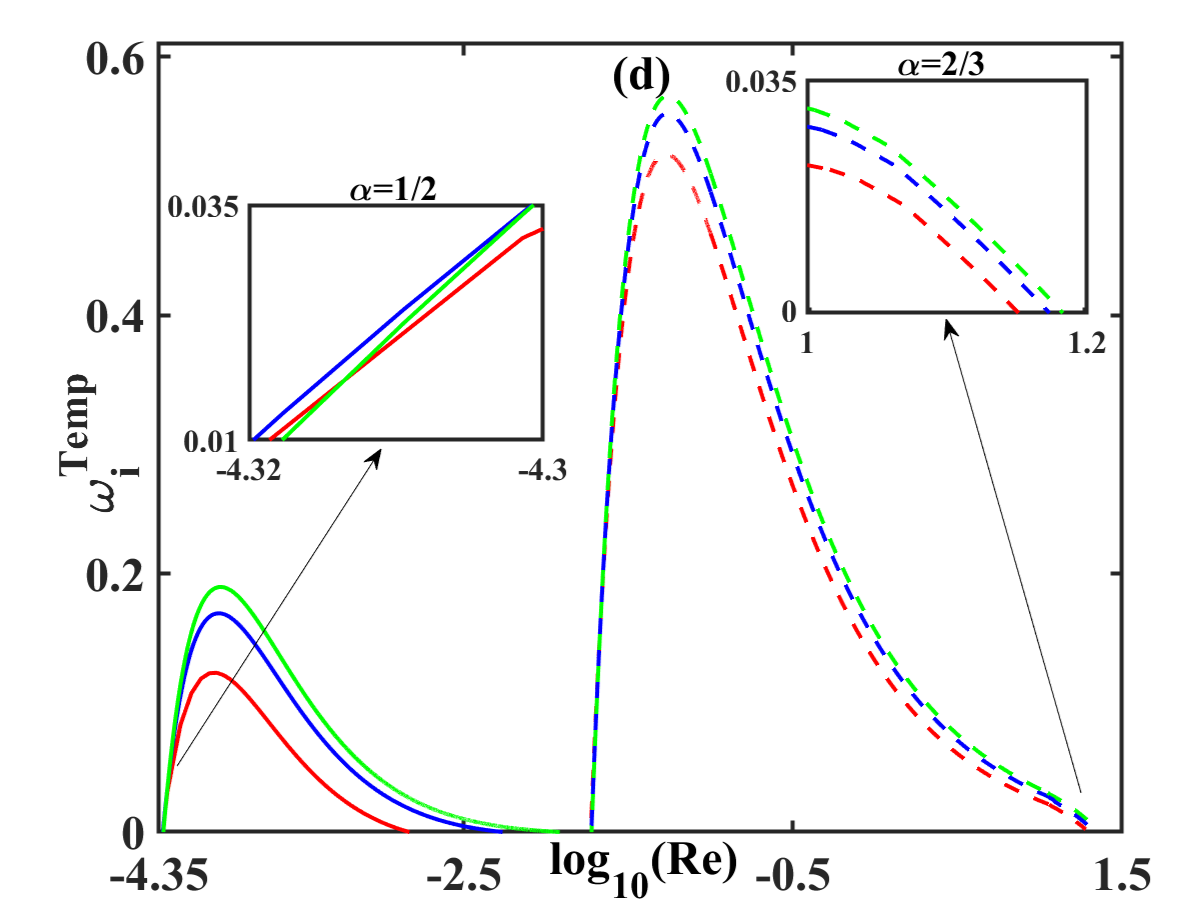}
\vskip -1pt
\includegraphics[width=0.45\linewidth, height=0.305\linewidth]{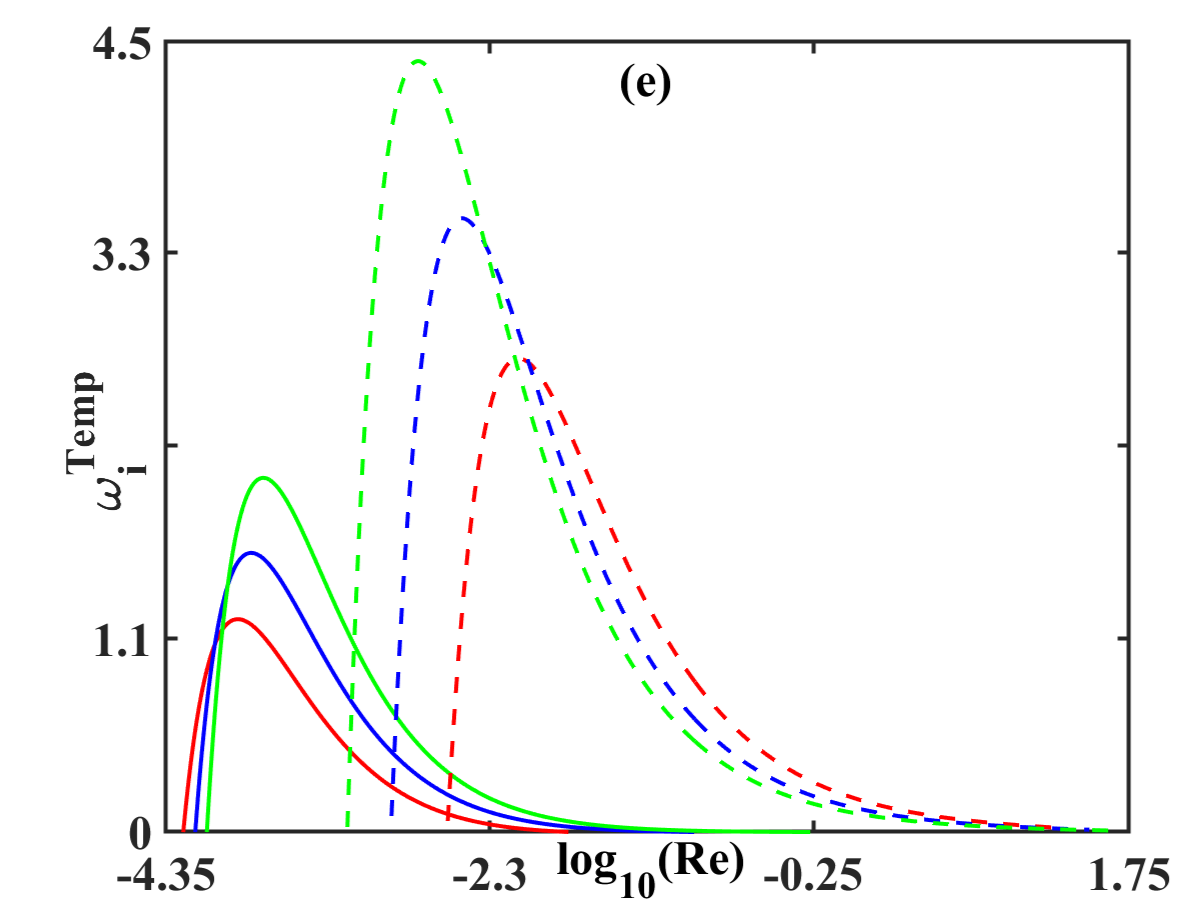}
\includegraphics[width=0.45\linewidth, height=0.305\linewidth]{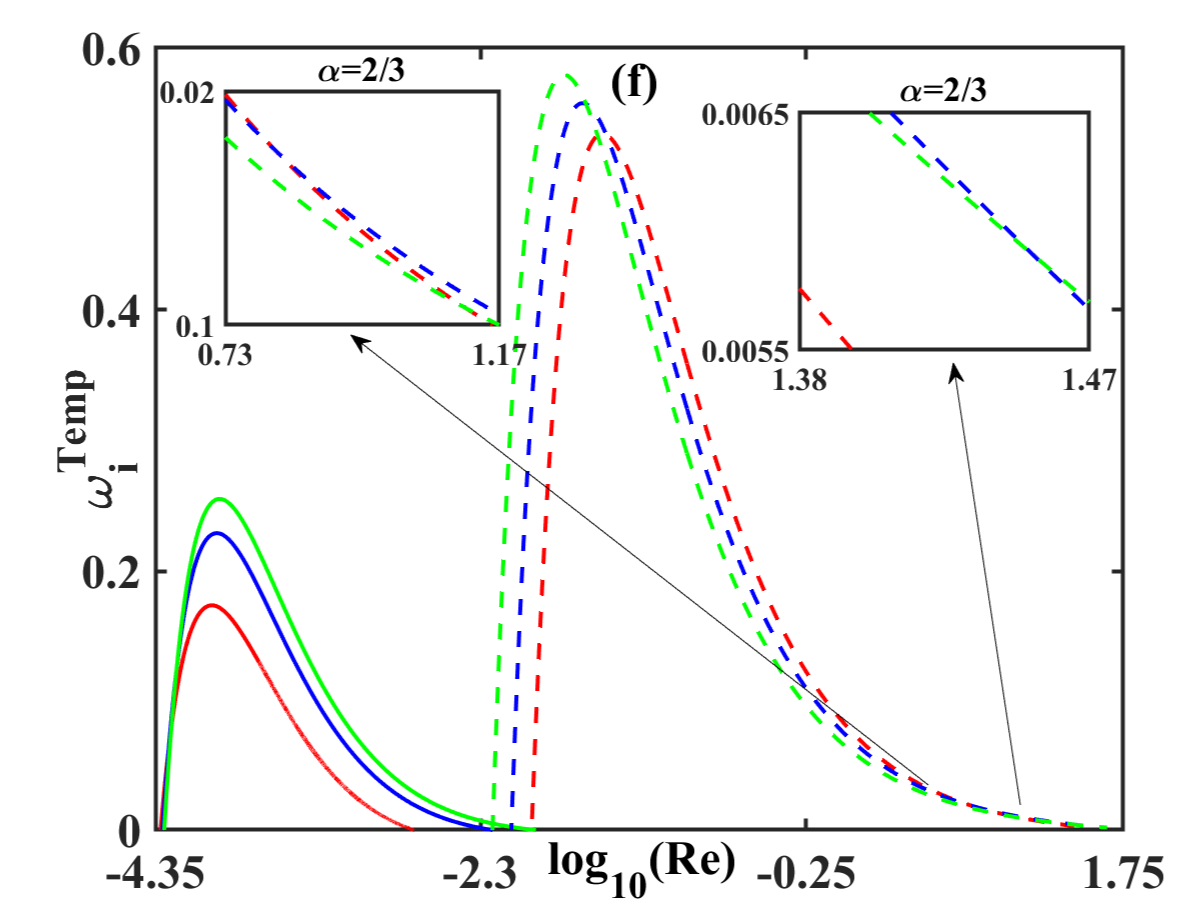}
\vskip -1pt
\includegraphics[width=0.45\linewidth, height=0.305\linewidth]{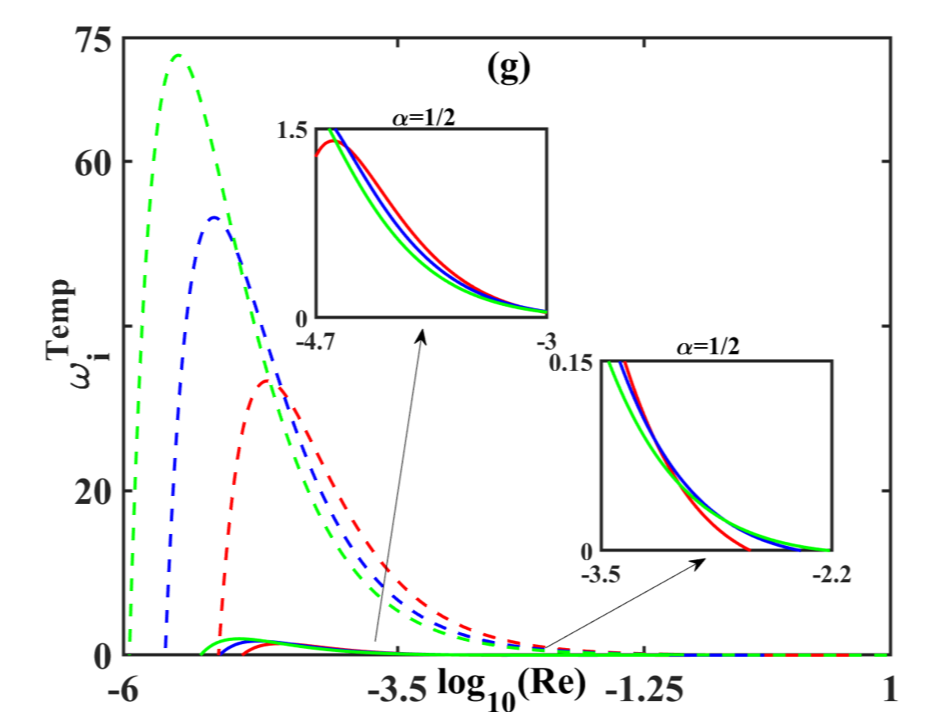}
\includegraphics[width=0.45\linewidth, height=0.305\linewidth]{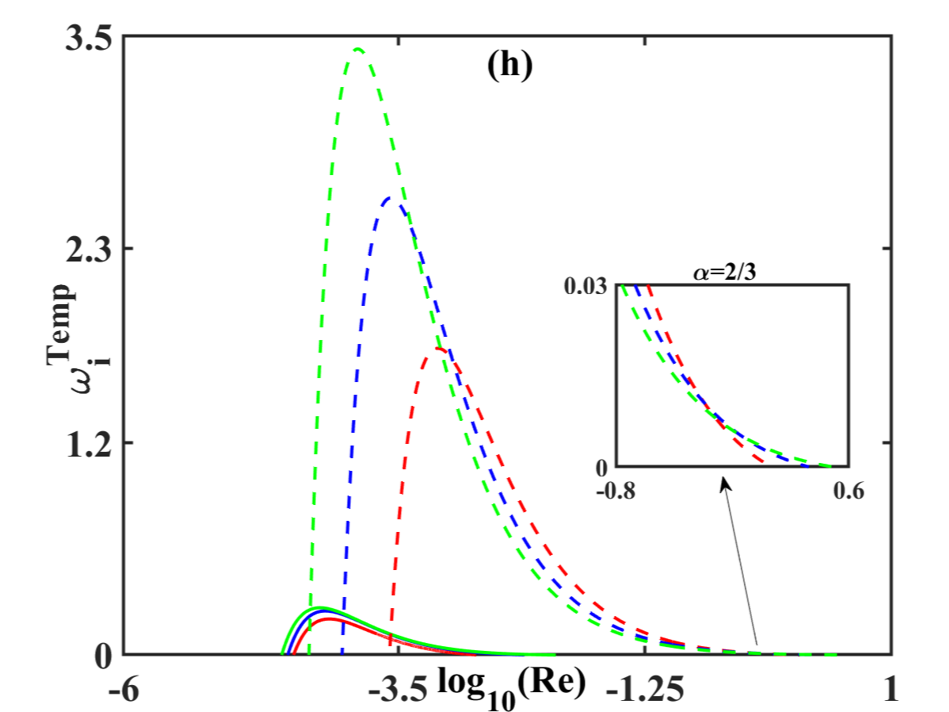}
\caption{Most unstable mode, $\omega^{\text{Temp}}_i$ for the {\it Poiseuille flow} case, vs. Reynolds number for the Rouse model (solid curves) and for the Zimm's model (dashed curves), evaluated at $We=15.0, 25.0, 35.0$ (red, blue and green curves, respectively), viscosity ratios, $\nu = 0.05$ (left column) and $\nu = 0.3$ (right column) and at transverse spatial locations: (a, b) $y = 0.2$, (c, d) $y = 0.5$, (e, f) $y = 0.7$ and (g, h) $y = 0.9$.}
\label{fig2}
\end{figure*}

\begin{figure*}[htbp]
\centering
\includegraphics[width=0.45\linewidth, height=0.305\linewidth]{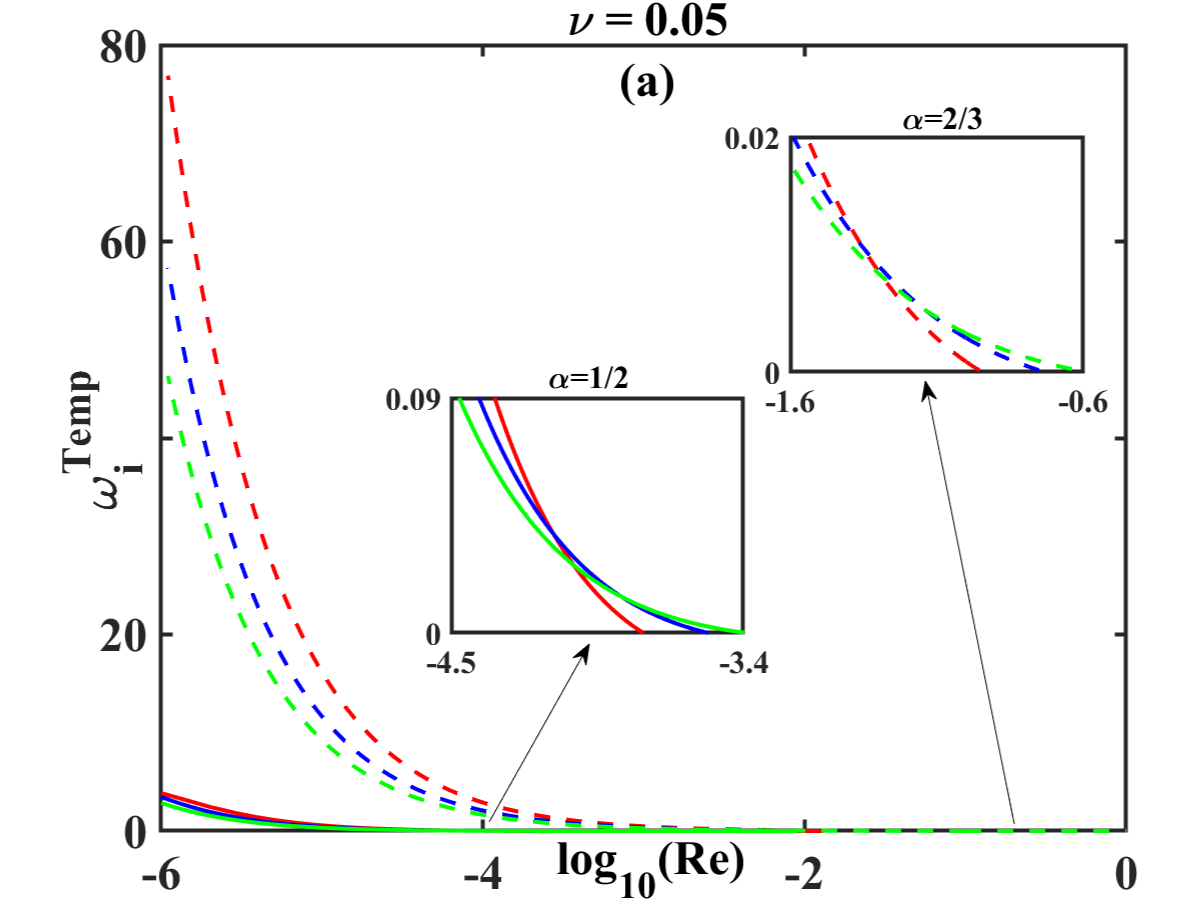}
\includegraphics[width=0.45\linewidth, height=0.305\linewidth]{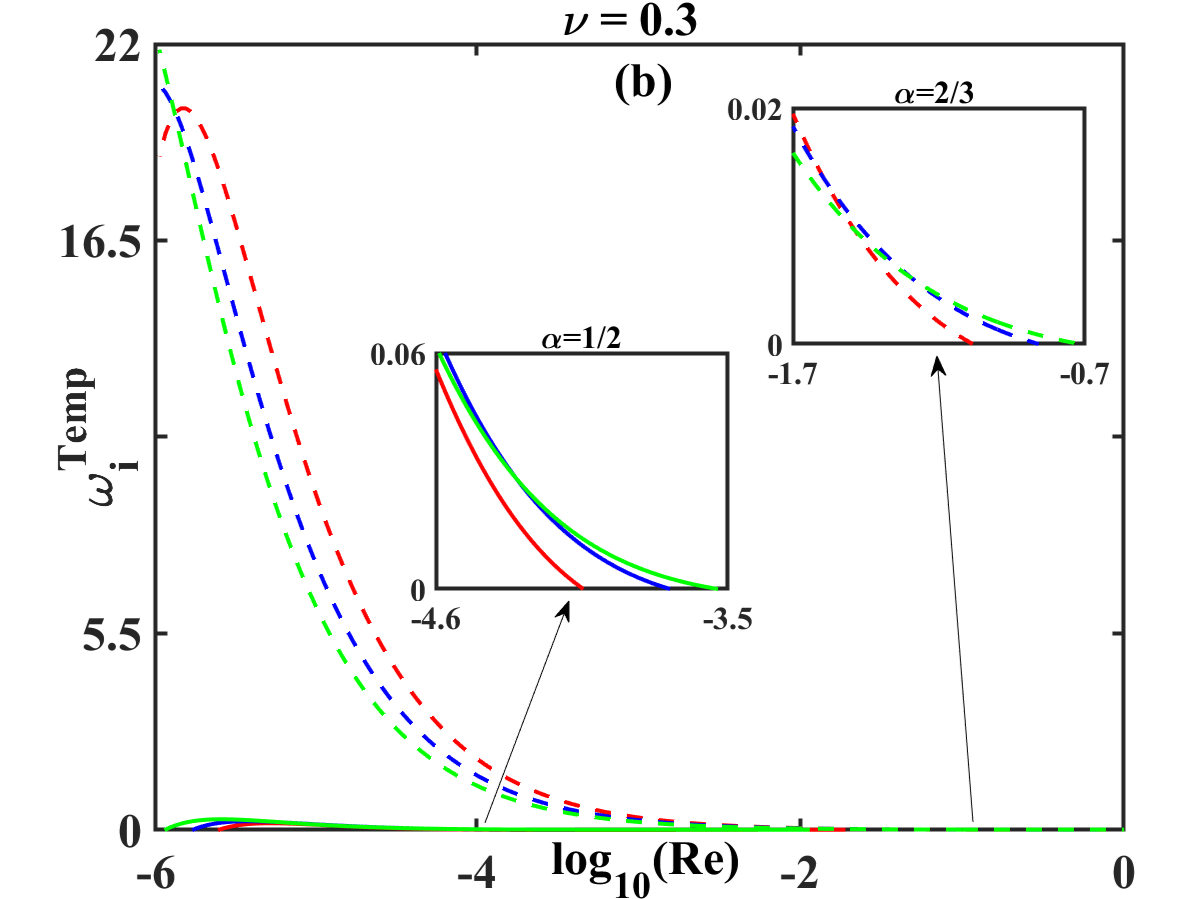}
\vskip -1pt
\includegraphics[width=0.45\linewidth, height=0.305\linewidth]{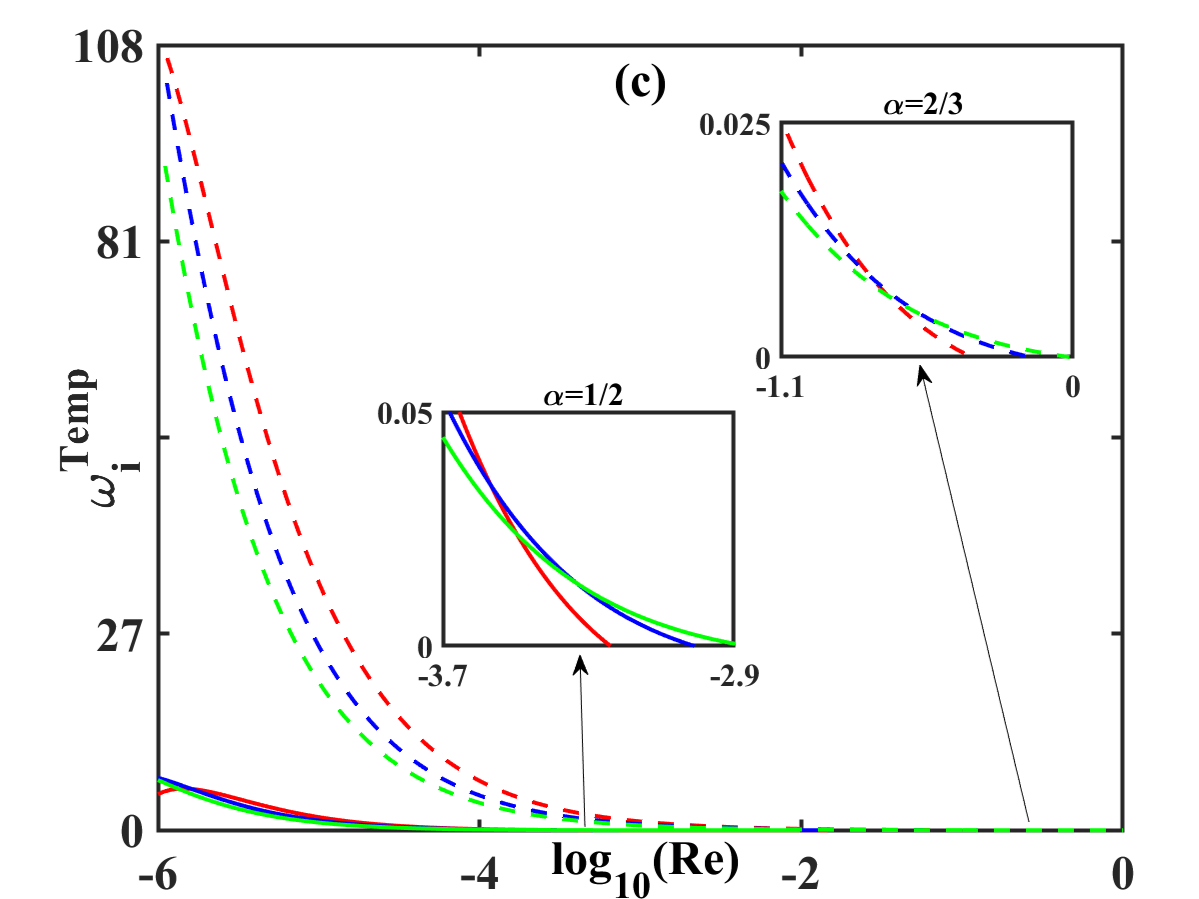}
\includegraphics[width=0.45\linewidth, height=0.305\linewidth]{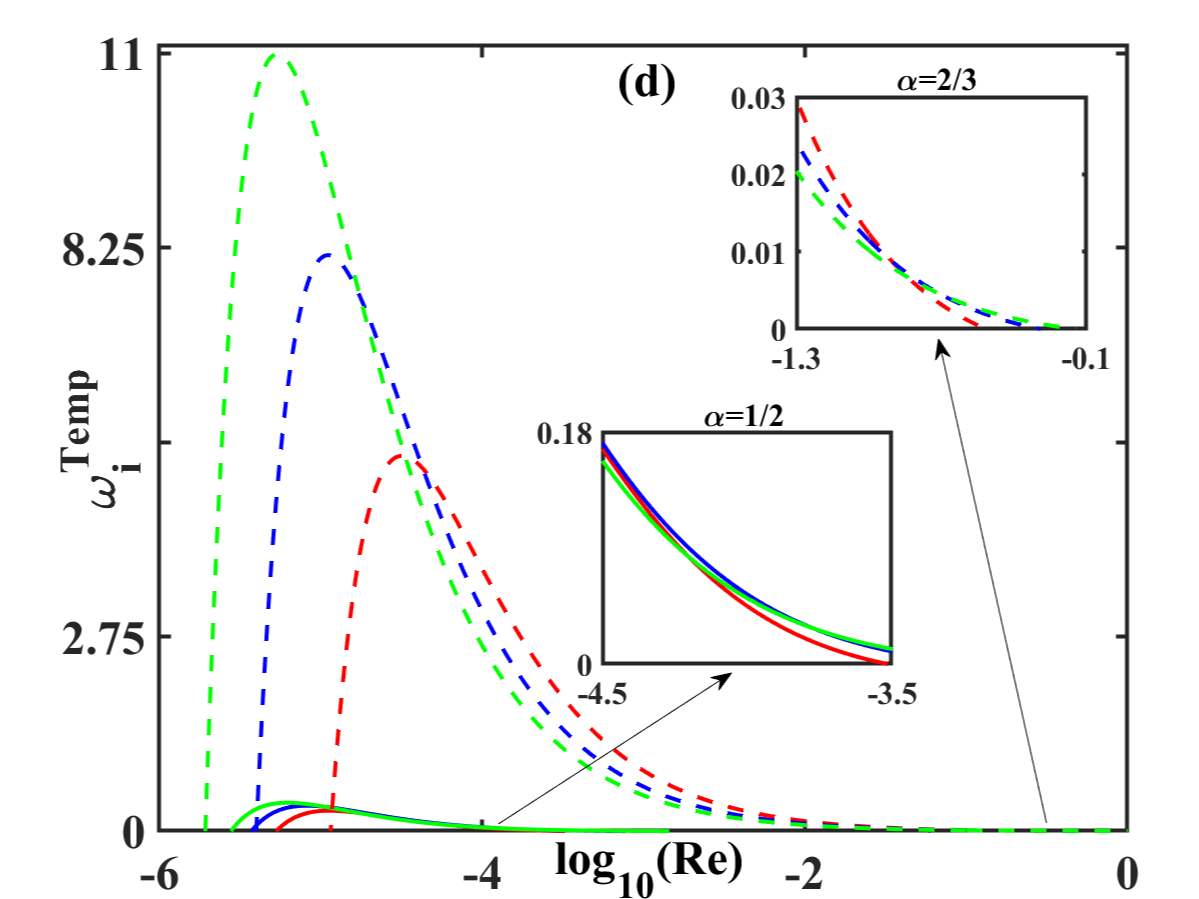}
\vskip -1pt
\includegraphics[width=0.45\linewidth, height=0.305\linewidth]{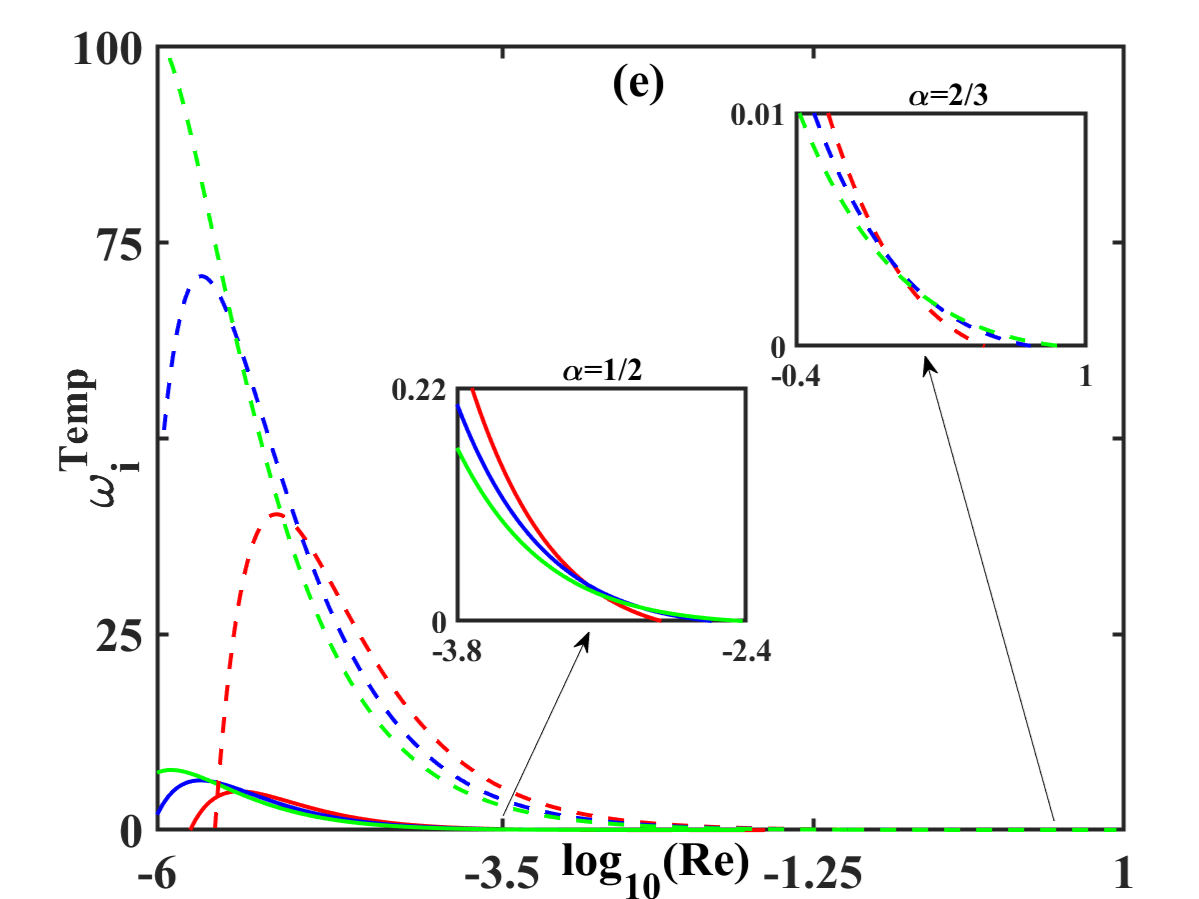}
\includegraphics[width=0.45\linewidth, height=0.305\linewidth]{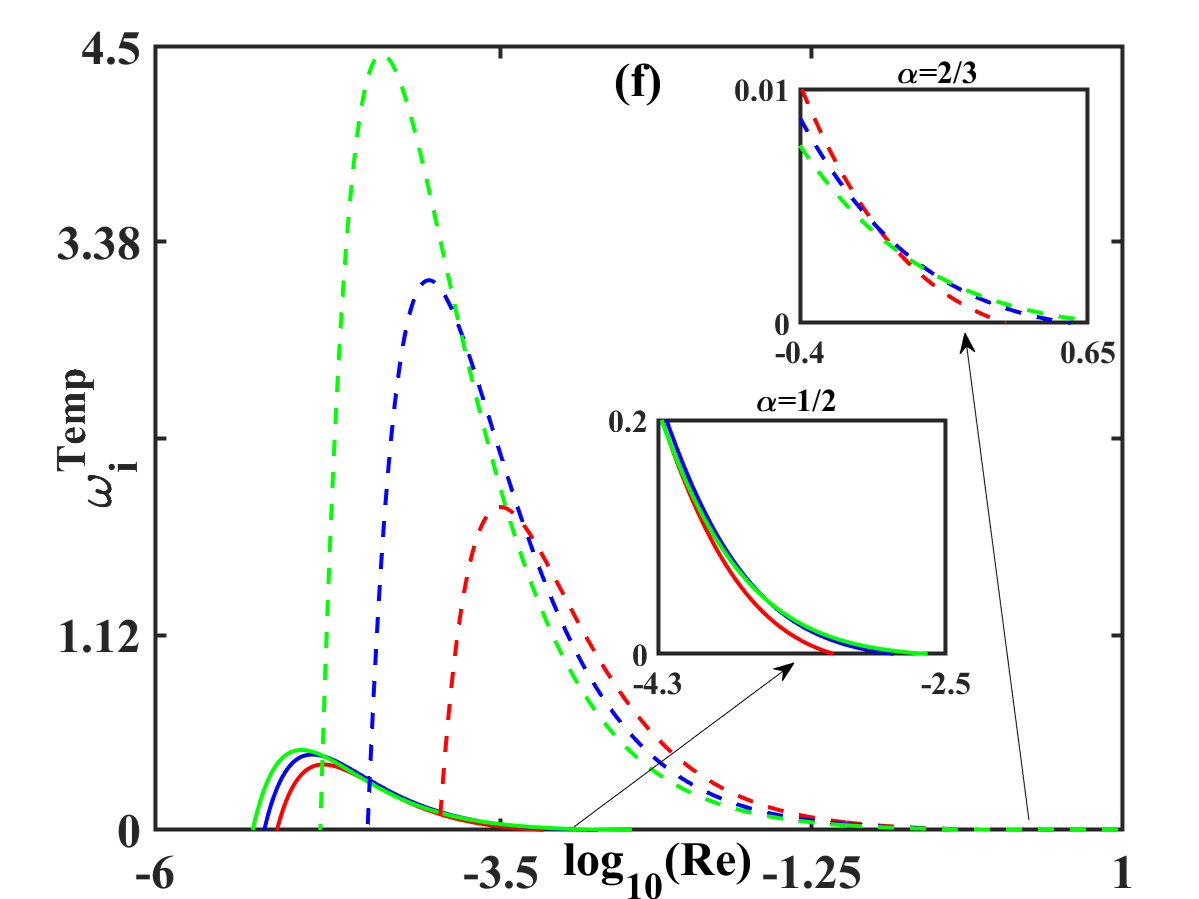}
\vskip -1pt
\includegraphics[width=0.45\linewidth, height=0.305\linewidth]{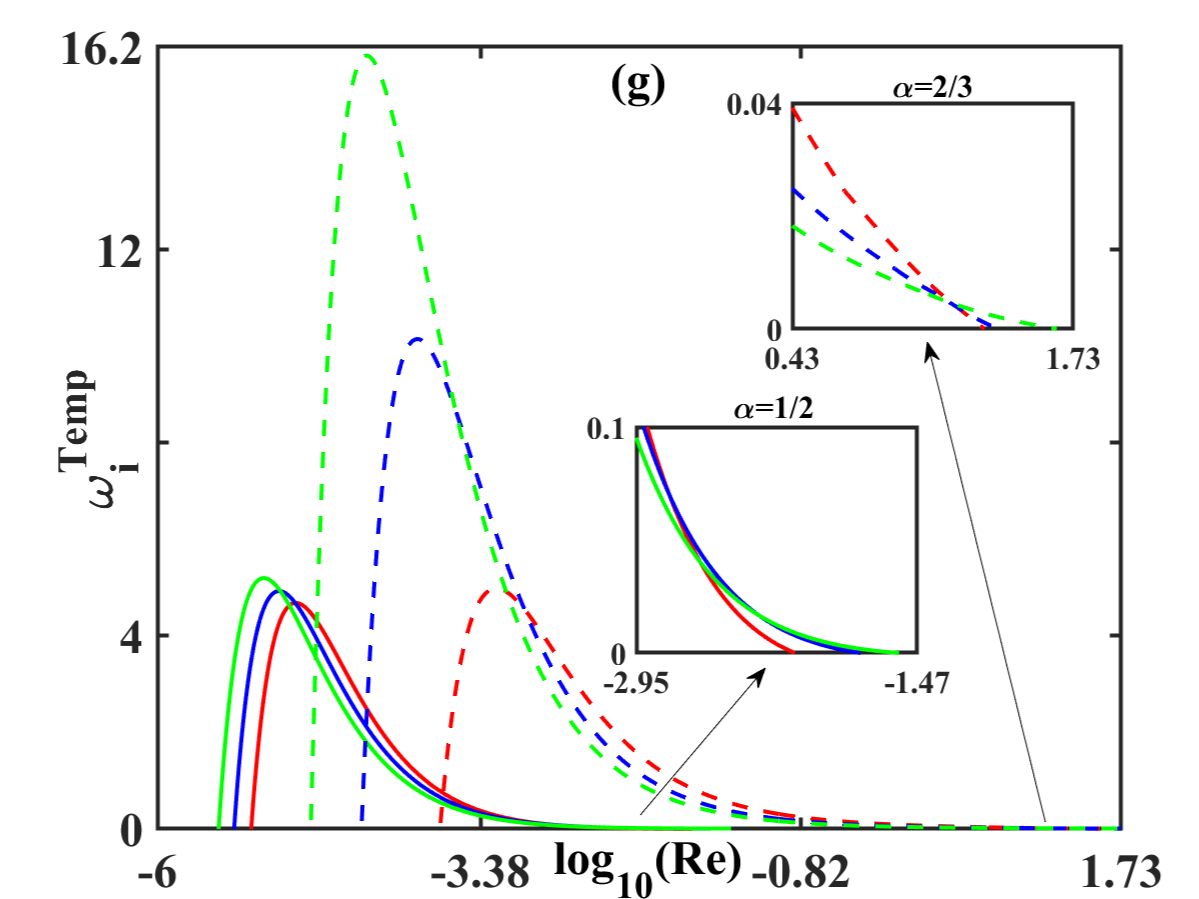}
\includegraphics[width=0.45\linewidth, height=0.305\linewidth]{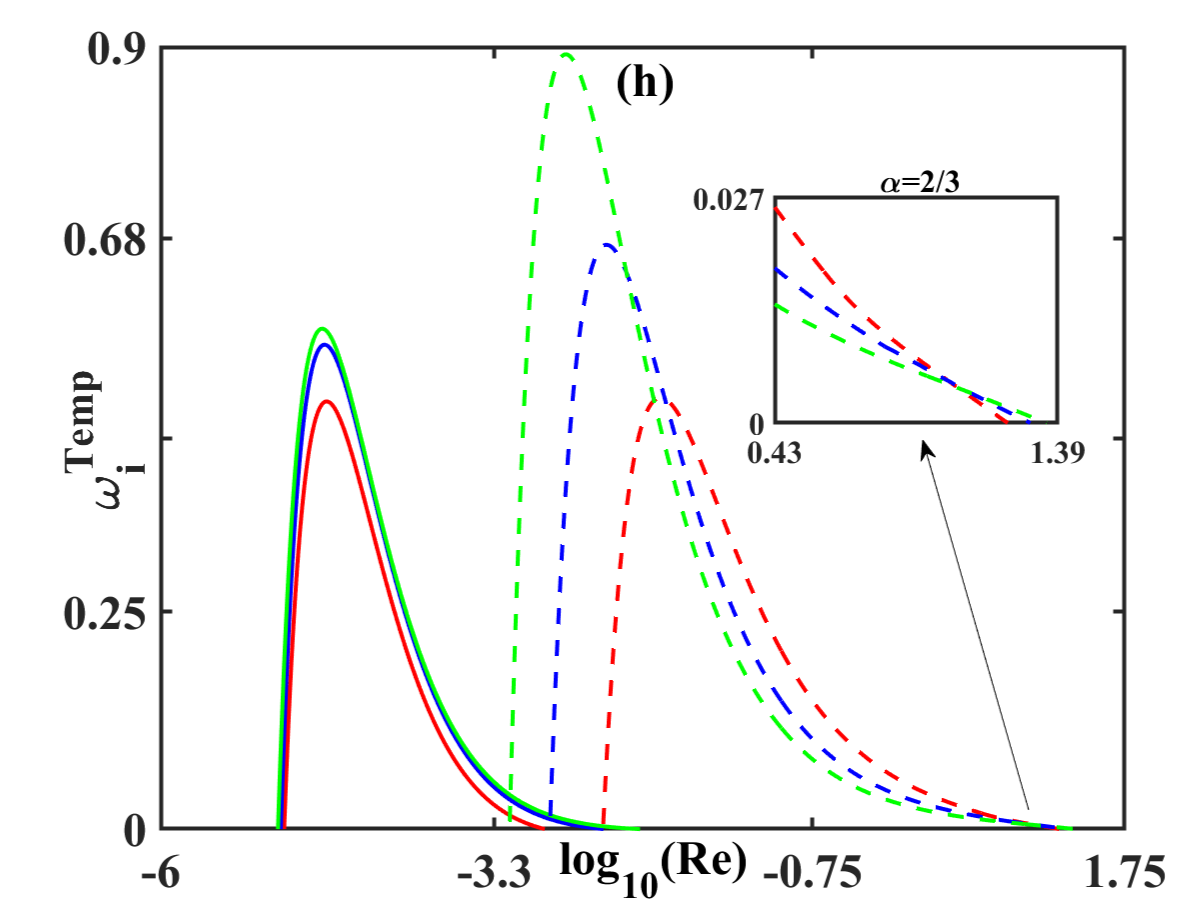}
\caption{Most unstable mode, $\omega^{\text{Temp}}_i$ for the {\it Couette flow} case, vs. Reynolds number for the Rouse model (solid curves) and for the Zimm's model (dashed curves), evaluated at $We=15.0, 25.0, 35.0$ (red, blue and green curves, respectively), viscosity ratios, $\nu = 0.05$ (left column) and $\nu = 0.3$ (right column) and at transverse spatial locations: (a, b) $y = 0.2$, (c, d) $y = 0.5$, (e, f) $y = 0.7$ and (g, h) $y = 0.9$.}
\label{fig3}
\end{figure*}

First, we explore the linear stability of the system~\eqref{eqn:DRPMatrix} by exclusively assigning $\omega$ to be a complex number. In earlier studies on wall-bounded viscoelastic flows, elasticity (characterized by the parameter, $We$) was found to have a destabilizing effect (for example{\color{black},} see~\cite{Khalid2021} and the references within). In this study, we partially extend some of these ideas for the subdiffusive, two-dimensional Poiseuille and Couette flows within a selected range of parameters, $Re, We, \nu$ and specially for the case of the Rouse chain melts and the Zimm chain solution{\color{black},} which corresponds to the fractional order derivatives, $\alpha=\nicefrac{1}{2}, \nicefrac{2}{3}$, respectively. Figures~\ref{fig2} and \ref{fig3} present the variation of the most unstable mode versus $Re$, and at fixed $We$ and $\nu$ for viscoelastic Poiseuille and Couette flows, respectively. 

Observe that the elastic stress-dominated case (or $\nu=0.05$ case) is temporally more unstable (i.~e., compare the maximum `y' value on the ordinate axis of the figures on the left column versus those on the right column in figures~\ref{fig2} and \ref{fig3}). Also{\color{black},} observe especially for the Zimm's case in Poiseuille flow{\color{black},} that not only the peak of the most unstable mode increases{\color{black},} but also the range of Reynolds number exhibiting temporal instability increases with increasing values of $We$ (i.~e., notice the dashed green, blue and the red curves in figure~\ref{fig2}). Also, analogous with the traditional (or integer order) viscoelastic channel flows, we find that for intermediate values of Reynolds number (or $1 \le Re \le 55$), elasticity is destabilizing (notice, from the dashed curves in figure insets in figures~\ref{fig2} and \ref{fig3}, that the most unstable mode is larger for larger values of $We$). These observations lead us to conclude that elasticity has a destabilizing impact, within the intermediate range of $Re$. This destabilization mechanism is the result of a complex interaction between the inertial forces (typically operative at larger Reynolds number) and the normal stress anisotropy through elasticity (proportional to $We$) and can be explained via an energy formalism: the stretching of the polymers with increasing elasticity brings about a normal stress anisotropy, leading to an elastically loaded fluid, that is, when the polymers stretch, elastic energy is stored in the sheared fluid. This energy is transferred and released after the fluid element has been adverted to other regions where the shear-induced stretching forces are smaller~\cite{Spagnolie2015}. However, for sufficiently larger values of $Re$ (or $Re > 55$), we find the emergence of the temporally stable state. The appearance of the temporally stable state at high fluid inertia, is a hallmark of subdiffusive flows and the details of the same are elaborated in Section~\ref{subsec:STSA}.

Regarding the near-wall effects, notice that the Zimm's case in Poiseuille flow is more unstable near wall (i.~e., comparing the maximum `y' value on the ordinate axis for $y=0.2, 0.9$, figures~\ref{fig2}a,b and \ref{fig2}g,h respectively) in comparison with the corresponding instability on channel centerline ($y=0.5$ case, figures~\ref{fig2}c,d). Further, within the intermediate range of Reynolds number (i.~e., $1 \le Re \le 55$), elasticity is destabilizing near the walls (e.~g., see the dashed curves in figure insets in figure~\ref{fig2}a,b,g,h). All these near-wall effects can be understood via a mechanism similar to the one proposed by Rabaud~\cite{Rabaud1988} for wall-bounded Newtonian flows: the boundary effects induce a local perturbation on the advancing interface which (when coupled with elasticity) destabilizes the flow.

Finally, we find that the order of the fractional derivative, $\alpha$ has a strong correlation with the temporal stability of the channel flows. For both types of flows, we deduce that the Zimm's model is temporally more unstable than the Rouse case. In a series of in silico studies, an investigation of the most unstable mode within the range, $\alpha \in [0.5\, 1.0]$, reveals: (a) the most unstable mode decreases with decreasing order of the fractional derivative, $\alpha$, (b) the peak of the most unstable mode shifts to lower values of $Re$ with decreasing values of $\alpha$, and (c) in particular, the peak of the most unstable mode, for the Rouse model (i.~e., the solid curves in figures~\ref{fig2}, \ref{fig3}), precipitates towards the limit $Re \rightarrow 0$. In other words, the transition pathway to flow turbulence in the Rouse polymer flows is characterized via elastic turbulence (appearing at vanishingly low values of $Re$ and at moderate to high values of $We$)~\cite{Larson2000}. To summarize, as $\alpha$ decreases, the nature of the transition pathway to flow turbulence changes from that of the elastoinertial turbulence (characterized by moderate values of $Re$ and $We$) to elastic turbulence.

We recapitulate the interplay of the inertial forces (characterized by the parameter $Re$), the elastic forces (represented by the parameter $We$) as well as the boundary effects and the order of the fractional derivative on the progression of the temporal instability (exemplified by the most unstable mode) of the viscoelastic subdiffusive channel flows as follows: elasticity combined with reasonably large fluid inertia has a destabilizing impact on the evolving flow front. The finite boundary is shown to have a destabilizing influence. Finally, the order of the subdiffusive timescale (alternatively, the order of the fractional derivative) impacts the nature of the transition pathway to turbulence (if any). In the next section, we outline a deeper characterization of these instabilities via the spatiotemporal analysis.

\section{Spatiotemporal stability analysis}\label{subsec:STSA}
Spatiotemporal analysis is typically relevant when one introduces an impulse excitation locally in a flow and observes how that disturbance evolves in time~\cite{Huerre1990}. More significantly, we evaluate the absolute growth rate (or $\omega^{\text{Cusp}}_i$, details on computing these points are elaborated in Section~ \ref{subsec:NM}) to identify the region of absolute instability, or the region indicating the topological reconfiguration and subsequent pinch-off of the advancing interface~\cite{Goldstein1993}. However, evanescent modes are also encountered in our analysis~\cite{Bansal2021}. These modes do not merely depend on the sign of the absolute growth rate and have to be found via the sufficient conditions proposed by Briggs~\cite{Kupfer1987} (refer Section~\ref{subsec:NM}). Evanescent modes are briefly refered in the description of the phase diagrams (figures \ref{fig6} and \ref{fig7}).

\begin{figure*}[htbp]
\centering
\includegraphics[width=0.45\linewidth, height=0.305\linewidth]{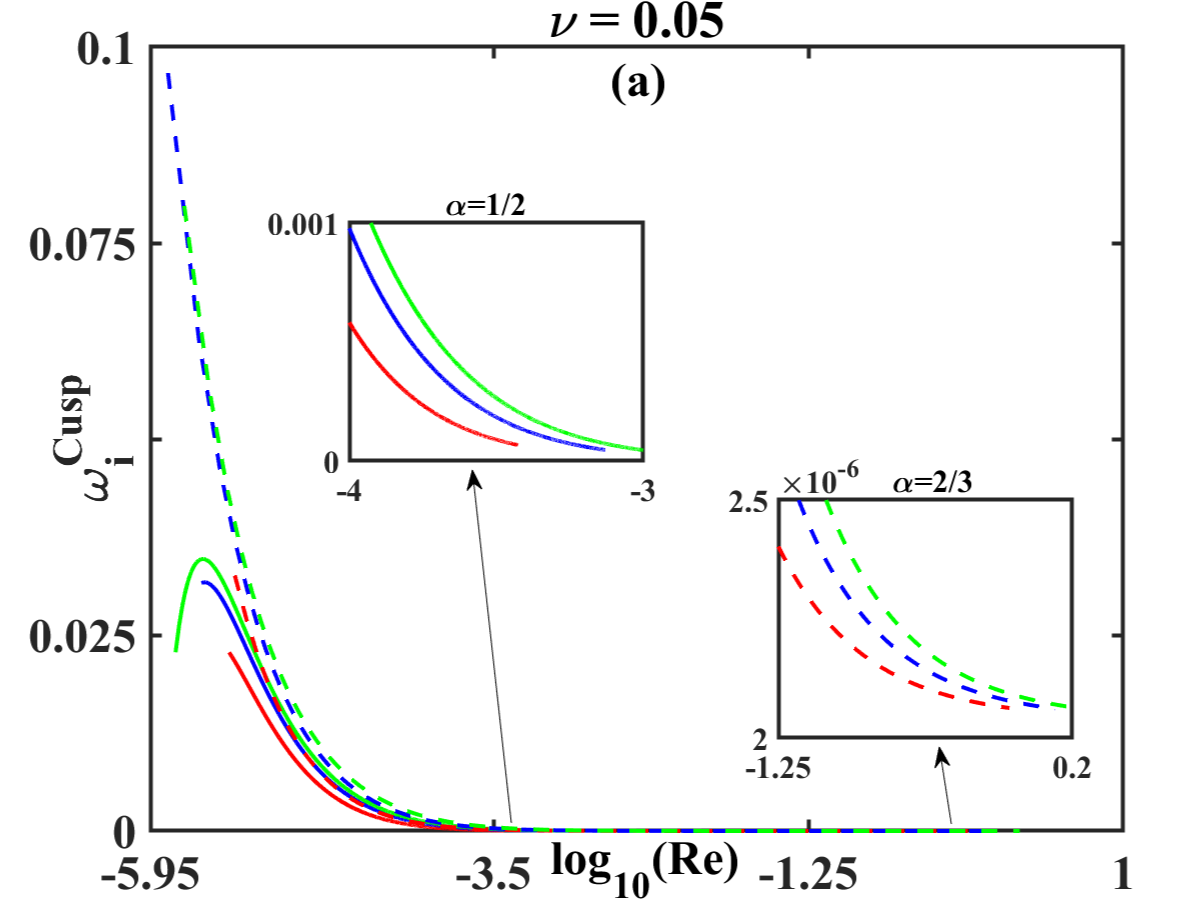}
\includegraphics[width=0.45\linewidth, height=0.305\linewidth]{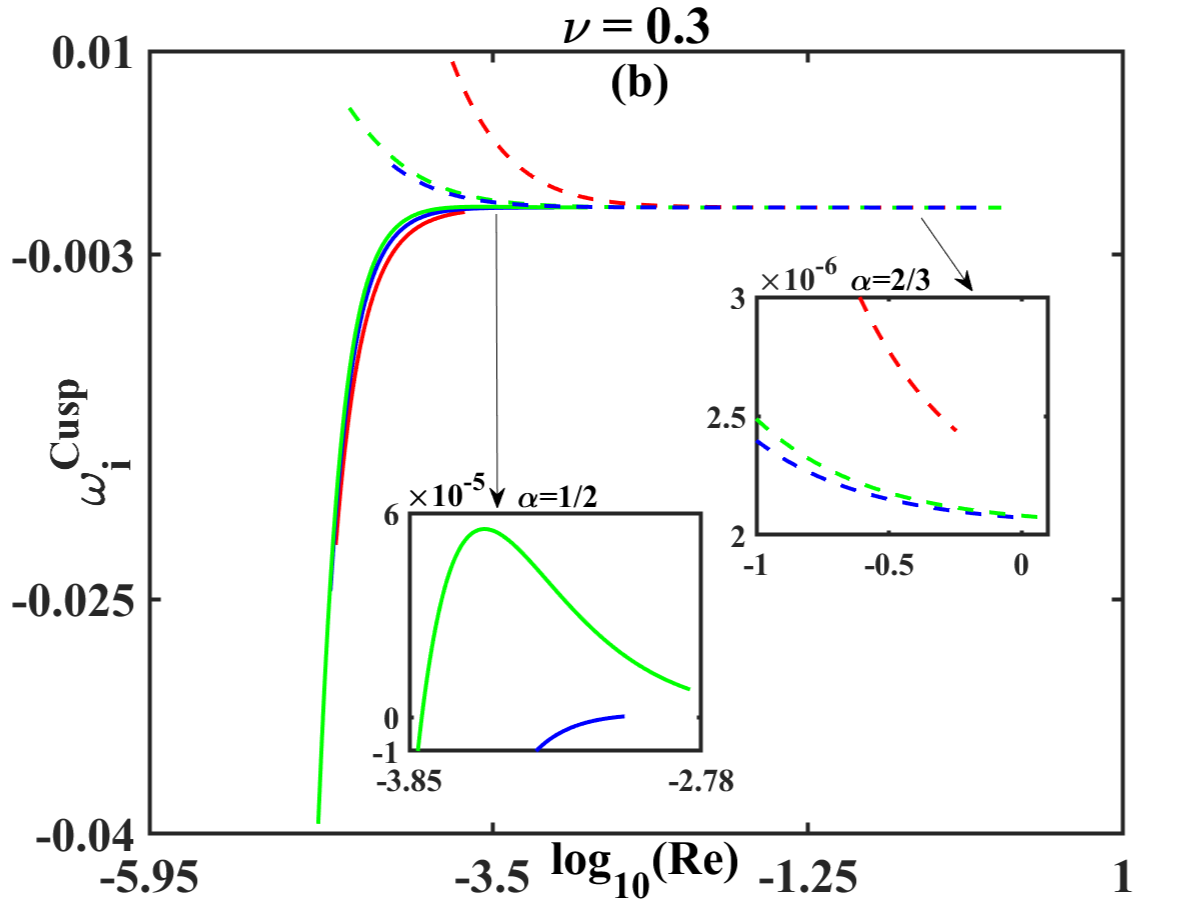}
\vskip -1pt
\includegraphics[width=0.45\linewidth, height=0.305\linewidth]{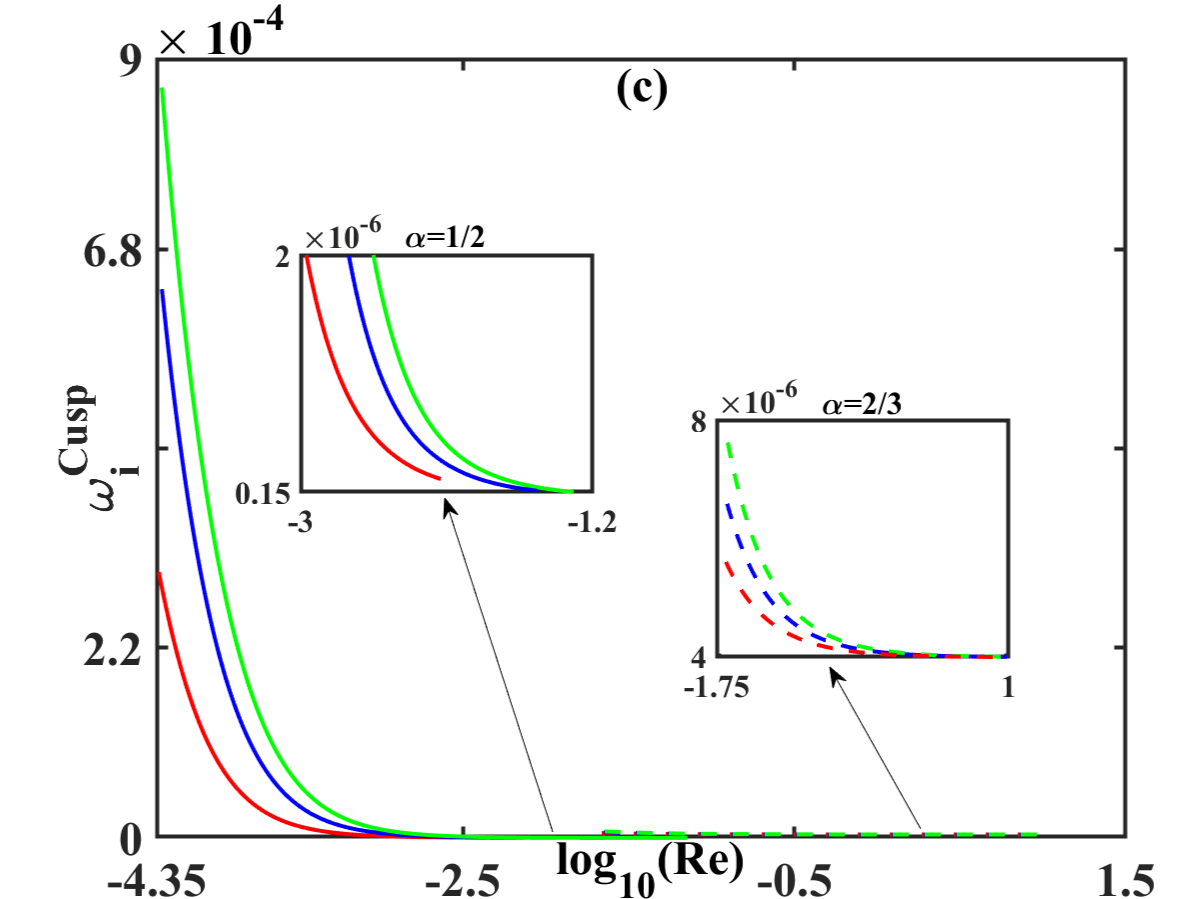}
\includegraphics[width=0.45\linewidth, height=0.305\linewidth]{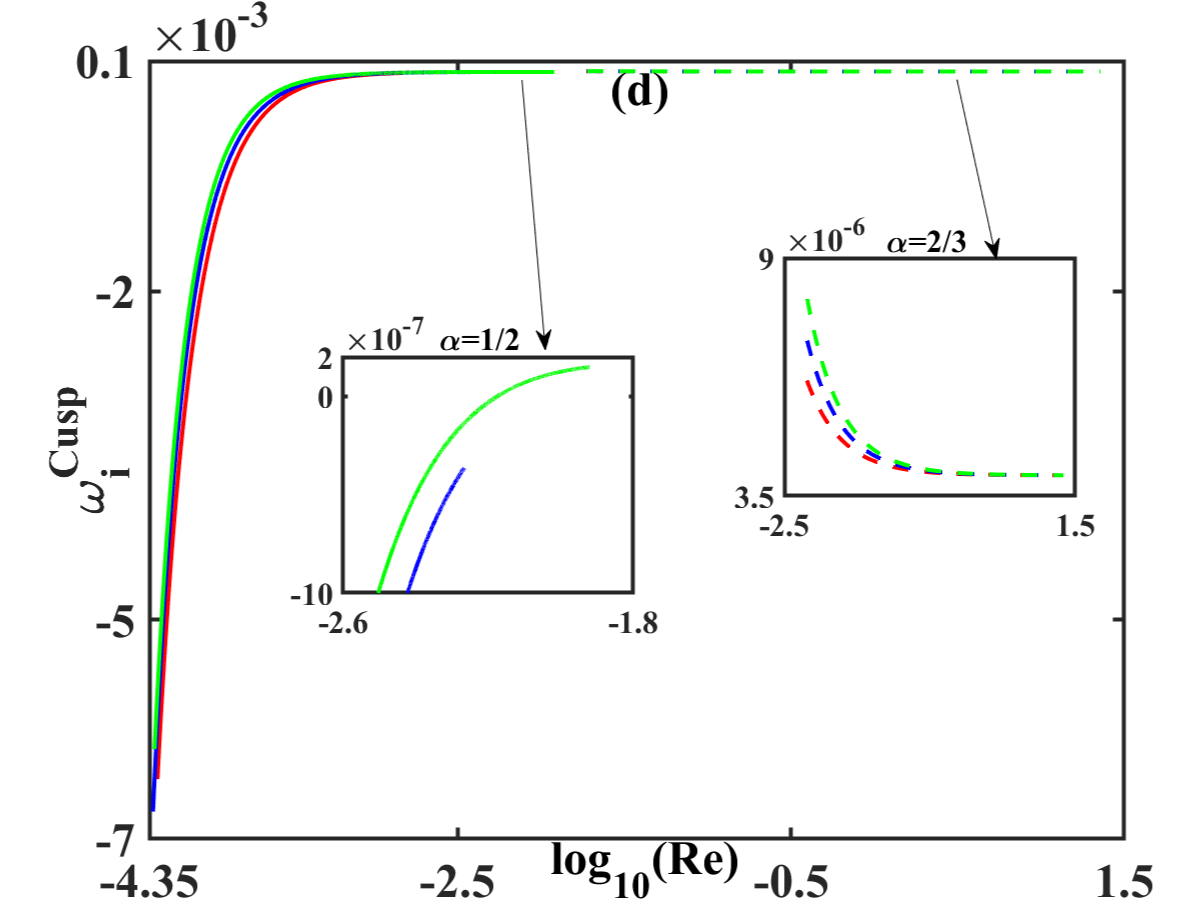}
\vskip -1pt
\includegraphics[width=0.45\linewidth, height=0.305\linewidth]{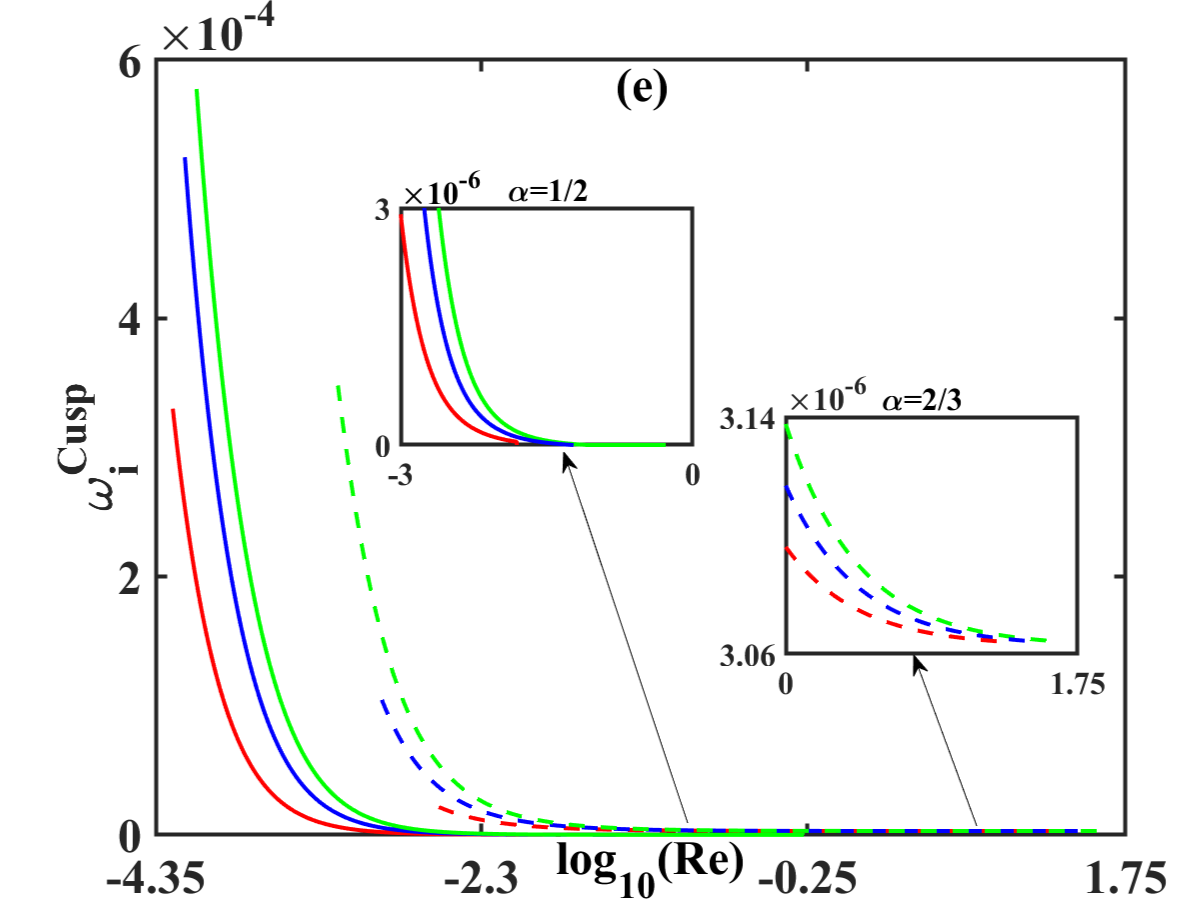}
\includegraphics[width=0.45\linewidth, height=0.305\linewidth]{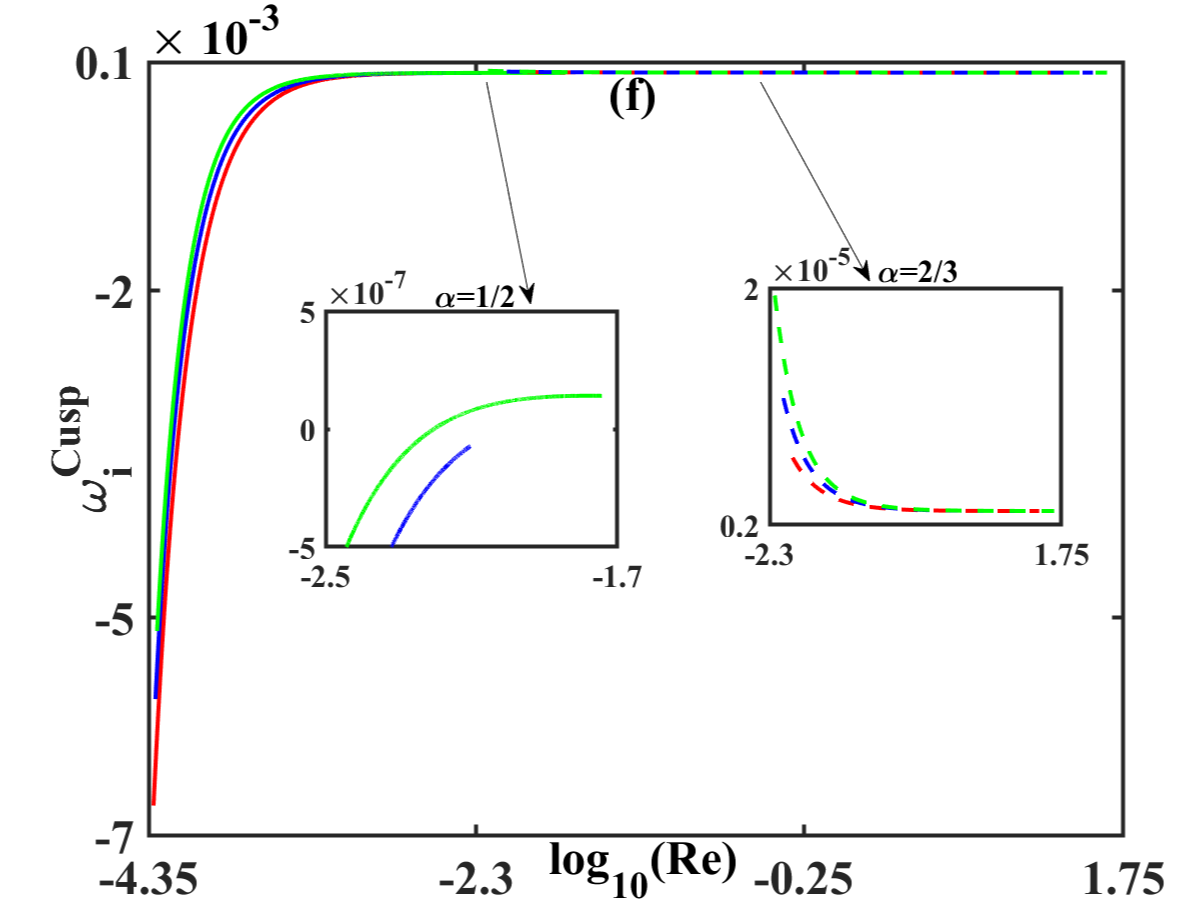}
\vskip -1pt
\includegraphics[width=0.45\linewidth, height=0.305\linewidth]{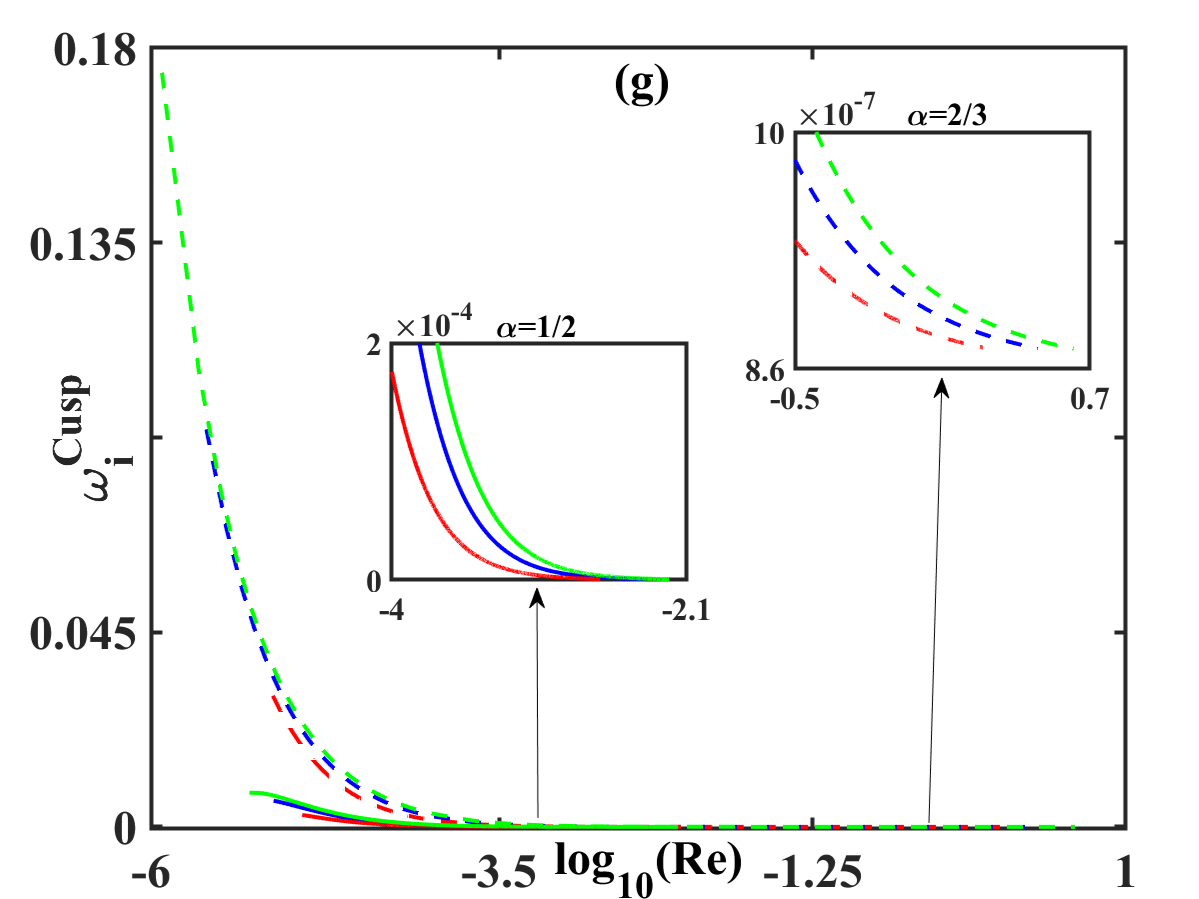}
\includegraphics[width=0.45\linewidth, height=0.305\linewidth]{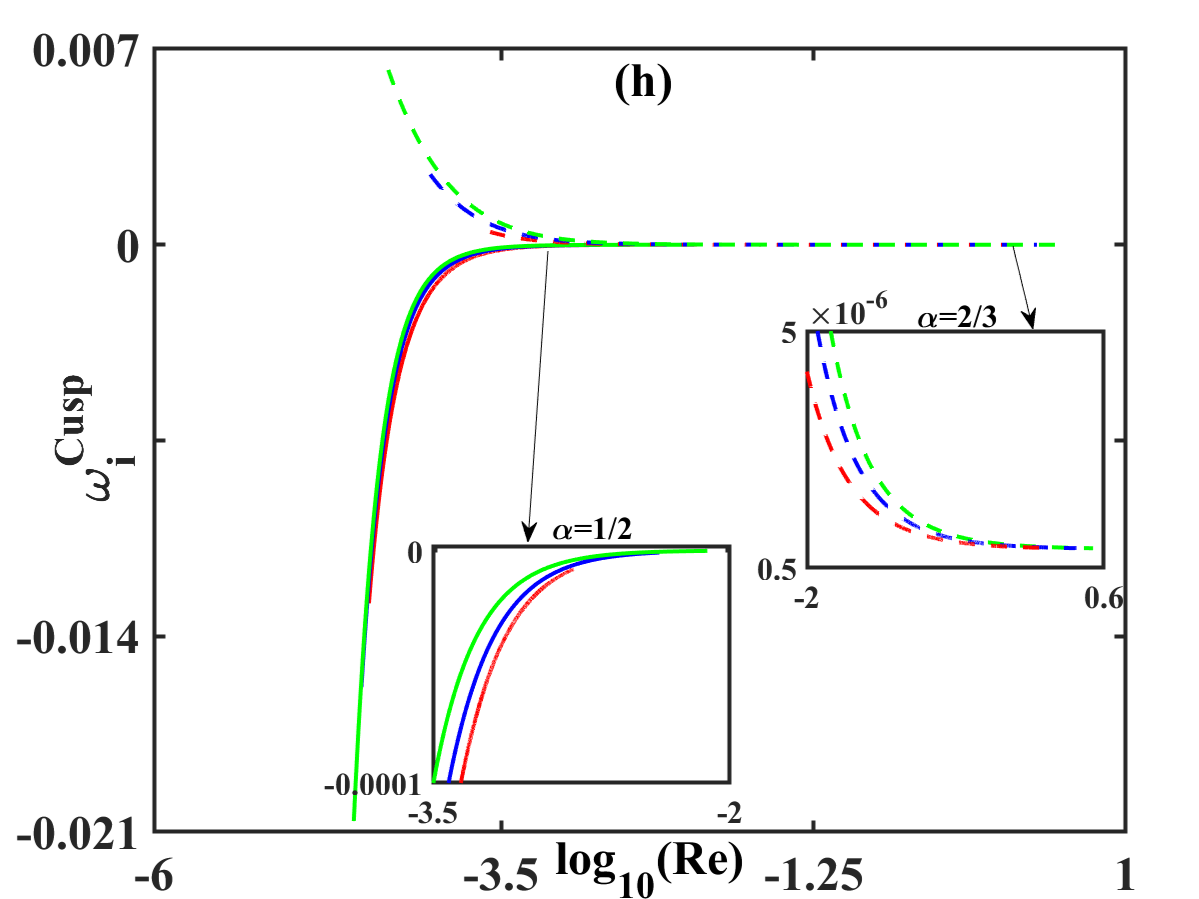}
\caption{Absolute growth rate, $\omega^{\text{Cusp}}_i$ for the {\it Poiseuille flow} case, vs. Reynolds number for the Rouse model (solid curves) and for the Zimm's model (dashed curves), evaluated at $We=15.0, 25.0, 35.0$ (red, blue and green curves, respectively), viscosity ratios, $\nu = 0.05$ (left column) and $\nu = 0.3$ (right column) and at transverse spatial locations: (a, b) $y = 0.2$, (c, d) $y = 0.5$, (e, f) $y = 0.7$ and (g, h) $y = 0.9$.}
\label{fig4}
\end{figure*}

Figure~\ref{fig4} represents the absolute growth rate curves versus $Re$ for Poiseuille flows, at three fixed values of Weissenberg number, $We = 15.0, 25.0, 35.0$, and at $\nu=0.05$ (the elastic stress{\color{black}-}dominated case) and $\nu=0.3$ (the viscous stress{\color{black}-}dominated case). For the selected values of $We$ and for the elastic stress{\color{black}-}dominated case, we find that the Rouse model exhibits a transition from absolute instability towards temporal stability at a critical value of Reynolds number, $Re_c < 10^{-3}$, at $y=0.2$ (figure~\ref{fig4}a). This critical Reynolds number increases as one moves closer to the upper plate and along the transverse spatial location, $y$  (i.~e., compare the $Re_c$ values from the inset in figures~\ref{fig4}a,c,e,g). For the viscous stress{\color{black}-}dominated case, the Rouse model indicates the following transition with increasing values of $Re$: convective instability$\rightarrow$ absolute instability$\rightarrow$ temporal stability. Again, the critical value of Reynolds number at these transition points, increases as one progressively moves towards the upper plate (refer figures~\ref{fig4}b,d,f,h). In contrast, the Zimm's model highlights a direct transition from absolute instability towards temporal stability, in the range of low to moderate values of $Re$ and for both the elastic as well as the viscous stress{\color{black}-}dominated case. However, both of these model reveal{\color{black}s} temporal stability in the limit of vanishingly small Reynolds number (or in the strongly elastic limit).

\begin{figure*}[htbp]
\centering
\includegraphics[width=0.45\linewidth, height=0.305\linewidth]{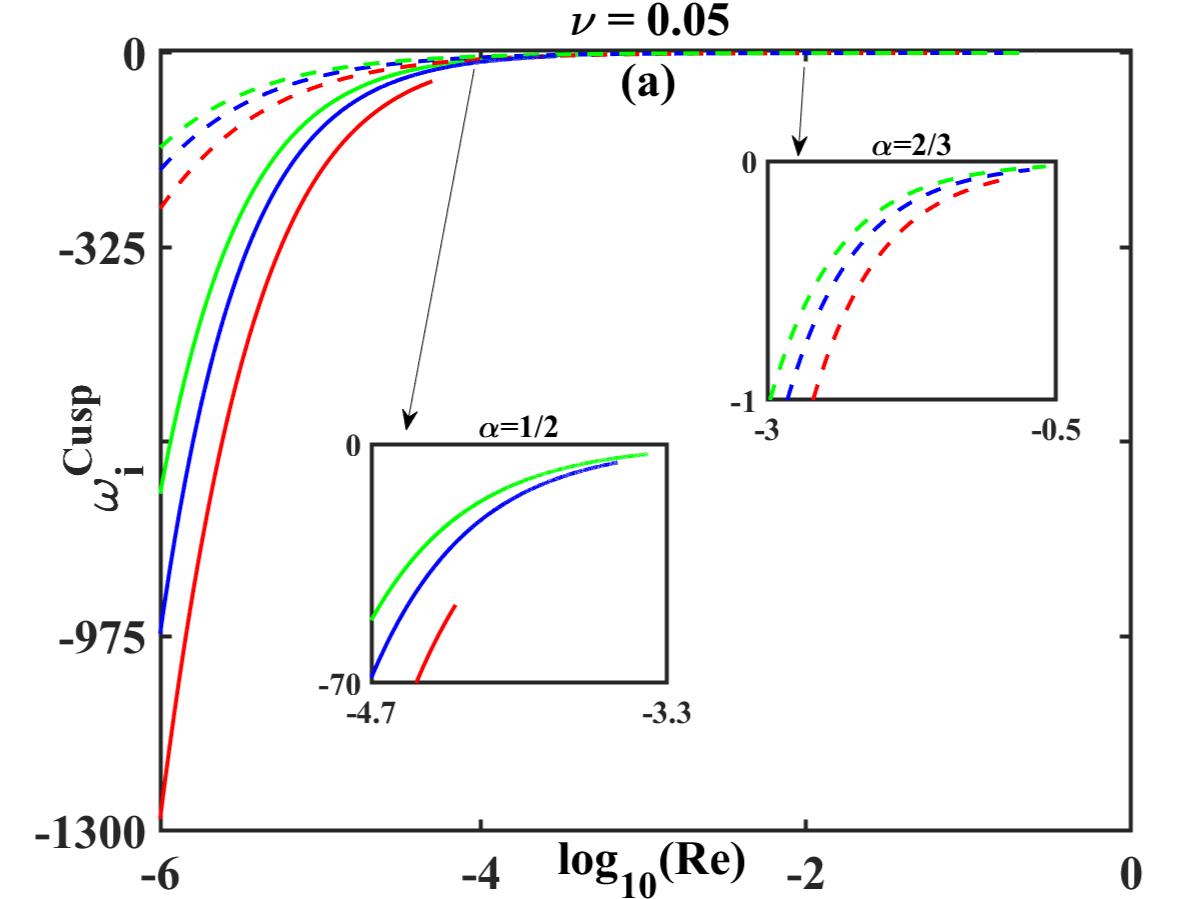}
\includegraphics[width=0.45\linewidth, height=0.305\linewidth]{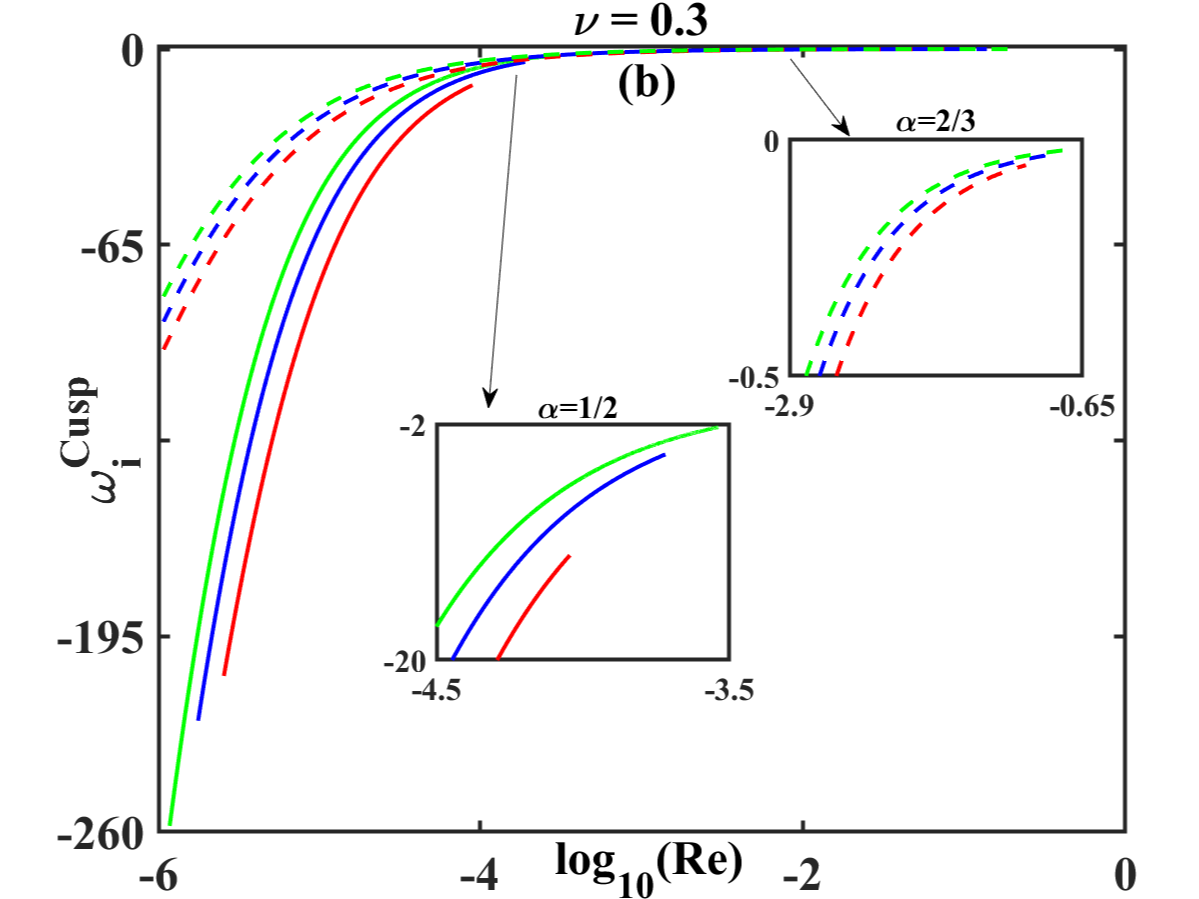}
\vskip -1pt
\includegraphics[width=0.45\linewidth, height=0.305\linewidth]{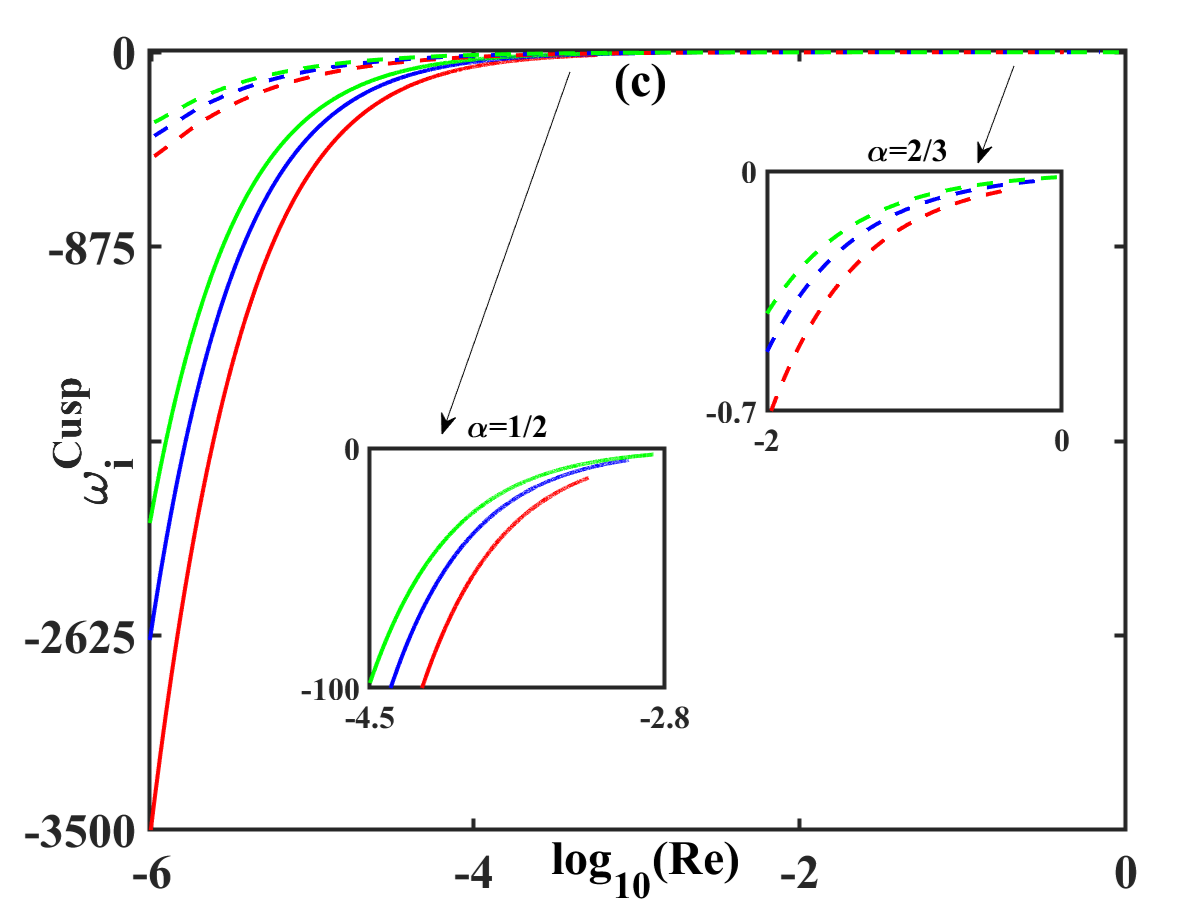}
\includegraphics[width=0.45\linewidth, height=0.305\linewidth]{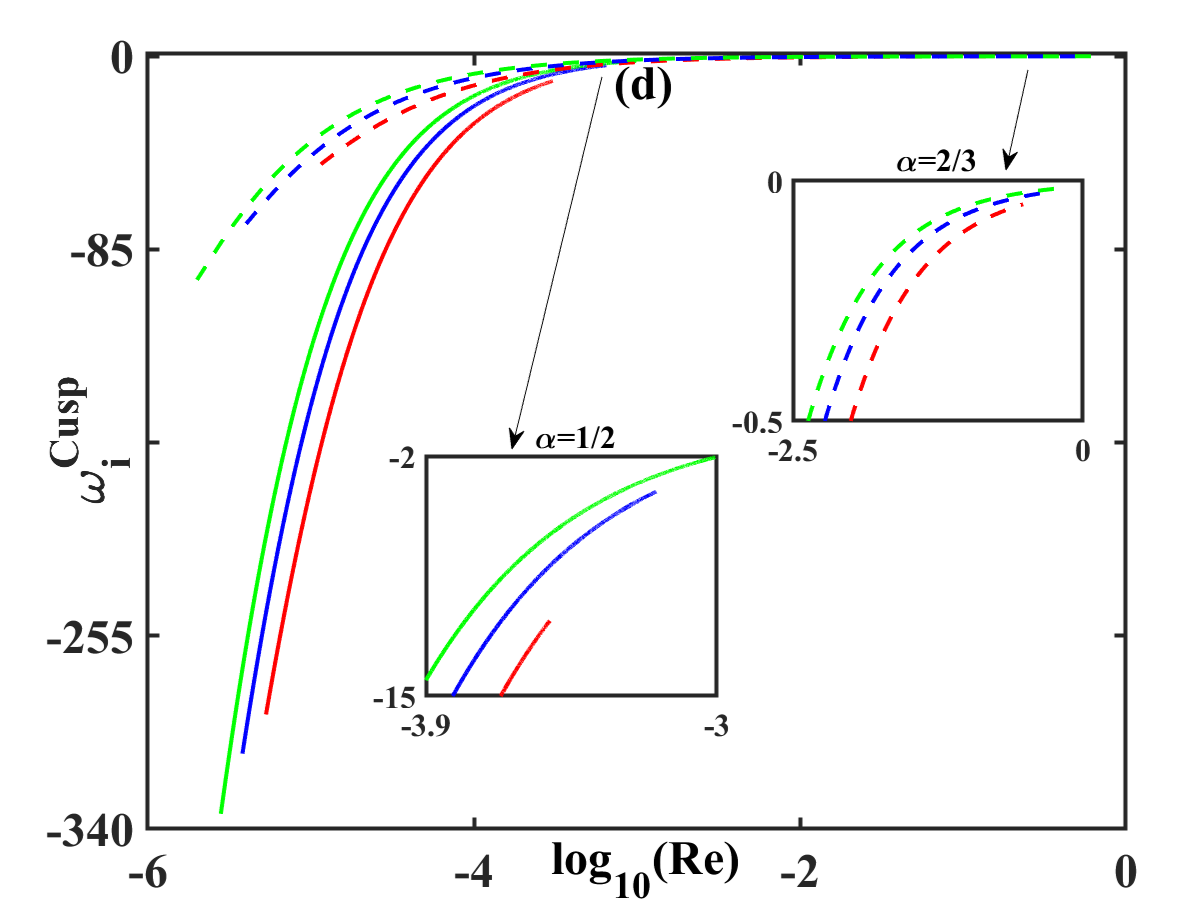}
\vskip -1pt
\includegraphics[width=0.45\linewidth, height=0.305\linewidth]{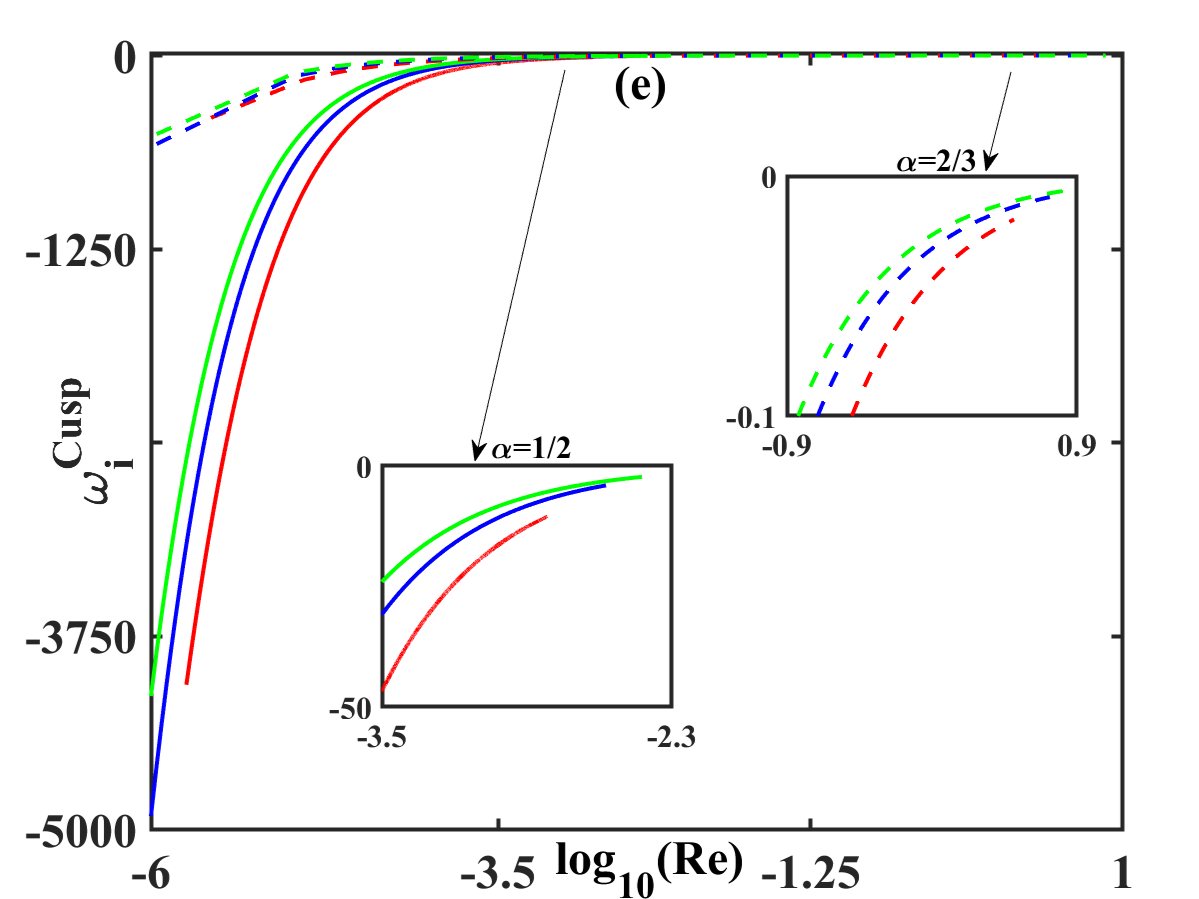}
\includegraphics[width=0.45\linewidth, height=0.305\linewidth]{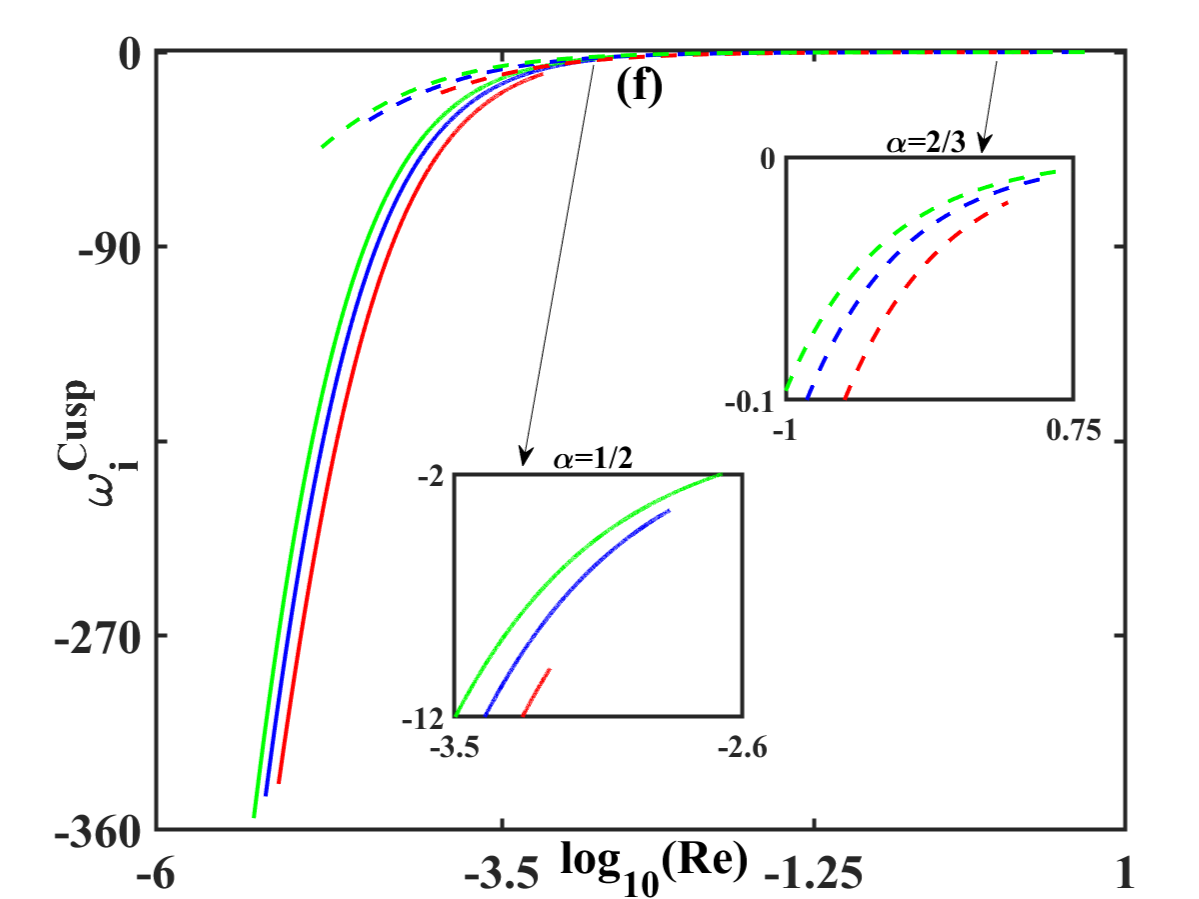}
\vskip -1pt
\includegraphics[width=0.45\linewidth, height=0.305\linewidth]{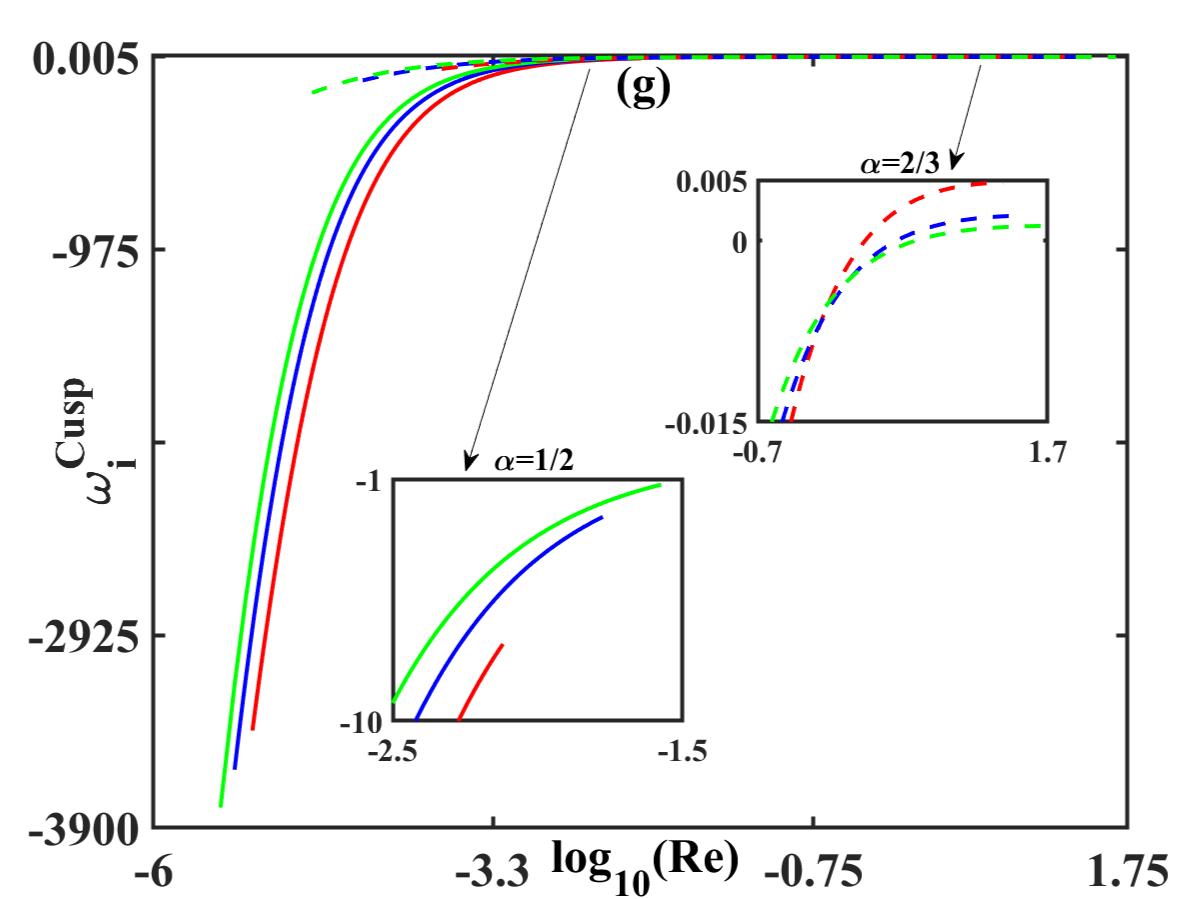}
\includegraphics[width=0.45\linewidth, height=0.305\linewidth]{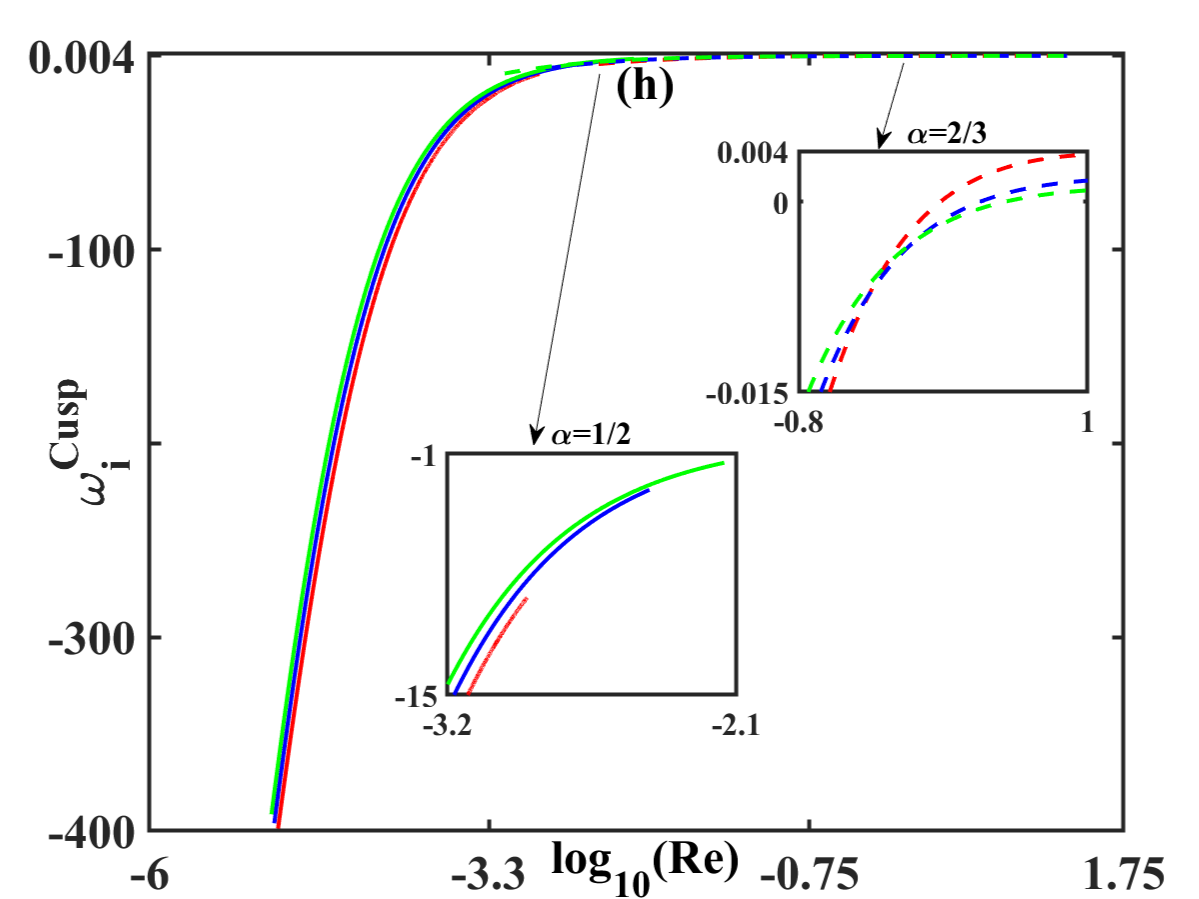}
\caption{Absolute growth rate, $\omega^{\text{Cusp}}_i$ for the {\it Couette flow} case, vs. Reynolds number for the Rouse model (solid curves) and for the Zimm's model (dashed curves), evaluated at $We=15.0, 25.0, 35.0$ (red, blue and green curves, respectively), viscosity ratios, $\nu = 0.05$ (left column) and $\nu = 0.3$ (right column) and at transverse spatial locations: (a, b) $y = 0.2$, (c, d) $y = 0.5$, (e, f) $y = 0.7$ and (g, h) $y = 0.9$.}
\label{fig5}
\end{figure*}

\begin{figure*}[htbp]
\centering
\includegraphics[width=0.45\linewidth, height=0.305\linewidth]{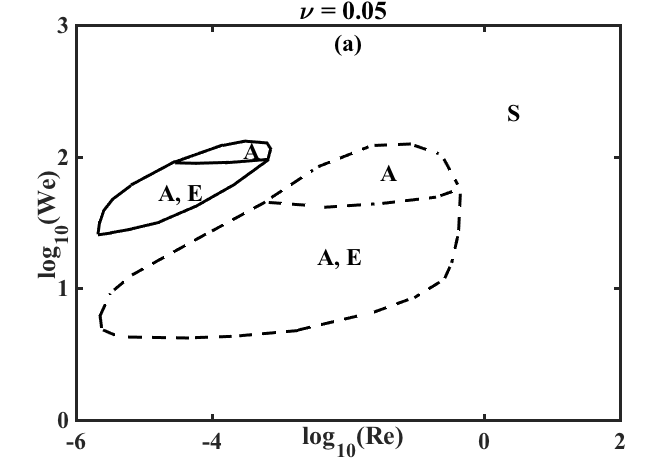}
\includegraphics[width=0.45\linewidth, height=0.305\linewidth]{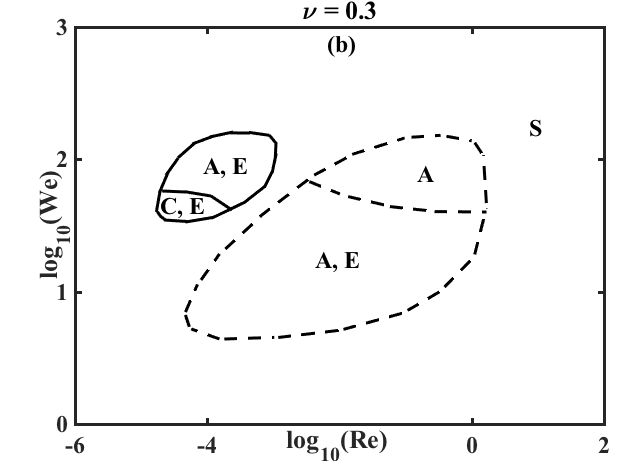}
\vskip -1pt
\includegraphics[width=0.45\linewidth, height=0.305\linewidth]{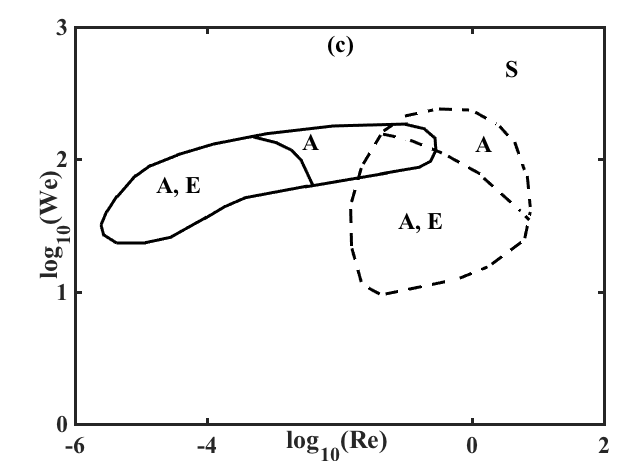}
\includegraphics[width=0.45\linewidth, height=0.305\linewidth]{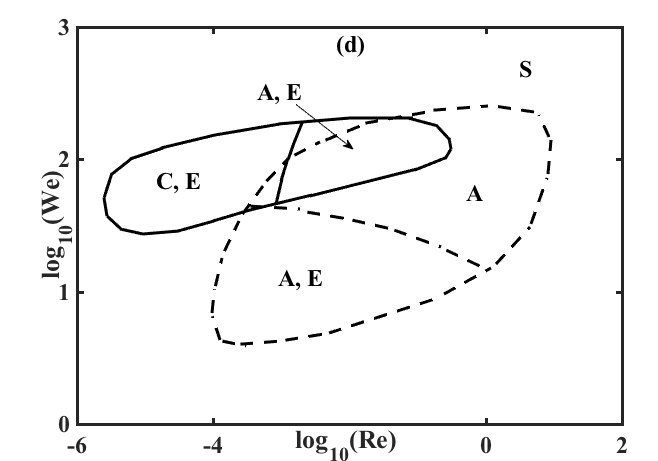}
\vskip -1pt
\includegraphics[width=0.45\linewidth, height=0.305\linewidth]{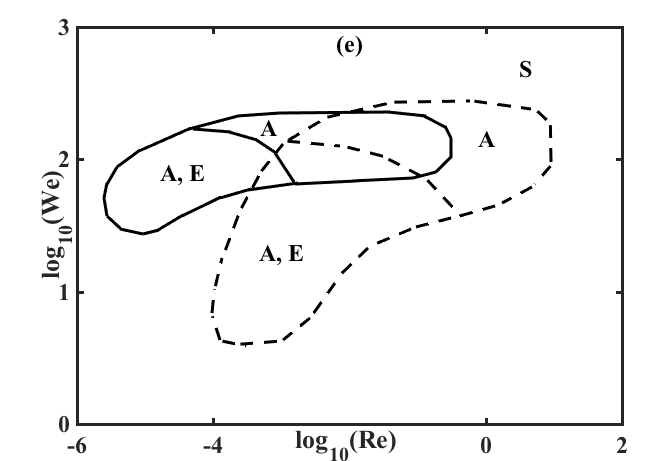}
\includegraphics[width=0.45\linewidth, height=0.305\linewidth]{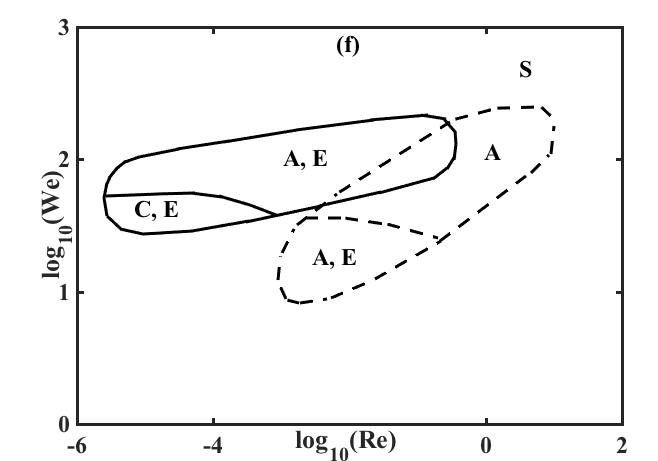}
\vskip -1pt
\includegraphics[width=0.45\linewidth, height=0.305\linewidth]{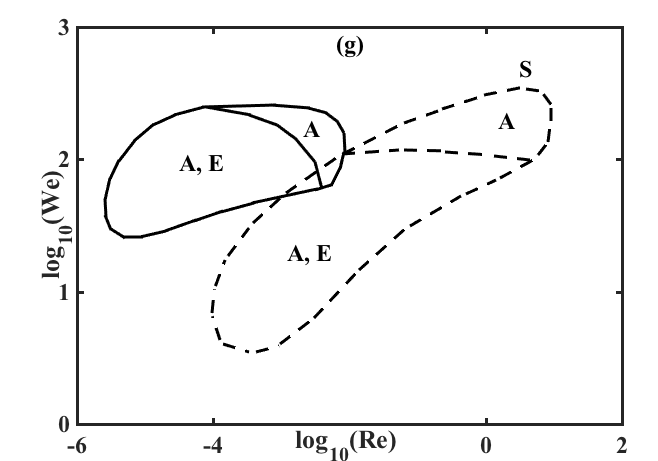}
\includegraphics[width=0.45\linewidth, height=0.305\linewidth]{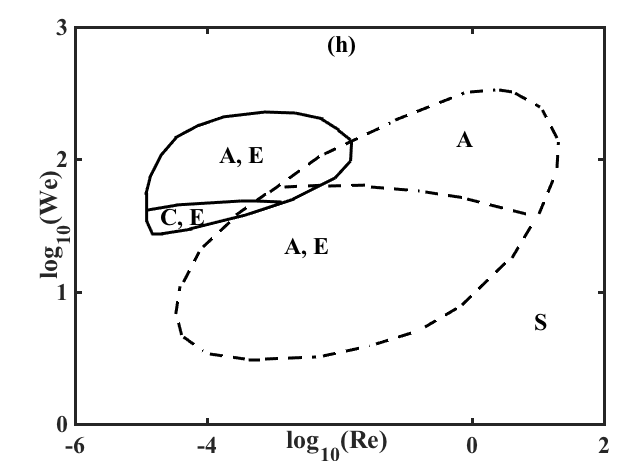}
\caption{Viscoelastic subdiffusive {\it Poiseuille flow} stability phase diagram in the $Re-We$ parametric space for the Rouse model (solid curves) and for the Zimm's model (dashed curves), evaluated at transverse spatial locations: (a, b) $y = 0.2$, (c, d) $y = 0.5$, (e, f) $y = 0.7$ and (g, h) $y = 0.9$ and at fixed values of viscosity ratios, $\nu = 0.05$ (left column) and $\nu = 0.3$ (right column).}
\label{fig6}
\end{figure*}

\begin{figure*}[htbp]
\centering
\includegraphics[width=0.45\linewidth, height=0.305\linewidth]{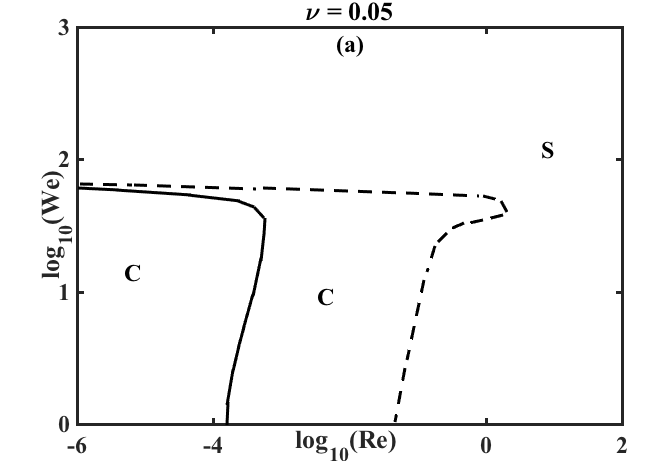}
\includegraphics[width=0.45\linewidth, height=0.305\linewidth]{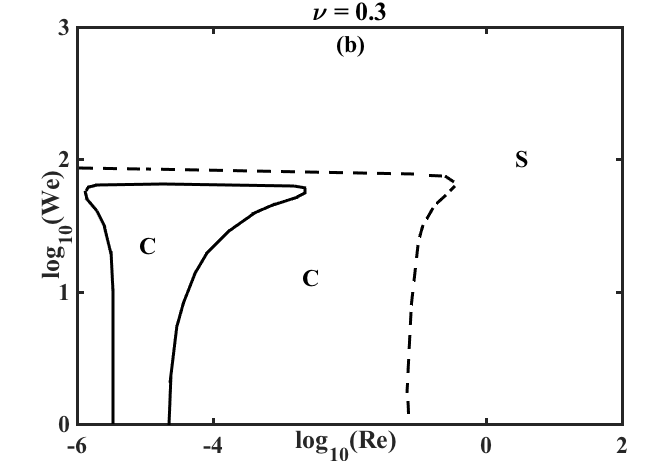}
\vskip -1pt
\includegraphics[width=0.45\linewidth, height=0.305\linewidth]{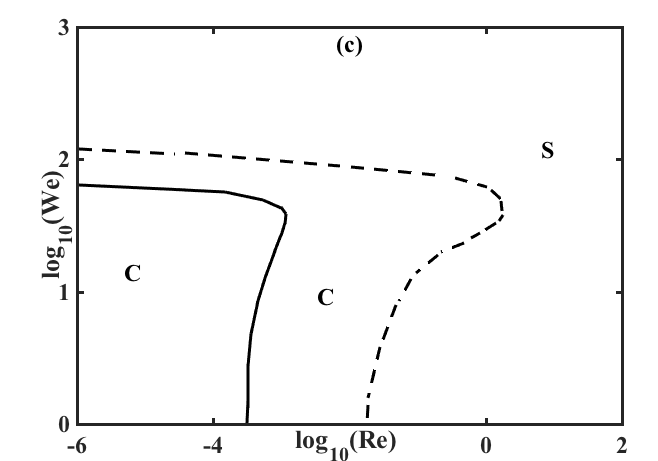}
\includegraphics[width=0.45\linewidth, height=0.305\linewidth]{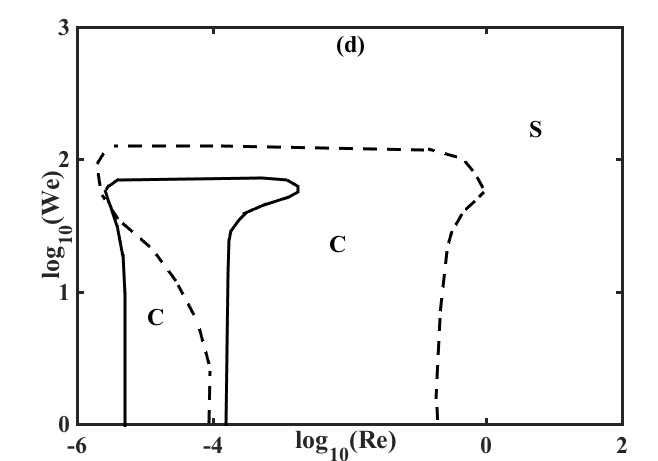}
\vskip -1pt
\includegraphics[width=0.45\linewidth, height=0.305\linewidth]{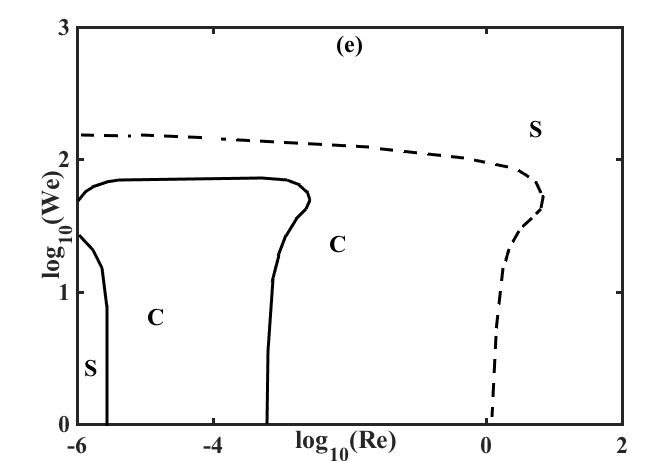}
\includegraphics[width=0.45\linewidth, height=0.305\linewidth]{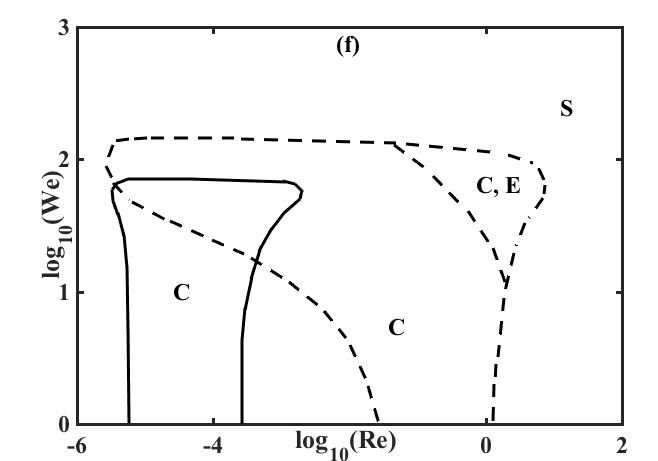}
\vskip -1pt
\includegraphics[width=0.45\linewidth, height=0.305\linewidth]{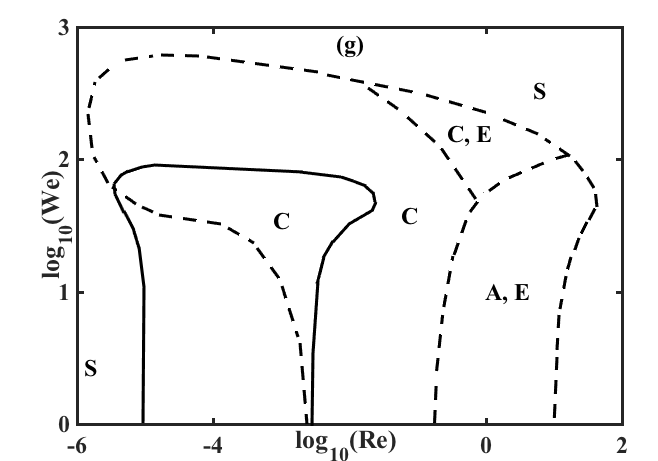}
\includegraphics[width=0.45\linewidth, height=0.305\linewidth]{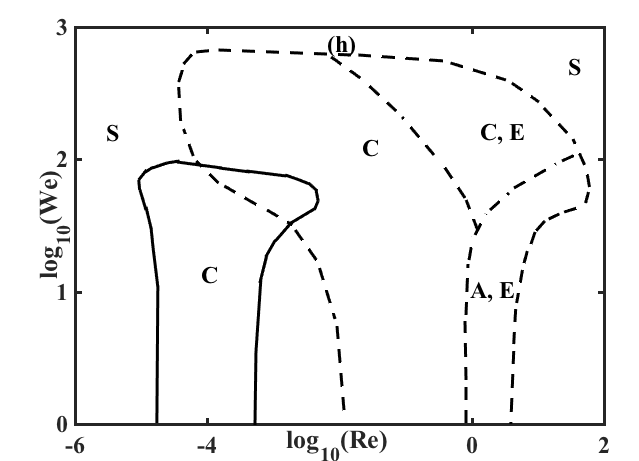}
\caption{Viscoelastic subdiffusive {\it Couette flow} stability phase diagram in the $Re-We$ parametric space for the Rouse model (solid curves) and for the Zimm's model (dashed curves), evaluated at transverse spatial locations: (a, b) $y = 0.2$, (c, d) $y = 0.5$, (e, f) $y = 0.7$ and (g, h) $y = 0.9$ and at fixed values of viscosity ratios, $\nu = 0.05$ (left column) and $\nu = 0.3$ (right column).}
\label{fig7}
\end{figure*}

For Couette flows (figure~\ref{fig5}), we find that the absolute growth rate curves display a transition from convective instability towards temporal stability versus $Re${\color{black},} such that the critical Reynolds number at the transition point increases with increasing transverse spatial coordinate, $y$, with an exception at $y=0.9$), irrespective of the selected values of $We, \nu$ or $\alpha$. At $y=0.9$, a transition sequence in the order: temporal stability$\rightarrow$ convective instability$\rightarrow$ temporal stability (temporal stability$\rightarrow$ convective instability$\rightarrow$ absolute instability$\rightarrow$ temporal stability) appears for the Rouse (Zimm's) model, with increasing values of $Re$. We remark that while some observations listed above follow from the well-established mechanisms seen in classical (or integer order) viscoelastic flows, namely the lack of symmetry in the flow-instability transition across the centerline (due to the anisotropy of the elastic stresses) as well as the appearance of absolute/convective instabilities at intermediate values of $Re$ (generated due to the instability via the polymer elasticity), other observations are relatively novel, specifically the flow induced (temporal) stabilization at higher values of $Re$.

Next, we classify the nature of these instabilities by computing the boundaries of the temporally stable regions ({\bf S}), evanescent modes ({\bf E}), the convectively unstable ({\bf C}) and the absolutely unstable regions ({\bf A}) within a selected range of the flow-elasticity-viscosity parameter space, i. e., $Re \in [10^{-6}, \, 100]$, $We \in [0,\,10^3]$ and $\nu \in \{ 0.05, 0.30 \}$. While convective instability grows in amplitude as it is swept along by the flow, absolute instability occurs at fixed spatial locations, leading to surface transitions (or pinch-off) of the advancing interface~\cite{Huerre1990}.
The flow stability phase diagram for Poiseuille flows, projected onto the $Re-We$ parameter space (figure~\ref{fig6}) divulge the presence of absolutely unstable and convectively unstable region at low to moderate values of $Re$ and $We$ ($Re \in [10^{-5}, \, 10]$ and $We \in [3, \, 350]$ for absolute instability, and $Re \in [10^{-5}, \, 10^{-2}]$ and $We \in [50, \, 150]$ for convective instability, respectively), as result of a complex tug-of-war between the inertial forces (proportional to $Re$) and the normal stress anisotropy through elasticity (proportional to $We$). Similarly, the flow stability phase diagram for Couette flows (figure~\ref{fig7}) disclose convectively unstable region at low to moderate values of $Re$ and moderately high values of $We$ ($Re \in [10^{-5}, \, 55]$ and $We \in [0, \, 900]$) and absolutely unstable region for moderate values of $Re$ and $We$, only near the upper plate (i.~e., $Re \in [0.1, \, 10]$ and $We \in [0, \, 100]$ at $y=0.9$, refer figure~\ref{fig7}h). To summarize, the parameter regions susceptible to topological transitions (or the parameter space which indicate absolute instability) in subdiffusive channel flows, are those driven by moderate inertia coupled with moderate to high elasticity.

A notably `abnormal' feature in the phase diagrams~(\ref{fig6}, \ref{fig7}) is the presence of temporal stability at high inertia (i.~e., $Re \ge 55$). While the in silico studies of the classical Oldroyd-B channel flows indicate the appearance of temporal instability for Reynolds number as low as $Re \sim 50$~\cite{Khalid2021}, temporal stability at high fluid inertia for viscoelastic flows is only recognized in experimental realizations (until now). For example, Riley~\cite{Riley1988} reported an elasticity induced flow stabilization of viscoelastic fluids coated over complaint surfaces at {\color{black}a} fairly high Reynolds number ($Re \sim 4000$). In a separate study involving ethanol gel fuels, elastic stabilization at {\color{black}a} high shear rate was attributed due to an abnormally high second normal stress difference~\cite{Nandagopalan2018}. Viscoelastic flow stabilization at higher values of $Re$, in tapered microchannels, was explained due to the presence of wall effects~\cite{Zarabadi2019}. In another in vitro study, a biofilm deacidification created a non-homogeneous environment for molecular diffusion, leading to a `subdiffusive effect' with hindered flow rates~\cite{Zarabadi2018}. These in vitro studies not only corroborate our numerical outcome, especially establishing the emergence of temporally stable region at high inertia, but also highlight the potential of fractional calculus in effectively capturing the flow-instability transition in subdiffusive flows.

\section{Concluding remarks}\label{sec:Conclusions}
This investigation addresses the temporal and the spatiotemporal linear stability analyses of viscoelastic, subdiffusive, plane Poiseuille and Couette flows in the limit of low to moderate Reynolds number and moderate to high Weissenberg number. Section~\ref{sec:math} presented the viscoelastic, subdiffusive channel flow model, the elements of linear stability analysis as well as the numerical method needed to solve the resulting dispersion relation. Section~\ref{sec:MV} validated the model for the classical planar Poiseuille flow obeying the Oldroyd-B stress constitutive equation~\cite{Atalik2002}. The temporal stability analysis in Section~\ref{subsec:TSA} indicates that with decreasing order of the fractional derivative: (a) the most unstable mode decreases, (b) the peak of the most unstable mode shifts to lower values of $Re$, and (c) in particular, the peak of the most unstable mode, for the Rouse model converges towards the limit $Re \rightarrow 0$. The spatiotemporal phase diagram in Section~\ref{subsec:STSA} indicates an abnormal region of temporal stability at high fluid inertia coupled with high elasticity, due to the presence of a non-homogeneous environment with hindered flow. Although we have shown how the exponent in the subdiffusive power{\color{black}-}law scaling of the mean square displacement of the tracer particle in the microscale is related to the fractional order of the corresponding non-linear stress constitutive equations in the continuum, the arguments presented herein are `phenomenological' in nature. A more rigorous effort involving the micro-to-macro upscaling via kinetic theory arguments~\cite{Spagnolie2015}, {\color{black}is} currently underway.

\section*{Acknowledgements}
T. C., D. B. and S.S. acknowledges the financial support of the Grant CSIR 09 /1117 (0012) /2020-EMR-I and DST ECR/2017/000632, respectively.

\begin{appendices}
\section{Viscoelastic dispersion relation}  \label{sec:appA}
The expressions, $M_i$, utilized in the viscoelastic dispersion relation outlined in Section~\ref{subsec:LSA}, is given as,
\begin{align}
& M_1 =  (y\!-\!y^2) \left(\!Re  (-i \omega)^\alpha \! +\! i k Re (y\!-\!y^2\! +\! \delta y) \! -\! {\color{black} 2} \nu(i k)^{2}\right)\! +\! {\color{black} 2} \nu , \nonumber\\
& M_2 = Re ( 1-2y+\delta) (y-y^2) - {\color{black} \nu} (1-2y) (i k), \nonumber\\
& M_3 = (y\!-\!y^2)\left(\! Re  (-i \omega)^\alpha \!+\! i k Re (y\!-\!y^2 \!+\! \delta y)\! -\! {\color{black} \nu}  (ik)^{2}\!\right) \!+\! {\color{black} 4} \nu , \nonumber\\
&  M_4 = (-i \omega)^\alpha  + i k (y-y^2+\delta y) + \frac{1}{We}, \nonumber \\
& M_5 = {\color{black} -i k} \left(1-2y+\delta \right)(y-y^2) - {\color{black} \frac{1}{ We}} \left( 1-2y \right),
\nonumber \\
& M_6 =- {\color{black} 2} (y-y^2) - {\color{black} 2} i k We (1-2y+\delta)^2 (y-y^2) - {\color{black}\left(1-
\right.} \nonumber \\
& \qquad {\color{black}\left.
 2y+\delta \right)} (1-2y) - {\color{black}\frac{1}{ We}} (y-y^2)(i k ) , \nonumber \\
&  M_7 =  - {\color{black}8 }  We \left(1-2y+\delta \right) (y-y^2) - {\color{black} 4} We (1-2y + \delta)^2 
\nonumber \\
& \qquad 
(1-2y)  -\frac{{\color{black} 2}(1-2y)}{We}.
\end{align}
\end{appendices}


\bibliography{sn-bibliography}


\begin{thebibliography}{68}
\ifx \bisbn   \undefined \def \bisbn  #1{ISBN #1}\fi
\ifx \binits  \undefined \def \binits#1{#1}\fi
\ifx \bauthor  \undefined \def \bauthor#1{#1}\fi
\ifx \batitle  \undefined \def \batitle#1{#1}\fi
\ifx \bjtitle  \undefined \def \bjtitle#1{#1}\fi
\ifx \bvolume  \undefined \def \bvolume#1{\textbf{#1}}\fi
\ifx \byear  \undefined \def \byear#1{#1}\fi
\ifx \bissue  \undefined \def \bissue#1{#1}\fi
\ifx \bfpage  \undefined \def \bfpage#1{#1}\fi
\ifx \blpage  \undefined \def \blpage #1{#1}\fi
\ifx \burl  \undefined \def \burl#1{\textsf{#1}}\fi
\ifx \doiurl  \undefined \def \doiurl#1{\url{https://doi.org/#1}}\fi
\ifx \betal  \undefined \def \betal{\textit{et al.}}\fi
\ifx \binstitute  \undefined \def \binstitute#1{#1}\fi
\ifx \binstitutionaled  \undefined \def \binstitutionaled#1{#1}\fi
\ifx \bctitle  \undefined \def \bctitle#1{#1}\fi
\ifx \beditor  \undefined \def \beditor#1{#1}\fi
\ifx \bpublisher  \undefined \def \bpublisher#1{#1}\fi
\ifx \bbtitle  \undefined \def \bbtitle#1{#1}\fi
\ifx \bedition  \undefined \def \bedition#1{#1}\fi
\ifx \bseriesno  \undefined \def \bseriesno#1{#1}\fi
\ifx \blocation  \undefined \def \blocation#1{#1}\fi
\ifx \bsertitle  \undefined \def \bsertitle#1{#1}\fi
\ifx \bsnm \undefined \def \bsnm#1{#1}\fi
\ifx \bsuffix \undefined \def \bsuffix#1{#1}\fi
\ifx \bparticle \undefined \def \bparticle#1{#1}\fi
\ifx \barticle \undefined \def \barticle#1{#1}\fi
\bibcommenthead
\ifx \bconfdate \undefined \def \bconfdate #1{#1}\fi
\ifx \botherref \undefined \def \botherref #1{#1}\fi
\ifx \url \undefined \def \url#1{\textsf{#1}}\fi
\ifx \bchapter \undefined \def \bchapter#1{#1}\fi
\ifx \bbook \undefined \def \bbook#1{#1}\fi
\ifx \bcomment \undefined \def \bcomment#1{#1}\fi
\ifx \oauthor \undefined \def \oauthor#1{#1}\fi
\ifx \citeauthoryear \undefined \def \citeauthoryear#1{#1}\fi
\ifx \endbibitem  \undefined \def \endbibitem {}\fi
\ifx \bconflocation  \undefined \def \bconflocation#1{#1}\fi
\ifx \arxivurl  \undefined \def \arxivurl#1{\textsf{#1}}\fi
\csname PreBibitemsHook\endcsname

\bibitem{Goychuk2017}
\begin{barticle}
\bauthor{\bsnm{Goychuk}, \binits{I.}},
\bauthor{\bsnm{Kharchenko}, \binits{V.O.}},
\bauthor{\bsnm{Metzler}, \binits{R.}}:
\batitle{Persistent {Sinai}-type diffusion in {Gaussian} random potentials with
  decaying spatial correlations}.
\bjtitle{Phys. Rev. E}
\bvolume{96}(\bissue{5}),
\bfpage{052134}
(\byear{2017}).
\doiurl{10.1103/PhysRevE.96.052134}
\end{barticle}
\endbibitem

\bibitem{Goychuk2020}
\begin{barticle}
\bauthor{\bsnm{Goychuk}, \binits{I.}},
\bauthor{\bsnm{P{\"o}schel}, \binits{T.}}:
\batitle{Hydrodynamic memory can boost enormously driven nonlinear diffusion
  and transport}.
\bjtitle{Phys. Rev. E}
\bvolume{102}(\bissue{1}),
\bfpage{012139}
(\byear{2020}).
\doiurl{10.1103/PhysRevE.102.012139}
\end{barticle}
\endbibitem

\bibitem{Goychuk2021}
\begin{barticle}
\bauthor{\bsnm{Goychuk}, \binits{I.}},
\bauthor{\bsnm{P{\"o}schel}, \binits{T.}}:
\batitle{Fingerprints of viscoelastic subdiffusion in random environments:
  {Revisiting} some experimental data and their interpretations}.
\bjtitle{Phys. Rev. E}
\bvolume{104}(\bissue{3}),
\bfpage{034125}
(\byear{2021}).
\doiurl{10.1103/PhysRevE.104.034125}
\end{barticle}
\endbibitem

\bibitem{Lai2009}
\begin{barticle}
\bauthor{\bsnm{Lai}, \binits{S.K.}},
\bauthor{\bsnm{Wang}, \binits{Y.Y.}},
\bauthor{\bsnm{Cone}, \binits{R.}},
\bauthor{\bsnm{Wirtz}, \binits{D.}},
\bauthor{\bsnm{Hanes}, \binits{J.}}:
\batitle{Altering {Mucus} {Rheology} to ``{Solidify}'' {Human} {Mucus} at the
  {Nanoscale}}.
\bjtitle{PLoS ONE}
\bvolume{4}(\bissue{1}),
\bfpage{4294}
(\byear{2009}).
\doiurl{10.1371/journal.pone.0004294}
\end{barticle}
\endbibitem

\bibitem{Coffey2004}
\begin{bbook}
\bauthor{\bsnm{Coffey}, \binits{W.T.}},
\bauthor{\bsnm{Kalmykov}, \binits{P.Y.}},
\bauthor{\bsnm{Waldron}, \binits{J.}}:
\bbtitle{The {Langevin} {Equation}: {With} {Applications} to {Stochastic}
  {Problems} in {Physics}, {Chemistry} and {Electrical} {Engineering}},
\bedition{2}nd edn.
\bsertitle{World {Scientific} {Series} in {Contemporary} {Chemical} {Physics}},
vol. \bseriesno{14}.
\bpublisher{World Scientific}, \blocation{???}
(\byear{2004}).
\doiurl{10.1142/5343}
\end{bbook}
\endbibitem

\bibitem{Rubenstein2003}
\begin{bbook}
\bauthor{\bsnm{Rubenstein}, \binits{M.}},
\bauthor{\bsnm{Colby}, \binits{R.H.}}:
\bbtitle{Polymer Physics}.
\bpublisher{Oxford University Press},
\blocation{New York}
(\byear{2003})
\end{bbook}
\endbibitem

\bibitem{Levine2001}
\begin{barticle}
\bauthor{\bsnm{Levine}, \binits{A.J.}},
\bauthor{\bsnm{Lubensky}, \binits{T.C.}}:
\batitle{Response function of a sphere in a viscoelastic two-fluid medium}.
\bjtitle{Phys. Rev. E}
\bvolume{63}(\bissue{4}),
\bfpage{041510}
(\byear{2001}).
\doiurl{10.1103/PhysRevE.63.041510}
\end{barticle}
\endbibitem

\bibitem{Kremer1990}
\begin{barticle}
\bauthor{\bsnm{Kremer}, \binits{K.}},
\bauthor{\bsnm{Grest}, \binits{G.S.}}:
\batitle{Dynamics of entangled linear polymer melts: {A} molecular‐dynamics
  simulation}.
\bjtitle{J. Chem. Phys.}
\bvolume{92}(\bissue{8}),
\bfpage{5057}--\blpage{5086}
(\byear{1990}).
\doiurl{10.1063/1.458541}
\end{barticle}
\endbibitem

\bibitem{Kou2004}
\begin{barticle}
\bauthor{\bsnm{Kou}, \binits{S.C.}},
\bauthor{\bsnm{Xie}, \binits{X.S.}}:
\batitle{Generalized {Langevin} {Equation} with {Fractional} {Gaussian}
  {Noise}: {Subdiffusion} within a {Single} {Protein} {Molecule}}.
\bjtitle{Phys. Rev. Lett.}
\bvolume{93}(\bissue{18}),
\bfpage{180603}
(\byear{2004}).
\doiurl{10.1103/PhysRevLett.93.180603}
\end{barticle}
\endbibitem

\bibitem{Fricks2009}
\begin{barticle}
\bauthor{\bsnm{Fricks}, \binits{J.}},
\bauthor{\bsnm{Yao}, \binits{L.}},
\bauthor{\bsnm{Elston}, \binits{T.C.}},
\bauthor{\bsnm{Forest}, \binits{M.G.}}:
\batitle{Time-{Domain} {Methods} for {Diffusive} {Transport} in {Soft}
  {Matter}}.
\bjtitle{SIAM J. Appl. Math.}
\bvolume{69}(\bissue{5}),
\bfpage{1277}--\blpage{1308}
(\byear{2009}).
\doiurl{10.1137/070695186}
\end{barticle}
\endbibitem

\bibitem{Morgado2002}
\begin{barticle}
\bauthor{\bsnm{Morgado}, \binits{R.}},
\bauthor{\bsnm{Oliveira}, \binits{F.A.}},
\bauthor{\bsnm{Batrouni}, \binits{G.G.}},
\bauthor{\bsnm{Hansen}, \binits{A.}}:
\batitle{Relation between {Anomalous} and {Normal} {Diffusion} in {Systems}
  with {Memory}}.
\bjtitle{Phys. Rev. Lett.}
\bvolume{89}(\bissue{10}),
\bfpage{100601}
(\byear{2002}).
\doiurl{10.1103/PhysRevLett.89.100601}
\end{barticle}
\endbibitem

\bibitem{Vainstein2008}
\begin{botherref}
\oauthor{\bsnm{Vainstein}, \binits{M.H.}},
\oauthor{\bsnm{Lapas}, \binits{L.C.}},
\oauthor{\bsnm{Oliveira}, \binits{F.A.}}:
Anomalous {Diffusion}.
Technical Report arXiv:0805.0270,
arXiv
(2008)
\end{botherref}
\endbibitem

\bibitem{Adelman1976}
\begin{barticle}
\bauthor{\bsnm{Adelman}, \binits{S.A.}}:
\batitle{Fokker--{Planck} equations for simple non-markovian systems}.
\bjtitle{J. Chem. Phys.}
\bvolume{64}(\bissue{1}),
\bfpage{124}--\blpage{130}
(\byear{1976}).
\doiurl{10.1063/1.431961}
\end{barticle}
\endbibitem

\bibitem{gemant1936}
\begin{barticle}
\bauthor{\bsnm{Gemant}, \binits{A.}}:
\batitle{A method of analyzing experimental results obtained from
  elasto-viscous bodies}.
\bjtitle{Physics}
\bvolume{7}(\bissue{8}),
\bfpage{311}--\blpage{317}
(\byear{1936}).
\doiurl{10.1063/1.1745400}
\end{barticle}
\endbibitem

\bibitem{gemant1938}
\begin{barticle}
\bauthor{\bsnm{Gemant}, \binits{A.}}:
\batitle{{XLV}. \textit{{On}} fractional differentials}.
\bjtitle{The London, Edinburgh, and Dublin Philosophical Magazine and Journal
  of Science}
\bvolume{25}(\bissue{168}),
\bfpage{540}--\blpage{549}
(\byear{1938}).
\doiurl{10.1080/14786443808562036}
\end{barticle}
\endbibitem

\bibitem{Blair1944}
\begin{barticle}
\bauthor{\bsnm{Scott-Blair}, \binits{G.W.}}:
\batitle{Analytical and {Integrative} {Aspects} of the {Stress}-{Strain}-{Time}
  {Problem}}.
\bjtitle{J. Sci. Instr.}
\bvolume{21}(\bissue{5}),
\bfpage{80}--\blpage{84}
(\byear{1944}).
\doiurl{10.1088/0950-7671/21/5/302}
\end{barticle}
\endbibitem

\bibitem{Blair1947}
\begin{barticle}
\bauthor{\bsnm{Scott-Blair}, \binits{G.W.}}:
\batitle{The role of psychophysics in rheology}.
\bjtitle{J. Coll. Sci.}
\bvolume{2}(\bissue{1}),
\bfpage{21}--\blpage{32}
(\byear{1947}).
\doiurl{10.1016/0095-8522(47)90007-X}
\end{barticle}
\endbibitem

\bibitem{caputo1967}
\begin{barticle}
\bauthor{\bsnm{M.~Caputo}, \binits{M.}}:
\batitle{Linear {Models} of {Dissipation} whose {Q} is almost {Frequency}
  {Independent}--{II}}.
\bjtitle{Geophys. J. Intern.}
\bvolume{13}(\bissue{5}),
\bfpage{529}--\blpage{539}
(\byear{1967}).
\doiurl{10.1111/j.1365-246X.1967.tb02303.x}
\end{barticle}
\endbibitem

\bibitem{bagley1983}
\begin{barticle}
\bauthor{\bsnm{Bagley}, \binits{R.L.}},
\bauthor{\bsnm{Torvik}, \binits{P.J.}}:
\batitle{A {Theoretical} {Basis} for the {Application} of {Fractional}
  {Calculus} to {Viscoelasticity}}.
\bjtitle{J. Rheol.}
\bvolume{27}(\bissue{3}),
\bfpage{201}--\blpage{210}
(\byear{1983}).
\doiurl{10.1122/1.549724}
\end{barticle}
\endbibitem

\bibitem{Rouse1953}
\begin{barticle}
\bauthor{\bsnm{Rouse}, \binits{P.E.}}:
\batitle{A {Theory} of the {Linear} {Viscoelastic} {Properties} of {Dilute}
  {Solutions} of {Coiling} {Polymers}}.
\bjtitle{J. Chem. Phys.}
\bvolume{21}(\bissue{7}),
\bfpage{1272}--\blpage{1280}
(\byear{1953}).
\doiurl{10.1063/1.1699180}
\end{barticle}
\endbibitem

\bibitem{Tan2002}
\begin{barticle}
\bauthor{\bsnm{Tan}, \binits{W.}},
\bauthor{\bsnm{Xu}, \binits{M.}}:
\batitle{Plane surface suddenly set in motion in a viscoelastic fluid with
  fractional maxwell model}.
\bjtitle{Acta Mech.}
\bvolume{18}(\bissue{4}),
\bfpage{342}--\blpage{349}
(\byear{2002})
\end{barticle}
\endbibitem

\bibitem{Qi2009}
\begin{barticle}
\bauthor{\bsnm{Qi}, \binits{M.}},
\bauthor{\bsnm{Xu}, \binits{M.}}:
\batitle{Some unsteady unidirectional flows of a generalized oldroyd-b fluid
  with fractional derivative}.
\bjtitle{Appl. Math. Model}
\bvolume{33},
\bfpage{4184}--\blpage{4191}
(\byear{2009})
\end{barticle}
\endbibitem

\bibitem{Fetecau2009}
\begin{botherref}
\oauthor{\bsnm{Fetecau}, \binits{C.}},
\oauthor{\bsnm{Fetecau}, \binits{C.}},
\oauthor{\bsnm{Kamran}, \binits{M.}},
\oauthor{\bsnm{Vieru}, \binits{D.}}:
Exact solutions for the flow of a generalized oldroyd-b fluid induced by a
  constantly accelerating plate between two side walls perpendicular to the
  plate.
J. non-Newt. Fluid Mech.,
189--201
(2009)
\end{botherref}
\endbibitem

\bibitem{Zheng2012}
\begin{barticle}
\bauthor{\bsnm{Zheng}, \binits{L.}},
\bauthor{\bsnm{Liu}, \binits{Y.}},
\bauthor{\bsnm{Zhang}, \binits{X.}}:
\batitle{Slip effects on mhd flow of a generalized oldroyd-b fluid with
  fractional derivative}.
\bjtitle{Nonlin. Anal. RWA}
\bvolume{13},
\bfpage{513}--\blpage{523}
(\byear{2012})
\end{barticle}
\endbibitem

\bibitem{Zhao2016}
\begin{barticle}
\bauthor{\bsnm{J.~Zhao}, \binits{X.Z.} \bsuffix{L.~Zheng}},
\bauthor{\bsnm{Liu}, \binits{F.}}:
\batitle{Unsteady natural convection boundary layer heat transfer of fractional
  maxwell viscoelastic fluid over a vertical plate}.
\bjtitle{Int. J. Heat Mass Trans.}
\bvolume{47},
\bfpage{760}--\blpage{766}
(\byear{2016})
\end{barticle}
\endbibitem

\bibitem{Ancey2019}
\begin{botherref}
\oauthor{\bsnm{Ancey}, \binits{C.}}:
Bedload transport: a walk between randomness and determinism. part 1. the state
  of the art.
J. Hydrau. Res.
\textbf{58}
(2020)
\end{botherref}
\endbibitem

\bibitem{Siedlik2015}
\begin{barticle}
\bauthor{\bsnm{Siedlik}, \binits{M.J.}},
\bauthor{\bsnm{Nelson}, \binits{C.M.}}:
\batitle{Regulation of tissue morphodynamics: an important role for actomyosin
  contractility}.
\bjtitle{Curr Opin Genet Dev.}
\bvolume{32},
\bfpage{80}--\blpage{85}
(\byear{2015})
\end{barticle}
\endbibitem

\bibitem{Zaks2018}
\begin{barticle}
\bauthor{\bsnm{Zaks}, \binits{M.A.}},
\bauthor{\bsnm{Nepomnyashchy}, \binits{A.}}:
\batitle{Subdiffusive and superdiffusive transport in plane steady viscous
  flows}.
\bjtitle{PNAS}
\bvolume{116}(\bissue{37}),
\bfpage{18245}--\blpage{18250}
(\byear{2018})
\end{barticle}
\endbibitem

\bibitem{Khalid2021}
\begin{botherref}
\oauthor{\bsnm{Khalid}, \binits{M.}},
\oauthor{\bsnm{Chaudhary}, \binits{I.}},
\oauthor{\bsnm{Garg}, \binits{P.}},
\oauthor{\bsnm{Shankar}, \binits{V.}},
\oauthor{\bsnm{Subramanian}, \binits{G.}}:
The centre-mode instability of viscoelastic plane poiseuille flow.
J. Fluid Mech.
\textbf{915}(A43)
(2021)
\end{botherref}
\endbibitem

\bibitem{zwanzig1970}
\begin{barticle}
\bauthor{\bsnm{Zwanzig}, \binits{R.}},
\bauthor{\bsnm{Bixon}, \binits{M.}}:
\batitle{Hydrodynamic {Theory} of the {Velocity} {Correlation} {Function}}.
\bjtitle{Phys. Rev. A}
\bvolume{2}(\bissue{5}),
\bfpage{2005}--\blpage{2012}
(\byear{1970}).
\doiurl{10.1103/PhysRevA.2.2005}
\end{barticle}
\endbibitem

\bibitem{Mason1995}
\begin{barticle}
\bauthor{\bsnm{Mason}, \binits{T.G.}},
\bauthor{\bsnm{Weitz}, \binits{D.A.}}:
\batitle{Optical {Measurements} of {Frequency}-{Dependent} {Linear}
  {Viscoelastic} {Moduli} of {Complex} {Fluids}}.
\bjtitle{Phys. Rev. Lett.}
\bvolume{74}(\bissue{7}),
\bfpage{1250}--\blpage{1253}
(\byear{1995}).
\doiurl{10.1103/PhysRevLett.74.1250}
\end{barticle}
\endbibitem

\bibitem{Mason1996}
\begin{barticle}
\bauthor{\bsnm{Mason}, \binits{T.G.}},
\bauthor{\bsnm{Gang}, \binits{H.}},
\bauthor{\bsnm{Weitz}, \binits{D.A.}}:
\batitle{Rheology of complex fluids measured by dynamic light scattering}.
\bjtitle{J. Mol. Struct.}
\bvolume{383},
\bfpage{81}--\blpage{90}
(\byear{1996})
\end{barticle}
\endbibitem

\bibitem{zimm1956}
\begin{barticle}
\bauthor{\bsnm{Zimm}, \binits{B.H.}}:
\batitle{Dynamics of {Polymer} {Molecules} in {Dilute} {Solution}:
  {Viscoelasticity}, {Flow} {Birefringence} and {Dielectric} {Loss}}.
\bjtitle{J. Chem. Phys.}
\bvolume{24}(\bissue{2}),
\bfpage{269}--\blpage{278}
(\byear{1956}).
\doiurl{10.1063/1.1742462}
\end{barticle}
\endbibitem

\bibitem{Kirkwood1954}
\begin{botherref}
\oauthor{\bsnm{Kirkwood}, \binits{J.G.}}:
The general theory of irreversible processes in solutions of macromolecules.
J. Poly. Sci.
\textbf{12}(1)
(1954)
\end{botherref}
\endbibitem

\bibitem{Makris2021}
\begin{botherref}
\oauthor{\bsnm{Makris}, \binits{N.}}:
A rheological analog for brownian motion with hydrodynamic memory.
Phys. Fluids
\textbf{33}(072014)
(2021)
\end{botherref}
\endbibitem

\bibitem{Brader2010}
\begin{botherref}
\oauthor{\bsnm{Brader}, \binits{J.M.}}:
Nonlinear rheology of colloidal dispersions.
J. Phys.: Condens. Matter
\textbf{22}(363101)
(2010)
\end{botherref}
\endbibitem

\bibitem{Mckinley2009}
\begin{barticle}
\bauthor{\bsnm{S.~McKinley}, \binits{L.Y.}},
\bauthor{\bsnm{Forest}, \binits{M.G.}}:
\batitle{Transient anomalous diffusion of tracer particles in soft matter}.
\bjtitle{J. Rheol.}
\bvolume{53}(\bissue{6}),
\bfpage{1487}--\blpage{1506}
(\byear{2009})
\end{barticle}
\endbibitem

\bibitem{Sircar2010}
\begin{barticle}
\bauthor{\bsnm{Sircar}, \binits{S.}},
\bauthor{\bsnm{Wang}, \binits{Q.}}:
\batitle{Transient rheological responses in sheared biaxial liquid crystals}.
\bjtitle{Rheo. Acta}
\bvolume{49}(\bissue{7}),
\bfpage{699}--\blpage{717}
(\byear{2010})
\end{barticle}
\endbibitem

\bibitem{Sircar2010eLC}
\begin{botherref}
\oauthor{\bsnm{Li}, \binits{J.}},
\oauthor{\bsnm{Sircar}, \binits{S.}},
\oauthor{\bsnm{Wang}, \binits{Q.}}:
A note on the kinematics of rigid molecules in linear flow fields and kinetic
  theory for biaxial liquid crystal polymers.
e-LC Commun. (DOI: https://citeseerx.ist.psu.edu/viewdoc/summary?
  doi=10.1.1.532.8779)
(2010)
\end{botherref}
\endbibitem

\bibitem{Sircar2010IJEFMS}
\begin{botherref}
\oauthor{\bsnm{Sircar}, \binits{S.}}:
A hydrodynamical kinetic theory for self-propelled ellipsoidal suspensions.
Int. J. Emerg. Multi. Flu. Sci.
\textbf{2}(4)
(2010)
\end{botherref}
\endbibitem

\bibitem{Sircar2015}
\begin{barticle}
\bauthor{\bsnm{Sircar}, \binits{S.}},
\bauthor{\bsnm{Younger}, \binits{J.G.}},
\bauthor{\bsnm{Bortz}, \binits{D.M.}}:
\batitle{Sticky surface: sphere--sphere adhesion dynamics}.
\bjtitle{J. Biol. Dyna.}
\bvolume{9},
\bfpage{79}--\blpage{89}
(\byear{2015})
\end{barticle}
\endbibitem

\bibitem{Sircar2015JTB}
\begin{barticle}
\bauthor{\bsnm{Sircar}, \binits{S.}},
\bauthor{\bsnm{Aisenbrey}, \binits{E.}},
\bauthor{\bsnm{Bryant}, \binits{S.J.}},
\bauthor{\bsnm{Bortz}, \binits{D.M.}}:
\batitle{Determining equilibrium osmolarity in poly (ethylene
  glycol)/chondrotin sulfate gels mimicking articular cartilage}.
\bjtitle{J. Theo. Biol.}
\bvolume{364},
\bfpage{397}--\blpage{406}
(\byear{2015})
\end{barticle}
\endbibitem

\bibitem{Sircar2016JMB}
\begin{barticle}
\bauthor{\bsnm{Sircar}, \binits{S.}},
\bauthor{\bsnm{Roberts}, \binits{A.J.}}:
\batitle{Surface deformation and shear flow in ligand mediated cell adhesion}.
\bjtitle{J. Math. Biol.}
\bvolume{73}(\bissue{4}),
\bfpage{1035}--\blpage{1052}
(\byear{2016})
\end{barticle}
\endbibitem

\bibitem{Sircar2016EPJE}
\begin{barticle}
\bauthor{\bsnm{Sircar}, \binits{S.}},
\bauthor{\bsnm{Nguyen}, \binits{G.}},
\bauthor{\bsnm{Kotousov}, \binits{A.}},
\bauthor{\bsnm{Roberts}, \binits{A.J.}}:
\batitle{Ligand-mediated adhesive mechanics of two static, deformed spheres}.
\bjtitle{Eur. Phys. J. E}
\bvolume{39}(\bissue{10}),
\bfpage{1}--\blpage{9}
(\byear{2016})
\end{barticle}
\endbibitem

\bibitem{Sircar2019}
\begin{barticle}
\bauthor{\bsnm{Sircar}, \binits{S.}},
\bauthor{\bsnm{Bansal}, \binits{D.}}:
\batitle{Spatiotemporal linear stability of viscoelastic free shear flows:
  Dilute regime}.
\bjtitle{Phys. Fluids}
\bvolume{31}(\bissue{8}),
\bfpage{084104}
(\byear{2019})
\end{barticle}
\endbibitem

\bibitem{Sircar2020}
\begin{botherref}
\oauthor{\bsnm{Singh}, \binits{S.}},
\oauthor{\bsnm{Bansal}, \binits{D.}},
\oauthor{\bsnm{Kaur}, \binits{G.}},
\oauthor{\bsnm{Sircar}, \binits{S.}}:
Implicit-explicit-compact methods for advection diffusion reaction equations.
Comp. Fluids
\textbf{212}(104709)
(2020)
\end{botherref}
\endbibitem

\bibitem{Glockle1991}
\begin{barticle}
\bauthor{\bsnm{Glockle}, \binits{W.G.}},
\bauthor{\bsnm{Nonnenmacher}, \binits{T.F.}}:
\batitle{Fractional integral operators and {Fox} functions in the theory of
  viscoelasticity}.
\bjtitle{Macromolecules}
\bvolume{24},
\bfpage{6426}--\blpage{6434}
(\byear{1991}).
\doiurl{10.1021/ma00024a009}
\end{barticle}
\endbibitem

\bibitem{Glockle1994}
\begin{barticle}
\bauthor{\bsnm{Glockle}, \binits{W.G.}},
\bauthor{\bsnm{Nonnenmacher}, \binits{T.F.}}:
\batitle{Fractional relaxation and the time-temperature superposition
  principle}.
\bjtitle{Rheo. Acta}
\bvolume{33},
\bfpage{337}--\blpage{343}
(\byear{1994}).
\doiurl{10.1007/BF00366960}
\end{barticle}
\endbibitem

\bibitem{Prodanov2016}
\begin{botherref}
\oauthor{\bsnm{Prodanov}, \binits{D.}}:
Some applications of fractional velocities.
Frac. Calc. Appl. Anal.
\textbf{19}(173-187)
(2016)
\end{botherref}
\endbibitem

\bibitem{Prodanov2017}
\begin{barticle}
\bauthor{\bsnm{Prodanov}, \binits{D.}}:
\batitle{Conditions for continuity of fractional velocity and existence of
  fractional taylor expansions}.
\bjtitle{Chaos Sol. Fractals}
\bvolume{102},
\bfpage{236}--\blpage{244}
(\byear{2017})
\end{barticle}
\endbibitem

\bibitem{Prodanov2018}
\begin{barticle}
\bauthor{\bsnm{Prodanov}, \binits{D.}}:
\batitle{Fractional velocity as a tool for the study of non-linear problems}.
\bjtitle{Fractal Fract.}
\bvolume{2}(\bissue{1}),
\bfpage{2}--\blpage{23}
(\byear{2018})
\end{barticle}
\endbibitem

\bibitem{Macosko1994}
\begin{bbook}
\bauthor{\bsnm{Macosko}, \binits{C.W.}}:
\bbtitle{Rheology: Principles, Measurements, and Applications},
\bedition{1}st edn.
\bpublisher{Wiley}, \blocation{???}
(\byear{1994})
\end{bbook}
\endbibitem

\bibitem{Spagnolie2015}
\begin{bbook}
\bauthor{\bsnm{Spagnolie}, \binits{S.E.}}:
\bbtitle{Complex Fluids in Biological Systems: Experiment, Theory, and
  Computation}.
\bpublisher{Springer}, \blocation{???}
(\byear{2015})
\end{bbook}
\endbibitem

\bibitem{Jimenez2002}
\begin{barticle}
\bauthor{\bsnm{A.~H.~Jim\'{e}nez}, \binits{A.M.G.} \bsuffix{J.~H.~Santiago}},
\bauthor{\bsnm{Gonz\'{a}les}, \binits{J.S.}}:
\batitle{Relaxation modulus in pmma and ptfe fitting by fractional maxwell
  model}.
\bjtitle{Polym. Testing}
\bvolume{21},
\bfpage{325}--\blpage{331}
(\byear{2002})
\end{barticle}
\endbibitem

\bibitem{Bansal2021}
\begin{botherref}
\oauthor{\bsnm{Bansal}, \binits{D.}},
\oauthor{\bsnm{Ghosh}, \binits{D.}},
\oauthor{\bsnm{Sircar}, \binits{S.}}:
Spatiotemporal linear stability of viscoelastic free shear flows: Nonaffine
  response regime.
Phys. Fluids
\textbf{33}(054106)
(2021)
\end{botherref}
\endbibitem

\bibitem{Jaishankar2014}
\begin{barticle}
\bauthor{\bsnm{Jaishankar}, \binits{A.}},
\bauthor{\bsnm{McKinley}, \binits{G.H.}}:
\batitle{A fractional k-bkz constitutive formulation for describing the
  nonlinear rheology of multiscale complex fluids}.
\bjtitle{J. Rheol.}
\bvolume{58}(\bissue{6}),
\bfpage{1751}--\blpage{1788}
(\byear{2014})
\end{barticle}
\endbibitem

\bibitem{Bistagnino2007}
\begin{barticle}
\bauthor{\bsnm{Bistagnino}, \binits{A.}},
\bauthor{\bsnm{Boffetta}, \binits{G.}},
\bauthor{\bsnm{Celani}, \binits{A.}},
\bauthor{\bsnm{Mazzino}, \binits{A.}},
\bauthor{\bsnm{Puliafito}, \binits{A.}},
\bauthor{\bsnm{Vergassola}, \binits{M.}}:
\batitle{Nonlinear dynamics of the viscoelastic kolmogorov flow}.
\bjtitle{J. Fluid Mech.}
\bvolume{590},
\bfpage{61}--\blpage{80}
(\byear{2007})
\end{barticle}
\endbibitem

\bibitem{Huerre1990}
\begin{barticle}
\bauthor{\bsnm{Huerre}, \binits{P.}},
\bauthor{\bsnm{Monkewitz}, \binits{P.A.}}:
\batitle{Local and global instabilities in spatially developing flows}.
\bjtitle{Ann. Rev. Fluid Mech.}
\bvolume{22},
\bfpage{473}--\blpage{537}
(\byear{1990})
\end{barticle}
\endbibitem

\bibitem{Briggs1964}
\begin{bbook}
\bauthor{\bsnm{Briggs}, \binits{R.J.}}:
\bbtitle{Electron-stream Interaction with Plasmas}.
\bpublisher{MIT Press},
\blocation{Cambridge}
(\byear{1964})
\end{bbook}
\endbibitem

\bibitem{Kupfer1987}
\begin{barticle}
\bauthor{\bsnm{Kupfer}, \binits{K.}},
\bauthor{\bsnm{Bers}, \binits{A.}},
\bauthor{\bsnm{Ram}, \binits{A.K.}}:
\batitle{The cusp map in the complex-frequency plane for absolute instability}.
\bjtitle{Phys. Fluids}
\bvolume{30}(\bissue{10}),
\bfpage{3075}--\blpage{3082}
(\byear{1987})
\end{barticle}
\endbibitem

\bibitem{Atalik2002}
\begin{barticle}
\bauthor{\bsnm{Atalik}, \binits{K.}},
\bauthor{\bsnm{Keunings}, \binits{R.}}:
\batitle{Non-linear temporal stability analysis of viscoelastic plane channel
  flows using a fully-spectral method}.
\bjtitle{J. non-Newt. Fluid Mech.}
\bvolume{102},
\bfpage{299}--\blpage{319}
(\byear{2002})
\end{barticle}
\endbibitem

\bibitem{Rabaud1988}
\begin{barticle}
\bauthor{\bsnm{Rabaud}, \binits{M.}},
\bauthor{\bsnm{Couder}, \binits{Y.}},
\bauthor{\bsnm{Gerard}, \binits{N.}}:
\batitle{Dynamics and stability of anomalous saffman-taylor fingers}.
\bjtitle{Phys. Rev. A}
\bvolume{37},
\bfpage{935}--\blpage{947}
(\byear{1988})
\end{barticle}
\endbibitem

\bibitem{Larson2000}
\begin{barticle}
\bauthor{\bsnm{Larson}, \binits{R.G.}}:
\batitle{Turbulence without inertia}.
\bjtitle{Nature}
\bvolume{405},
\bfpage{27}--\blpage{28}
(\byear{2000})
\end{barticle}
\endbibitem

\bibitem{Goldstein1993}
\begin{barticle}
\bauthor{\bsnm{Goldstein}, \binits{R.E.}},
\bauthor{\bsnm{Pesci}, \binits{A.I.}},
\bauthor{\bsnm{Shelley}, \binits{M.J.}}:
\batitle{Topological transitions and singularities in viscous flows}.
\bjtitle{Phys. Rev. Lett.}
\bvolume{70}(\bissue{20}),
\bfpage{3043}--\blpage{3047}
(\byear{1993})
\end{barticle}
\endbibitem

\bibitem{Riley1988}
\begin{barticle}
\bauthor{\bsnm{Riley}, \binits{J.J.}},
\bauthor{\bsnm{Hak}, \binits{M.G.}},
\bauthor{\bsnm{Metcalfe}, \binits{R.W.}}:
\batitle{Complaint coatings}.
\bjtitle{Ann. Rev. Fluid Mech.}
\bvolume{20},
\bfpage{393}--\blpage{420}
(\byear{1988})
\end{barticle}
\endbibitem

\bibitem{Nandagopalan2018}
\begin{barticle}
\bauthor{\bsnm{Nandagopalan}, \binits{P.}},
\bauthor{\bsnm{John}, \binits{J.}},
\bauthor{\bsnm{Baek}, \binits{S.W.}},
\bauthor{\bsnm{Miglani}, \binits{A.}},
\bauthor{\bsnm{Ardhianto}, \binits{K.}}:
\batitle{Shear-flow rheology and viscoelastic instabilities of ethanol gel
  fuels}.
\bjtitle{Exp. Thermal Fluid Sci.}
\bvolume{99},
\bfpage{181}--\blpage{189}
(\byear{2018})
\end{barticle}
\endbibitem

\bibitem{Zarabadi2019}
\begin{botherref}
\oauthor{\bsnm{Zarabadi}, \binits{M.}}:
Development of a robust microfluidic electrochemical cell for biofilm study in
  controlled hydrodynamic conditions.
PhD thesis,
Univ. Laval
(2019)
\end{botherref}
\endbibitem

\bibitem{Zarabadi2018}
\begin{barticle}
\bauthor{\bsnm{Zarabadi}, \binits{M.P.}},
\bauthor{\bsnm{Charette}, \binits{S.J.}},
\bauthor{\bsnm{Greener}, \binits{J.}}:
\batitle{Flow-based deacidification of geobacter sulfurreducens biofilms
  depends on nutrient conditions: a microfluidic bioelectrochemical study}.
\bjtitle{Chem. Electrochem.}
\bvolume{5}(\bissue{23}),
\bfpage{3645}--\blpage{3653}
(\byear{2018})
\end{barticle}
\endbibitem

\end{thebibliography}

\end{document}